\documentclass[preprint,3p,12pt, authoryear
]{elsarticle}

\journal{Preventive Veterinary Medicine}
\usepackage[T1]{fontenc}
\usepackage{mathtools}

\usepackage{booktabs}
\usepackage{array}
\usepackage{tabularx}
\usepackage{longtable}
\newcolumntype{L}[1]{>{\raggedright\arraybackslash}p{#1}}
\usepackage{threeparttable} 

\usepackage{algorithm}
\usepackage{algorithmic}

\usepackage{graphicx}
\usepackage{float}
\usepackage{pdflscape}
\usepackage{multirow}

\usepackage[symbol]{footmisc}
\usepackage{enumitem}
\usepackage{changepage}
\usepackage{xcolor}
\usepackage{placeins}
\usepackage{appendix}

\usepackage[hidelinks]{hyperref}

\biboptions{authoryear} 
\usepackage{booktabs}
\usepackage{tabularx}
\usepackage{siunitx} 

\begin{document}

\begin{frontmatter}

\title{Descriptive and risk analysis of vehicle movements linked to porcine reproductive and respiratory syndrome and porcine epidemic diarrhea transmission in US commercial swine farms}

\author[1]{Jason A. Galvis} 
\author[1]{Taylor B. Parker} 
\author[2]{Cesar A. Corzo} 
\author[1]{Juliana B. Ferreira} 
\author[1]{Kelly A. Meiklejohn} 
\author[1]{Gustavo Machado\corref{cor2}}
\ead{gmachad@ncsu.edu}

\cortext[cor2]{Corresponding Author.}

 \affiliation[1]{organization={Department of Population Health and Pathobiology},
             addressline={North Carolina State University}, 
             city={Raleigh},
             state={NC},
             country={USA}}

\affiliation[2]{organization={Veterinary Population Medicine Department, College of Veterinary Medicine},
             addressline={University of  Minnesota}, 
             city={St. Paul},
             state={MN},
             country={USA}}
        
\begin{abstract}

Vehicle movements, including truck cabs and trailers, play a role in disseminating swine diseases. However, there are significant information gaps regarding vehicle movement patterns and their role in disease transmission, which are crucial for developing better preventive strategies. In this study, we described the movement pattern of truck cabs and trailers and identified risk factors for porcine reproductive and respiratory syndrome (PRRS) and porcine epidemic diarrhea (PED) farms' infectious status. We collected global positioning system (GPS) movement data from truck cabs and trailers for 18 months and infection information of 6621 commercial swine farms in the U.S. For the vehicle movement data, we calculated 66 variables and evaluated their association with farms' PRRS and PED status. Our univariate analysis showed that 56 variables were significantly associated (p < 0.05) with PRRS and PED farm status. Within these variables, vehicle visit frequency and previous exposure to positive farms were the main risk factors for both diseases. Otherwise, increased truck cab and trailer network loyalty for farm shipments and vehicle cleaning and disinfection events were protective factors. In the multivariate model, each additional weekly visit by a truck cab that had been exposed to a positive farm one day before shipments were associated with a 234\% and 243\% increase in the odds of a farm testing PRRS- and PED-positive, respectively. Our analysis revealed that vehicle contact history played a crucial role in the transmission of PRRS and PED. These findings contribute to  the development of strategies aimed at reducing the transmission and outbreaks linked to vehicle movements in swine production.

\end{abstract}


\end{frontmatter}

\section*{Introduction}

Similar to the movement of pigs between farms, the movement of vehicles, including truck cabs and trailers (Figure \ref{fig:vehicle_scheme}), have been associated with the transmitting of diseases between swine production systems not only via trucks moving pigs, but also feed, humans, and other farm-related materials \citep{yoo_transmission_2021, galvis_modelling_2022, vanderwaal_role_2018, martinez_truck_2025}. Vehicles travel to infected farms, slaughterhouses, and other swine facilities, exposing the vehicle's exterior (e.g., tires and fender) and interior (e.g., floor mats) to infectious pathogens \citep{parker_evaluation_2025, dee_mechanical_2002, bartolome_environmental_2025}, and potentially transporting them across a large number of farms \citep{galvis_role_2024}. In the U.S., porcine epidemic diarrhea (PED) and porcine respiratory and reproductive syndrome (PRRS) cause substantial economic losses to the swine industry \citep{schulz_assessment_2015, weng_economic_2016, neumann_assessment_2005, nathues_cost_2017} and their transmission has been associated with several dissemination routes, including vehicle movements \citep{vanderwaal_role_2018, galvis_modelling_2022, galvis_modeling_2022}. Thus, identifying the role of trucks and trailer and risk factors associated with the introduction of PED and PRRS into swine farms could lead to strengthened biosecurity protocols and develop new interventions to reduce disease propagation.

In previous studies describing the role of vehicle movements, it was estimated that vehicles contributed up to 42\% for the propagation of PED and up to 20\% for PRRS \citep{ galvis_modelling_2022, galvis_modeling_2022}. Additionally, other studies have identified transportation vehicles as a significant risk factor for disease transmission in animal production systems. For example, in the case of foot-and-mouth disease, the odds of infection were 5.1 times higher for farms visited by feed vehicles compared to uninfected farms \citep{muroga_risk_2013}. For African swine fever (ASF) in Europe, the return of trucks from affected regions was associated with a 65\% higher risk of disease introduction \citep{mur_risk_2012}. In the case of PRRS and PED, limited on-farm biosecurity related to vehicle management was also associated with an increased likelihood of disease presence \citep{sasaki_development_2020}. Despite the critical role vehicles play in disease transmission, there are still several knowledge gaps regarding events that may increase the risk of disease introduction, such as previous contact with infected farms, frequency of visits, and contact patterns within vehicle movement networks. Similarly, uncertainty remains about the events that may reduce vehicle risks, such as the frequency of cleaning and disinfection (C\&D) and the consistency of farm visits by the same vehicles during shipments \citep{parker_evaluation_2025, martinez_truck_2025, galvis_role_2024}. Furthermore, the interactions between truck cabs and trailers have been poorly explored. Truck cabs and trailers may each play a distinct role in the transmission of PED and PRRS, potentially presenting different challenges for the design of control measures. These uncertainties surrounding the role of vehicles in the spread of infectious diseases may limit the effectiveness of current disease control practices. 

In our current understanding of PRRS and PED infection dynamics, factors such as farm density, production type, and a farm's pig capacity have been associated with an increased likelihood of infection \citep{mortensen_risk_2002, sanhueza_spatial_2020, sykes_interpretable_2022}. In contrast, stricter biosecurity measures, such as minimizing the entry of visitors and vehicles, have been shown to reduce this risk \citep{alarcon_biosecurity_2021, fleming_enhancing_2025}. This knowledge is essential for estimating the probability of farm infection and for designing targeted control strategies to minimize disease transmission. Similarly, identifying vehicle movement patterns associated with the higher risk of PRRS and PED transmission can complement our current understanding of risk factors and strengthen control strategies. To this end, we conducted a descriptive and risk factor analysis of vehicle cabs and trailer movements associated with PED and PRRS infection on swine farms in the U.S. This analysis included events associated with farm visits, C\&D, and interactions between infected farms through the contact network via truck cabs and trailer movements.

\section*{Methodology}

\subsection*{Data}
We collected commercial swine farm-level data from 6621 farms across 30 swine companies in 26 U.S. states. For each farm, we obtained the unique identification, farm type, and farm capacity (number of pigs space available). In addition, we collected the perimeter area of each barn, referred to as the line of separation (LOS), using the Rapid Access Biosecurity application (RABapp™)\citep{fleming_enhancing_2025}. The LOS data were obtained from farm maps included in enhanced on-farm Secure Pork Supply (SPS) biosecurity plans \citep{machado_rapid_2023}, which are part of an initiative improving business continuity by helping swine producers implement robust biosecurity measures \citep{fleming_enhancing_2025}. We also collected 1.1 billion global positioning system (GPS) records from truck cabs and trailers operated by four swine companies (Figure \ref{fig:vehicle_scheme} and Table \ref{tab:data_summary}). These records were collected every five seconds between January 1, 2024, and July 30, 2025, for both truck cabs and trailers. Each GPS record includes the vehicle ID, date, time, speed, latitude, longitude, and external temperature for vehicle cabs. To recover trailer temperature, we downloaded daily temperature raster layers between January 1, 2024, and July 30, 2025, with 1 km$^{2}$ resolution from Google Earth Engine \citep{abatzoglou_development_2013}. In addition, we collected the centroid coordinates and perimeter area from 37 swine companies' own clean stations (CCS) and 32 slaughterhouses serving these four companies. For PRRS and PED outbreak records, reports were obtained from the Morrison Swine Health Monitoring Program (MSHMP) for the period from January 1, 2020, to July 30, 2025, along with data directly shared by companies' outbreak monitoring reports. Each outbreak record contains the farm's unique identification and the date the farm became infected \citep{mshmp_prrs_2025}. 

\begin{table}[htbp]
\centering
\caption{Summary of truck cabs and trailers contacts from January 1st, 2024 to July 30, 2025}
\label{tab:data_summary}
\begin{tabularx}{\textwidth}{>{\centering\arraybackslash}p{0.7cm} p{3.0cm} *{6}{>{\centering\arraybackslash}X} | >{\centering\arraybackslash}X}
\toprule
& & \multicolumn{6}{c|}{\textbf{Truck cabs}} & \multicolumn{1}{c}{\textbf{Trailers}} \\
\cmidrule(lr){3-8}
& Metric & Crew Vehicle & Truck Feed & Truck Market & Truck Pig & Truck Undefined & Total & Total \\
\midrule
\multirow{5}{*}{\rotatebox[origin=c]{90}{General}} 
& GPS records & $\sim$57M & $\sim$78M & $\sim$146M & $\sim$520M & $\sim$116M & $\sim$919M & $\sim$224M \\
& Vehicles & 72 & 44 & 112 & 330 & 203 & 761 & 584 \\
& Farm visits & 38283 & 53021 & 58540 & 368774 & 39696 & 558314 & 560345 \\
& C\&D visits & 13731 & 255 & 57036 & 72898 & 24548 & 168468 & 207213 \\
& Market visits & 3147 & 941 & 70052 & 2261 & 18731 & 95132 & 95769 \\
\midrule
\multirow{6}{*}{\rotatebox[origin=c]{90}{PRRS network}} 
& Nodes & 1224 & 414 & 1223 & 1856 & 1866 & 2770 & 1491 \\
& Nodes - C\&D & 1223 & 414 & 1223 & 1856 & 1866 & 2770 & 1488 \\
& Edges & 482120 & $\sim$1.5M & $\sim$1.6M & $\sim$16.8M & 887.261 & $\sim$21.3M & $\sim$13.7M \\
& Edges - C\&D & 448226 & $\sim$1.5M & $\sim$1.5M & $\sim$15.9M & 806155 & $\sim$20.2M & $\sim$11.6M \\
& Median edges per vehicle & 6692 & 42611 & 15936 & 13437 & 3581 & 9704 & 1358 \\
& Median edges per vehicle - C\&D  & 5977 & 42611 & 14224 & 13291 & 3122 & 9319 & 770 \\
\midrule
\multirow{6}{*}{\rotatebox[origin=c]{90}{PED network}} 
& Nodes & 1215 & 413 & 1213 & 1853 & 1852 & 2758 & 1484 \\
& Nodes - C\&D & 1215 & 413 & 1213 & 1853 & 1852 & 2758 & 1481 \\
& Edges & 257583 & 818674 & 927053 & $\sim$9.4M & 532982 & $\sim$11.9M & $\sim$8.1M \\
& Edges - C\&D & 240575 & 818362 & 865646 & $\sim$9.0M & 485569 & $\sim$11.5M & $\sim$7.0M \\
& Median edges per vehicle & 3425 & 23458 & 9487 & 7586 & 2172 & 5505 & 814 \\
& Median edges per vehicle - C\&D & 3207 & 23458 & 8944 & 7586 & 1882 & 5297 & 524 \\
\bottomrule
\end{tabularx}
\end{table}

\subsection*{Disease status definition}

PRRS and PED outbreak records included positive laboratory results, indicating that the farm was positive based on laboratory result date \citep{mshmp_prrs_2025}. However, the exact timing of when a farm transitioned to a negative status is often uncertain \citep{mshmp_prrs_2025, paiva_description_2024}. When multiple reports were available for the same farm over time, we defined positive and negative infectious dates based on the sequence and timing of these disease records \citep{mshmp_prrs_2025}. First, we classified each disease record as either an outbreak or a monitoring event. An outbreak was defined when two consecutive positive reports were separated by more than a disease-specific time threshold, described later. Conversely, if the interval between consecutive reports was shorter than the threshold, the record was considered part of the same monitoring period. Then, we defined farms with negative status when the elapsed time since the last positive record exceeded the disease-specific threshold. These thresholds were determined by production type as follows: breeding farms, these thresholds reflect the estimated times to achieve disease stability, based on recent regional literature that reported 38 weeks for PRRS \citep{paiva_description_2024} and 14 weeks for PED \citep{corzo_developing_2024}. For growing pig farms, the thresholds reflected typical all-in/all-out schedules and the standard time required to move animals through the commercial production system: seven weeks for nursery farms and 25 weeks for finisher farms. For example, a sow farm with a positive PRRS test result on January 1, 2024, remained classified as positive for the following 38 weeks unless a subsequent positive result was recorded sooner. If no additional positive results were reported during the period, the farm’s status was changed to negative on September 23, 2024. Using this classification, we estimated the daily disease status of each farm from January 1, 2020, to July 30, 2025, labeling farms as positive if the most recent record was classified as outbreak or monitoring, and as negative if the last qualifying record indicated a negative status.

\subsection*{Vehicle visit events}

We defined a premises visit as a risk event in which a vehicle, truck cab, or trailer has a probability of introducing PRRS and or PED. A visit was registered when a vehicle's GPS tracking system registered a geolocation within 50 meters of any premises' barn(s). This threshold was based on the typical length of swine transportation vehicles \citep{whiting_minimum_2024}. Additionally, we considered only stops lasting at least 1 minute within a 50-meter radius, excluding transient movements (i.e., brief pass-through events in which a vehicle entered the 50-meter buffer but stopped for less than one minute). 

\subsection*{Between-farm contact network by vehicle movements with C\&D events}

\begin{figure}[H]
\centering
\includegraphics[width=0.99\textwidth]{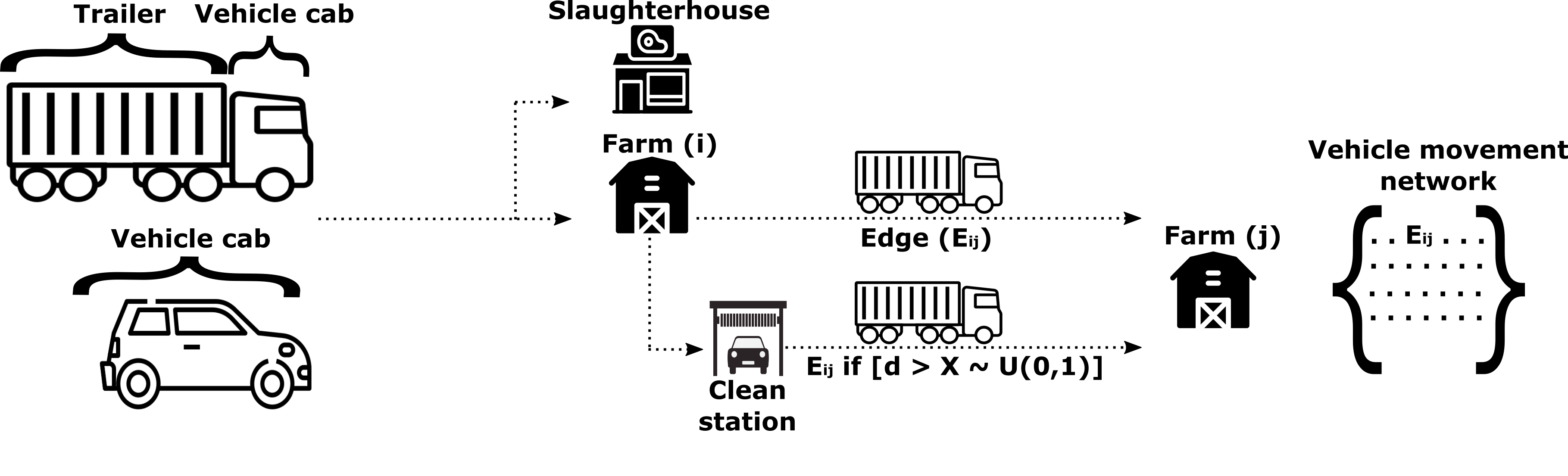}
\caption{Vehicles can visit farms, slaughterhouses, or clean stations. A subsequent visit to farm $j$ creates a network edge ($E_{i,j}$) between farms, with a weight determined by the estimated pathogen stability. The edge weight was calculated as $E_{i,j} = \exp\!\left(-\lambda \, \omega_{i,j} \, \Gamma_{i,j}\right)$ if $d > X \sim U(0,1)$, where $\Gamma_{i,j}$ is the cumulative elapsed time between farms $i$ and $j$, $\omega_{i,j}$ is the average temperature between farms $i$ and $j$, and $\lambda$ is the pathogen stability decay rate for each $\omega$. If a movement to a cleaning station was recorded before visiting farm $j$, the edge in the network was recorded only if the cleaning effectiveness $d$ was higher than a random value $X$ drawn from a uniform distribution $U(0,1)$.}
\label{fig:vehicle_scheme}
\end{figure}

\begin{figure}[ht]
  \centering
  \includegraphics[width=\linewidth]{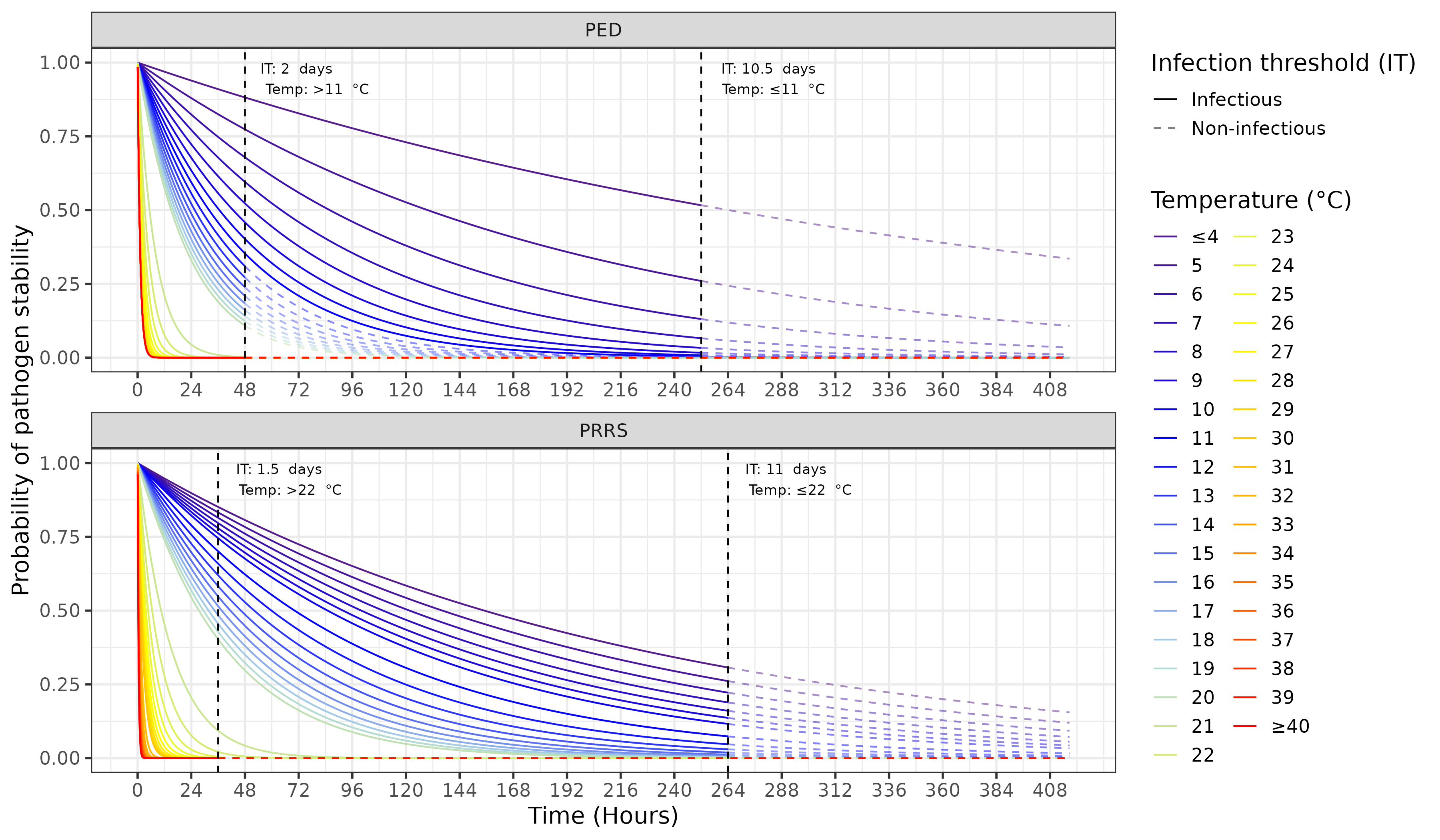}  
  \caption{Assumed stability curves for porcine epidemic diarrhea (PED) and porcine reproductive and respiratory syndrome (PRRS) viruses across time and temperature. Each line represents the decay in pathogen stability at a specific temperature, modeled using an exponential distribution. Line types (solid and dashed) indicate the infectious period, marked by a vertical line that denotes the threshold beyond which the pathogens are considered non-infectious.}
  \label{fig:stability}
\end{figure}

We assumed that vehicles became contaminated after visiting a premises' barn with positive PRRS and/or PED status \citep{dee_mechanical_2002, bernini_when_2019}, thus transferring pathogen(s) to subsequently visited premises. To represent these risk movement events, we computed the sequence of farm-to-farm contacts made by each vehicle and trailer in chronological order and defined these vehicles as edges ($E$) in each vehicle's temporal network (Figure \ref{fig:vehicle_scheme}). Each edge was assigned a weight ($E_w$), representing the pathogen's stability on the vehicle surface. Here we estimated $E_w$ for PRRS and PED, assuming the $E_w$ decayed over time following an exponential distribution (Figure \ref{fig:stability}). The decay rate was time- and temperature-dependent, with faster degradation at higher temperatures and longer elapsed time between farm contacts. Edge weights ranged from 1 (high pathogen stability) to 0 (no pathogen stability). For negligible edge probability values, we set edge weights below 0.001 to 0 for computational efficiency. In addition, Parker et. al., 2025 reported detecting PED viral RNA on vehicle surfaces via reverse transcription-quantitative PCR and subsequently exposing pigs to such samples to estimate infection probability \citep{parker_evaluation_2025_b}. While viral RNA was often detected, not all contaminated vehicles were able to cause infection in pigs, likely due to loss of viral integrity or insufficient quantity over time \citep{parker_evaluation_2025_b}. To capture this outcome, beyond the exponential decay of stability, we incorporated a maximum infectious time threshold. This threshold assumes that even if viral RNA remained detectable on vehicle surfaces, the virus may no longer retain sufficient integrity or concentration to cause infection. Reference values for pathogen stability decay and infectious time thresholds are provided in Figure \ref{fig:stability} and Table \ref{tab:data_stability}.

We simulated C\&D effectiveness, defined as the proportion of vehicles successfully disinfected after visiting a CCS, which ranged from 0\% to 100\%. To identify vehicles contacting a CCS, we defined a visit as a complete stop (0 km/h) within 50 meters of a CCS perimeter area, lasting at least 60 minutes for feed truck cabs, and 45 minutes for other vehicle cabs and trailers \citep{parker_evaluation_2025} (Figure \ref{fig:vehicle_scheme}). Based on previous estimates \citep{parker_evaluation_2025_b}, we assumed C\&D effectiveness of 13\% for truck cabs and 40\% for trailers. The 13\% estimate originated from feed truck tire samples, which evaluated animal infection linked to contaminated vehicles after C\&D. However, in the absence of effectiveness data for other truck cab types, we extrapolated the 13\% efficacy to all truck cabs. Similarly, the 40\% estimate was derived from trailer samples previously exposed to PED-positive farms. Although Parker et. al., 2025 provided reference values for specific trailer types, this label was not available for the trailer data used in this study. Therefore, we assumed 40\% for all trailers, reasoning that it reflects C\&D performance in trailers known to come from infected farms, which are typically subject to more intensive cleaning protocols \citep{parker_evaluation_2025_b}.

Ultimately, we reconstructed the between-farm vehicle movement network, with and without C\&D, using static and temporal contact network representations. The static network aggregates vehicle movements over the entire study period into a graph, providing an overview of long-term connectivity among farms. In contrast, a temporal network retains the order and timing of movements, capturing day-to-day contact dynamics and the potential for sequential disease transmission. 

\begin{table}[ht]
\centering
\caption{Summary of virus stability and infectious thresholds.}
\label{tab:data_stability}
\begin{tabularx}{\textwidth}{@{}lL{5cm}L{9cm}@{}}
\toprule
\textbf{Pathogen} & \textbf{Stability decay metric} & \textbf{Assumed infectious threshold} \\
\midrule
PRRS virus & Half-life: 6.5 days at $4^\circ$C, 3.5 days at $10^\circ$C, 27.4 hours at $20^\circ$C, 1.6 hours at $30^\circ$C \citep{jacobs_stability_2010}, and 16 minutes at $40^\circ$C \citep{mil-homens_assessment_2024}. & Recovered from fomites for 36 hours at $22$–$25^\circ$C \citep{quinonez-munoz_comparative_2024}. We assumed infectivity lasts 36 hours at temperatures $>22^\circ$C. In the absence of data for lower temperatures, we assumed up to 11 days of infectivity at $\leq22^\circ$C. \\
PED virus & Half-life: 11 days at $4^\circ$C (based on SARS-CoV-2) \citep{guang_determining_2023}, 15 hours at $20^\circ$C, 47 minutes at $30^\circ$C and 46 minutes at $40^\circ$C \citep{mil-homens_assessment_2024} & Described as infectious up to 14 days at 4°C, 7 days at 12°C and less than 7 days at 20-22°C, but more than 24 hours at 20°C \citep{pujols_survivability_2014, thomas_evaluation_2015}. We assumed a threshold at 11°C: temperatures $\leq11^\circ$C allow 10.5 days of infectivity, while $>11^\circ$C allow only 2 days.\\
\bottomrule
\end{tabularx}
\end{table}

\subsection*{Network connectivity analysis}

We characterized the connectivity of the vehicle's movement network using its temporal representation. First, we quantified the number of edges and edge saturation through the temporal density metric, defined as the ratio of the sum of daily observed directed edges to the sum of the maximum possible directed edges among active nodes on each day. Second, we evaluated reachability by measuring the probability that a farm could reach other farms via time-respecting paths, using earliest-arrival paths constrained by the chronological order of edges. Reachability was assessed across the total study period using the total average reachability (TAR) and the monthly average reachability (MAR). These metrics describe the fraction of farms connected through vehicle movements, either cumulatively or monthly, while accounting for the temporal order of contacts. Together, the temporal contact network metrics used provided insight into the potential speed and extent of pathogen spread across the network.

\subsection*{Risk factor analysis}

Given the limited number of PRRS-reported positive farms in growing pig production types (e.g., finisher, nursery) (Supplementary Material Figure S1), descriptive and risk factor analyses were restricted to sow farms for PRRS and PED. Thus, our binary outcome was whether a sow farm was classified as PRRS- or PED-positive at least once between January 1, 2024, and July 30, 2025. To identify farm-level risk and protective factors associated with farms' PRRS and PED status, we conducted a descriptive analysis using the variables listed in Table \ref{tab:farm_variables}.

We assessed the relationship between predictors listed in Table \ref{tab:farm_variables} and the farms' PRRS and PED status using a univariate logistic regression model. Predictors with a p-value < 0.2 in the univariate analysis were included in a multivariable logistic regression model. The selection of predictors in the multivariate model was based on statistical significance (p < 0.05) and model fit, measured by the Akaike Information Criterion (AIC), following a backward elimination approach. Farm density (Table \ref{tab:farm_variables}) was retained in the final multivariate models as a fixed effect regardless of its statistical significance, given its known relevance as a variable that explains PRRS and PED transmission dynamics \citep{machado_identifying_2019}. Multicollinearity among predictors was evaluated using the variance inflation factor (VIF), with VIF values > 5 considered indicative of collinearity and predictors with such values excluded from the multivariate model. Ultimately, to enhance the interpretability of effect sizes, odds ratios (OR) from the logistic regression models were converted to percentage changes using the transformation \textit{(OR - 1) x 100}. This approach provides an intuitive understanding of the magnitude and direction of associations, particularly for OR values < 1 or close to 1. The original OR values and their confidence intervals were provided in the Supplementary Material for reference. All data processing and analyses were conducted using R version 4.2.3 \citep{r_core_team_r_2023}.

\begin{table}[htbp]
\centering
\caption{Farm-level variables used in the descriptive and risk factor analyses}
\label{tab:farm_variables}
\renewcommand{\arraystretch}{1.1} 
\small 
\begin{tabularx}{\textwidth}{|p{0.24\textwidth}|X|}
\hline
\textbf{Variable} & \textbf{Description and metric} \\
\hline
Farm capacity & The maximum number of pigs a farm can accommodate. \\
\hline
Farm neighbor density & Number of farms located within a 10 km radius. \\
\hline
Vehicle visit frequency & Frequency of vehicle cabs and trailer arrivals at the farm. Measured as the median number of visits per week and month, and cumulative visits over 36 months. \\
\hline
Exposed vehicle visit frequency & Frequency of vehicles that had previously visited positive farms before arriving at the target farm. Exposure was tracked for 1, 7, and 11 days prior to arrival and measured as the median number of visits per week and per month, and as cumulative visits over 36 months. \\
\hline
C\&D vehicle visits & Proportion of visits where the vehicle was cleaned and disinfected prior to arrival. Calculated as (visits with prior C\&D) / (total farm visits). \\
\hline
Slaughterhouse visits & Number of visits where the vehicle had visited a slaughterhouse 1, 7, or 11 days prior to arrival. Measured as cumulative visits over 36 months. \\
\hline
Grower visits & Number of visits where the vehicle had visited a finisher or wean-to-finisher farm 1, 7, or 11 days prior to arrival. Measured as cumulative visits over 36 months. \\
\hline
Vehicle–farm loyalty & Repeated visits by the same vehicle (vehicle cab or trailer) to the same farm. Measured as the median number of visits per day, week, and month, and cumulative visits over 36 months. \\
\hline
Cab–trailer–farm loyalty & Repeated visits by the same vehicle cab–trailer pair to the same farm. Measured as the median number of visits per day, week, and month, and cumulative visits over 36 months. \\
\hline
In-degree & Cumulative number of distinct farms that connected to the target farm in the static vehicle movement network with and without C\&D.\\
\hline
Infected in-degree & Cumulative number of disease positive farms that connected to the target farm in the static vehicle movement network with and without C\&D.. \\
\hline
Community size & Number of farms in the same vehicle network (with and without C\&D) community as the target farm. Community structures were identified using the Louvain modularity optimization algorithm \citep{blondel_fast_2008}. This algorithm partitions the network into communities by maximizing modularity, a measure of the density of links within communities relative to those between communities.\\
\hline
Infected community size & Number of disease-positive farms in the same vehicle network community as the target farm, using the same approach applied to community size  \\
\hline
Median pathogen stability from disease-positive farms & Median pathogen stability received by a farm through connections from disease-positive farms through the temporal vehicle movement network. \\
\hline
\end{tabularx}
\end{table}

\section*{Results}

Between January 1st, 2024, and July 30th, 2025, we analyzed 919 million GPS records from $761$ truck cabs and $224$ million from 584 trailers (Table \ref{tab:data_summary}). These vehicles completed over $558000$ farm visits, with truck pig cabs accounting for the majority (66\%). Additional visits included over $168000$ truck cab and $207000$ trailer visits to C\&D stations, and over $95000$ visits to slaughterhouses by both truck cabs and trailers.

Among the $6621$ farms evaluated for interactions with vehicle movements, truck cabs connected 40\% of farms in the PRRS and PED networks, while trailers connected 22\% (Table \ref{tab:data_summary}). When C\&D events were incorporated using the assumed efficacy of vehicle disinfection, truck cabs connected the same number of farms, while for trailers, this number was reduced by three farms (Table \ref{tab:data_summary}).

The PRRS network contained approximately 21.3 million truck cab-mediated edges and 13.7 million trailer-mediated edges, whereas the PED network included 11.9 million and 8.1 million edges, respectively (Table \ref{tab:data_summary}). Most truck cab-mediated edges were associated with pig transport between farms, accounting for more than 79\% of truck cab-related movements in the PRRS and PED networks. After accounting for C\&D, over 95\% of PRRS and 97\% of PED cab-mediated edges remained, compared with only 85\% and 86\% of trailer-mediated edges, respectively (Table \ref{tab:data_summary}). Similarly, the median number of edges generated by each vehicle decreased. Overall, this reduction was 4\% for truck cabs in the PRRS and PED networks, while for trailers it reached 43\% in the PRRS network and 35\% in the PED network.

Overall, the temporal network density of truck cabs and trailers was low, with only about 1\% of the possible farm-to-farm edges observed on any given day (Table \ref{tab:data_metrics}). When evaluating the proportion of farms that could be reached through time-respecting paths across the entire study period, an average of 73–75\% of farms were reachable via truck cab movements and 81–82\% via trailer movements. After incorporating C\&D, this reachability decreased 1\% for truck cabs in the PRRS network, while no reduction was observed for truck cabs in the PED network and for trailers in the PRRS and PED networks. At monthly intervals, the average reachability was 29–30\% for truck cabs and 71–73\% for trailers, without declining after accounting for C\&D (Table \ref{tab:data_metrics}).

\begin{table}[H]
\centering
\begin{threeparttable}
\caption{Temporal network connectivity metrics by truck cabs and trailers movements from January 1st, 2024 to July 30, 2025}
\label{tab:data_metrics}
\begin{tabularx}{\textwidth}{>{\centering\arraybackslash}p{0.8cm} p{3.0cm} *{6}{>{\centering\arraybackslash}X} | >{\centering\arraybackslash}X}
\toprule
& & \multicolumn{6}{c|}{\textbf{Vehicle cabs}} & \multicolumn{1}{c}{\textbf{Trailers}} \\
\cmidrule(lr){3-8}
 & Network metric & Crew Vehicle & Truck Feed & Truck Market & Truck Pig & Truck Undefined & Total & Total \\
\midrule
\multirow{6}{*}{\rotatebox{90}{PRRS}} 
& Density & 0.01 & 0.05 & 0.02 & 0.01 & 0.01 & 0.01 & 0.01 \\
& Density - C\&D & 0.01 & 0.05 & 0.02 & 0.01 & 0.01 & 0.01 & 0.01 \\
& TAR & 0.88 & 0.66 & 0.53 & 0.60 & 0.23 & 0.75 & 0.81 \\
& TAR - C\&D & 0.88 & 0.65 & 0.53 & 0.60 & 0.23 & 0.74 & 0.81 \\
& MAR & 0.05 & 0.54 & 0.08 & 0.51 & 0.01 & 0.30 & 0.73 \\
& MAR - C\&D & 0.05 & 0.54 & 0.07 & 0.51 & 0.01 & 0.30 & 0.73 \\
\midrule
\multirow{6}{*}{\rotatebox{90}{PED}} 
& Density & 0.01 & 0.04 & 0.02 & 0.01 & 0.01 & 0.01 & 0.01 \\
& Density - C\&D & 0.01 & 0.04 & 0.02 & 0.01  & 0.01 & 0.01 & 0.01 \\
& TAR & 0.68 & 0.65 & 0.52 & 0.60 & 0.22 & 0.73 & 0.82 \\
& TAR - C\&D & 0.68 & 0.65 & 0.52 & 0.60 & 0.22 & 0.73 & 0.82 \\
& MAR & 0.03 & 0.54 & 0.06 & 0.50 & 0.01 & 0.29 & 0.71 \\
& MAR - C\&D & 0.03 & 0.54 & 0.06 & 0.50 & 0.01 & 0.29 & 0.71 \\
\bottomrule
\end{tabularx}
\begin{tablenotes}[flushleft]
\small
\item TAR = Total average reachability; MAR = Monthly average reachability.
\end{tablenotes}
\end{threeparttable}
\end{table}

\subsection*{Univariate analysis}

After analyzing farms and companies with available disease records and vehicle movement data, we identified only one company with sufficient variability in PRRS and PED status to conduct the association analysis. For PRRS, the dataset comprised 191 sow farms, including 76 classified as negative and 115 as positive. For PED, the analysis included 190 sow farms, of which 122 were classified as negative and 68 as positive.

The univariate logistic regression analysis identified several variables significantly associated with PRRS- and PED-positive farm status (Tables \ref{tab:uni_prrs_table} and \ref{tab:uni_ped_table}, and Supplementary Material Tables S1–S4 and Figures S2-S68). Overall, both direct visit frequency and indirect exposure to infected premises through shared vehicle movements emerged as key drivers of infection risk in this univariate assessment for both diseases. Variables reflecting indirect exposure through truck cab and trailer movements to positive farms over the previous one to 11 days of the observed visit showed the strongest associations. For example, the median weekly number of farm visits by trailers exposed to positive farms one day prior was associated with a 359\% and 299\% increase in the odds of being PRRS- and PED-positive, respectively. Similarly, trailer visits were followed closely by exposure through truck cabs, with corresponding increases of 251\% and 348\% in the odds of PRRS and PED positivity, respectively. Furthermore, truck cabs exposed to slaughterhouses seven and 11 days before a shipment were associated with a 3\% to 4\% increase in the odds of being PRRS- and PED-positive. Network metrics, such as in-degree and community, also showed a significant association, increasing the odds of PRRS and PED positivity by 0.1\%-2\%.

In contrast, truck cab and trailer loyalty metrics were negatively associated with PRRS and PED positivity (Table \ref{tab:uni_prrs_table} and \ref{tab:uni_ped_table}). Median cab loyalty was associated with 80\% and 85\% reductions in the odds of PRRS and PED positivity, respectively, suggesting a protective effect when farms consistently received the same truck cabs during shipments. Although the associations were weaker, the use of C\&D procedures prior to farm visits was also negatively associated with infection status. The proportion of visits in which vehicle cabs underwent C\&D before arriving at a farm was linked to a 5\% and 6\% reduction in the odds of PRRS and PED positivity, respectively (Supplementary Material Tables S1–S4).


\begin{table}[H]
\centering
\small 
\caption{Odds ratio percentage change (ORPC) $(OR - 1) \times 100$ and significance values from the univariate analysis for the top 20 variables associated with PRRS positive farm status}
\label{tab:uni_prrs_table}
\begin{tabularx}{\textwidth}{X l r} 
\toprule
\textbf{Variable} & \textbf{ORPC (95\% CI)} & \textit{p}-value \\
\midrule
Median weekly farm visit by trailers exposed to positive farms 1 day before & 489.28 (263.41 to 941.85) & <0.001 \\
Median weekly farm visit by cabs exposed to positive farms 1 day before & 435.39 (246.80 to 794.95) & <0.001 \\
Median weekly farm visit by trailers exposed to positive farms 7 days before & 171.73 (99.47 to 285.50) & <0.001 \\
Median weekly farm visit by trailers exposed to positive farms 11 days before & 163.88 (94.31 to 272.47) & <0.001 \\
Median weekly farm visit by cabs exposed to positive farms 7 days before & 116.03 (69.98 to 182.97) & <0.001 \\
Median weekly farm visit by cabs exposed to positive farms 11 days before & 100.86 (59.41 to 160.06) & <0.001 \\
Median cab loyalty & $-$85.92 ($-$93.79 to $-$70.83) & <0.001 \\
Median cab-trailer loyalty & $-$85.69 ($-$96.56 to $-$48.09) & 0.010 \\
Median trailer loyalty & $-$78.61 ($-$89.74 to $-$59.37) & <0.001 \\
Median monthly farm visit by trailers exposed to positive farms 1 day before & 48.59 (34.17 to 67.55) & <0.001 \\
Median monthly farm visit by cabs exposed to positive farms 1 day before & 39.59 (28.23 to 54.38) & <0.001 \\
Median monthly farm visit by trailers exposed to positive farms 7 days before & 26.20 (18.24 to 35.85) & <0.001 \\
Median monthly farm visit by trailers exposed to positive farms 11 days before & 24.43 (16.91 to 33.47) & <0.001 \\
Median monthly farm visit by cabs exposed to positive farms 7 days before & 20.02 (13.73 to 27.53) & <0.001 \\
Median monthly farm visit by cabs exposed to positive farms 11 days before & 18.95 (12.78 to 26.32) & <0.001 \\
Median monthly cabs loyalty & $-$18.49 ($-$29.09 to $-$8.68) & <0.001 \\
Median weekly farm visit by trailers & 18.47 ($-$4.26 to 48.91) & 0.130 \\
Median monthly trailer loyalty & $-$15.39 ($-$22.49 to $-$8.51) & <0.001 \\
Median weekly farm visit by cabs & 15.04 ($-$3.87 to 39.30) & 0.140 \\
Median monthly cab-trailer loyalty & $-$13.31 ($-$20.57 to $-$6.48) & <0.001 \\
\bottomrule
\end{tabularx}
\end{table}


\begin{table}[H]
\centering
\caption{Odds ratio percentage change (ORPC) $(OR - 1) \times 100$ and significance values from the univariate analysis for top 20 variables associated with PED positive farm status}
\label{tab:uni_ped_table}
\begin{tabular}{p{9cm}p{5cm}p{1cm}}
\toprule
\textbf{Variable} & \textbf{ORPC (95\% CI)} & \textit{p} \\
\midrule
Median weekly farm visit by cabs exposed to positive farms one day before the visit & 348.60 (165.33–713.87) & 0.00 \\
Median weekly farm visit by trailers exposed to positive farms one day before the visit & 299.40 (135.43–610.29) & 0.00 \\
Median weekly farm visit by trailers exposed to positive farms seven days before the visit & 164.20 (81.71–298.08) & 0.00 \\
Median weekly farm visit by trailers exposed to positive farms 11 days before the visit & 112.70 (51.97–205.31) & 0.00 \\
Median weekly farm visit by cabs exposed to positive farms seven days before the visit & 97.80 (46.90–173.73) & 0.00 \\
Median weekly farm visit by cabs exposed to positive farms 11 days before the visit & 92.80 (46.00–161.30) & 0.00 \\
Median cab loyalty & -85.30 (-94.42–-66.10) & 0.00 \\
Median trailer loyalty & -65.40 (-83.59–-34.33) & 0.00 \\
Median cab-trailer loyalty & -47.80 (-85.27–-6.40) & 0.24 \\
Median monthly farm visit by trailers exposed to positive farms one day before the visit & 47.30 (28.55–71.38) & 0.00 \\
Median monthly farm visit by cabs exposed to positive farms one day before the visit & 45.80 (28.76–67.74) & 0.00 \\
Median weekly farm visit by trailers & 21.70 (-1.25–51.48) & 0.07 \\
Median monthly farm visit by trailers exposed to positive farms seven days before the visit & 21.10 (12.16–31.58) & 0.00 \\
Median monthly farm visit by trailers exposed to positive farms 11 days before the visit & 20.00 (11.59–29.83) & 0.00 \\
Median monthly farm visit by cabs exposed to positive farms seven days before the visit & 16.90 (9.90–24.94) & 0.00 \\
Median monthly farm visit by cabs exposed to positive farms 11 days before the visit & 15.60 (9.06–23.19) & 0.00 \\
Median weekly farm visit by cabs & 14.60 (-4.05–37.58) & 0.14 \\
Median monthly cab loyalty & -14.30 (-26.06–-4.03) & 0.02 \\
Median monthly trailer loyalty & -11.30 (-19.25–-4.15) & 0.01 \\
Median monthly cab-trailer loyalty & -9.20 (-17.11–-2.48) & 0.02 \\
\bottomrule
\end{tabular}
\end{table}

\subsection*{Multivariate analysis}

A total of 64 and 60 variables met the $p < 0.2$ threshold in the univariate analysis for PRRS and PED, respectively. For PRRS, 53 variables were excluded due to collinearity with other predictors, while 47 were excluded for PED. Consequently, 11 variables for PRRS and 13 variables for PED were retained in the multivariate logistic regression (Supplementary Material Table S1-S4). In these models, the median weekly number of farm visits by truck cabs that had visited positive farms one day earlier remained strongly associated with increased odds of infection for both diseases (Table \ref{tab:multi_table}). This exposure was associated with a 234\% increase in the odds of being PRRS-positive and a 243\% increase in the odds of being PED-positive. Conversely, a higher proportion of shipments conducted with previously disinfected truck cabs showed a protective effect against PED, with a 4.88\% reduction in the odds of infection (Table \ref{tab:multi_table}). Finally, farm neighbor density was statistically associated with PRRS,  with a 1.4\% increase in the odds of being PRRS-positive, whereas no association was observed with PED infection ($p > 0.05$) (Table \ref{tab:multi_table}).


\begin{table}[H]
\centering
\caption{Odds ratio percentage change (ORPC) $(OR - 1) * 100$ and significance values from the multivariate analysis}
\label{tab:multi_table}
\begin{tabularx}{\textwidth}{>{\raggedright\arraybackslash}Xcc|cc}
\toprule
\textbf{Variable} & \multicolumn{2}{c|}{\textbf{PPRS}} & \multicolumn{2}{c}{\textbf{PED}} \\
 & \textbf{ORPC (95\% CI)} & \textbf{p value} & \textbf{ORPC (95\% CI)} & \textbf{p value} \\
\midrule
Neighbors within 10km radius & 1.37 (0.15–2.66) & 0.03 & -0.30 (-1.47–0.86) & 0.61 \\
Median weekly farm visit by vehicle cabs exposed to positive farms one day before the visit & 233.91 (137.39–395.95) & 0.00 & 243.06 (98.78–521.32) & 0.00 \\
Proportion of farm visits with previous vehicle cab C\&D & & & -4.88 (-8.14–-1.98) & 0.00 \\
\bottomrule
\end{tabularx}
\end{table}

\section*{Discussion}

This study leveraged more than 1.1 billion GPS records from truck cabs and trailers to reconstruct vehicle visits to swine facilities and the connectivity among them through a temporal directed contact network, which was then associated with farms’ PRRS and PED status. Over an 18-month period, the truck cab connected between $2758$ and $2770$ farms in the PRRS and PED movement network, while the trailer movement network connected between $1484$ and $1491$ farms. Truck cabs transporting pigs between farms were central to network connectivity, with the highest number of between-farm connections and enabling 60\% farm reachability via temporal paths in the network. Similarly, temporal paths in the trailer movement network allowed more than 80\% farm reachability. Incorporating C\&D effectiveness reduced 3–5\% of edges from the truck cab network and 14–15\% from the trailer network. Shipments of vehicles with recent exposure to positive farms were identified as risk factors for PRRS and PED infection. In contrast, greater vehicle loyalty (i.e., repeated visits of the same vehicles or trailers to the same farms) and the implementation of C\&D protocols were protective factors, lowering the risk of infection. Overall, these findings underscore the critical role of vehicle-mediated indirect contacts in PRRS and PED transmission and highlight opportunities for targeted interventions to disrupt these pathways.

In the vehicle movement network, truck cabs transporting pigs between farms played a dominant role in maintaining connectivity, accounting for up to 79\% of connections in the total temporal network. Contrary to the findings of \citep{galvis_role_2024}, pig-transport truck cabs were more influential than feed truck cabs in sustaining network connectivity. This discrepancy could be attributed to methodological differences. In \citep{galvis_role_2024} dataset, a large proportion of vehicle cabs classified as “undefined” were likely transporting feed; therefore, combining feed and undefined truck cabs would yield results more consistent with those reported by Galvis and Machado, 2024. Another key methodological difference lies in how networks were constructed. The previous study defined vehicle visits using perimeter buffer areas (PBA) drawn around groups of barns \citep{galvis_role_2024}. In some farms, these PBAs encompassed large areas \citep{fleming_enhancing_2025}, which may have increased the risk of misclassifying vehicle visits and, consequently, overestimating farm–vehicle contacts. By contrast, our current methodology relied on barn-level perimeter areas, offering greater precision in detecting true vehicle visits to farms. Finally, the previous study modeled pathogen stability using ASF, allowing vehicles to generate subsequent connections with other farms for more than 30 days \citep{galvis_role_2024}, whereas here, connections were restricted to 11 days (Table \ref{tab:data_stability}). Together, these methodological differences likely explain the variation observed in the relative role of different vehicle types in shaping network connectivity.

One of the most important findings from the multivariate models for PRRS and PED was that truck cabs and trailers that visited disease-positive farms a few days earlier significantly increased the odds of infection for subsequently visited farms. For vehicles exposed to disease-positive farms one day before a shipment, each additional weekly visit was associated with a 243\% increase in the odds of PED infection and a 234\% increase in the odds of PRRS infection. Similar patterns were observed in the univariate analyses, in which vehicle cabs and trailers exposed to positive farms before a shipment were also associated with increased odds of infection. It is important to note that swine companies in the United States often segregate vehicle movements by farm health status \citep{galvis_mitigating_2025}, aiming to avoid contact between positive and negative farms, especially for vehicles and trailers serving breeding farms. Thus, the observed association between prior exposure to positive farms and higher infection risk may be partially influenced by this practice of movement segregation. Nevertheless, truck cabs and trailers are frequently contaminated during shipments \citep{parker_evaluation_2025, boniotti_porcine_2018}, and a short interval before their next visit may provide suitable conditions for the pathogen to remain stable and infectious on vehicle surfaces \citep{mil-homens_assessment_2024}. These findings highlight that vehicle-mediated contacts likely play a role in environmental contamination and subsequent transmission to negative farms. Future studies incorporating environmental sampling from parking areas, vehicle access routes, and loading zones within farms could provide complementary insights into how contaminated surfaces contribute to farm-level transmission and help design more effective biosecurity protocols.

The number of farms infected in close proximity in the truck cab and trailer network (infected in-degree) and a higher number of infected farms within network communities, although not retained in the multivariate models, were significant in the univariate analysis. For each additional infected farm neighbor in the network, the odds of PRRS and PED infection increased by 0.3\% to 0.7\%. At the community level, each additional infected farm within a network community was associated with a 0.5\% to 1.2\% increase in the odds of PRRS and PED infection. Although other variables better explained farm infectious status in the regression models, the influence of infected neighbors in the vehicle movement network, either directly connected or within the same community, cannot be dismissed. Similar findings have been reported for live animal movements in swine production systems, where in-degree and community structure also represent risk factors \citep{sanchez_spatiotemporal_2023, machado_identifying_2019, gomez-vazquez_evaluation_2019}. Therefore, strategies that reduce overall connectivity between infected and susceptible farms via truck cabs and trailer movements are likely to affect overall PRRS and PED transmission dynamics.

Truck and trailer movement loyalty and C\&D emerged as protective factors in our analysis. Higher vehicle movement loyalty, defined as truck cabs and trailers repeatedly visiting the same farms rather than moving among many different sites, was associated with a reduced odds of PRRS and PED infection. This pattern suggests that limiting the diversity of farm contacts reduces opportunities for cross-farm transmission, a finding consistent with studies of live-animal movements, where greater connectivity across farms increases the probability of pathogen spread \citep{schulz_network_2017, cardenas_analyzing_2024}. Similarly, previous reports have highlighted that restricting vehicle movement to smaller, well-defined farm networks can reduce the risk of disease introduction \citep{galvis_mitigating_2025}. In addition, C\&D between visits significantly reduced the risk of infection, underscoring the critical role of hygiene practices in interrupting indirect transmission pathways. Several experimental studies have demonstrated that PRRS and PED can survive for extended periods on contaminated surfaces \citep{jacobs_stability_2010, guang_determining_2023, mil-homens_assessment_2024}, and field investigations have shown that systematic C\&D protocols can markedly reduce contamination risk \citep{parker_evaluation_2025_b}. Together, these findings indicate that biosecurity practices focused on enhancing vehicle movement loyalty and enforcing strict C\&D protocols were not only protective but may be among the most practical interventions available to reduce the risk of disease spread through vehicle movements.

\subsection*{Limitations and further remarks}

Our study has several limitations that should be considered when interpreting the results. We collected GPS records from four participating companies and cross-referenced these data with $6621$ farms from 30 companies in the U.S. This approach allowed us to evaluate overall interactions among farms through vehicle movements, acknowledging that cross-company contacts occur occasionally. However, including vehicle movements from all companies could yield different associations than those identified here. In addition, although we were informed that the dataset represented a high proportion of the truck cabs and trailers used by the four participating companies, it is likely that some vehicles visiting farms without tracking devices were not captured. The lack of such movement edges may have underestimated the true extent of farm connectivity and vehicle-mediated risk. To strengthen future analyses, it will be important to expand data collection to include vehicle movements from all swine companies and ensure more comprehensive coverage of truck cabs and trailers. Such efforts would provide a more complete representation of vehicle networks in the U.S.

The PRRS and PED stability and maximum infectious thresholds were assumed from the literature, primarily based on experimental studies evaluating pathogen decay through PCR (Table \ref{tab:data_stability}). For pathogen decay, data were generally available for only a limited range of temperatures and time periods. Thus, we extrapolated the missing values using an exponential function (Figure \ref{fig:stability}). In addition, the maximum infectious threshold was estimated from studies reporting the duration of pathogen detectability on different surfaces (Table \ref{tab:data_stability}). However, PCR detection does not necessarily indicate that the virus remains infectious to pigs, and as a result, our estimates of vehicle–farm connectivity may be under- or overestimated \citep{parker_evaluation_2025_b, lugo_mesa_survival_2024}. In the absence of more detailed information to reconstruct vehicle movement networks using pathogen stability, our approach provides a conservative estimate. Specifically, we assumed a higher probability of pathogen stability within the first two days after a vehicle visited a farm (Figure \ref{fig:stability}), followed by continuous exponential decay over an 11-day period. This implies that transmission is most likely from vehicles recently contaminated and progressively less likely toward the end of the 11 days. Thus, future studies should use bioassays to directly assess the infectivity of PRRS and PED over time under varying environmental conditions \citep{parker_evaluation_2025_b}. Such data would provide the basis for defining the maximum period during which contaminated vehicle surfaces could pose an infection risk to pigs.

We assumed C\&D effectiveness of 13\% (for truck cabs) and 40\% (for trailers) based on estimates from Parker et al. (2025), who used a bioassay to detect infected animals from samples collected from truck cabs and trailers contaminated with PED. While this study provided a useful baseline, their study had several limitations for estimating C\&D effectiveness. In particular, the link between contaminated vehicle samples and subsequent infection in pigs was not directly observed, and the estimates relied on a probabilistic model rather than on direct measurements of transmission events. Consequently, the true effect of C\&D could have been either under- or overestimated. Nevertheless, similar results were reported by Boniotti et al. (2018), who found that 46\% of samples collected from truck cabs and trailers remained positive after C\&D \citep{boniotti_porcine_2018}. Taken together, these findings suggest that the C\&D effectiveness assumed in this study may not be far from field conditions. Future research should aim to quantify C\&D effectiveness through controlled experimental trials and longitudinal field studies that directly assess pathogen transmission to pigs.

Despite these limitations, our study provides one of the most comprehensive evaluations to date of how vehicle movements contribute to the spread of PRRS and PED across U.S. swine production systems, particularly to sow farms. By identifying high-risk vehicle types, quantifying the role of prior exposure to positive farms, and highlighting the protective effects of vehicle loyalty and C\&D practices, our findings offer actionable insights that can inform biosecurity planning and risk management, ultimately reducing the risk of disease introduction and improving the resilience of the swine industry.

\section*{Conclusion}

This study demonstrates that vehicle movements, particularly those involving truck cabs transporting pigs between farms, play a critical role in shaping the connectivity of swine production networks and influencing the risk of PRRS and PED transmission. Our results show that prior exposure of vehicles to positive farms significantly increases the odds of PRRS and PED infections, while protective factors such as vehicle loyalty and effective cleaning and disinfection can reduce them. By integrating vehicle movement data into risk assessment and biosecurity planning, swine companies can better target interventions, optimize resource allocation, and reduce the probability of costly outbreaks.

\section*{Acknowledgments}


\section*{Authors’ contributions}				
JAG and GM conceived the study. JAG and GM participated in the study design. CAC collected disease data. JAG prepared the data and developed the analysis. JAG and GM wrote and edited the manuscript. All authors discussed the results and critically reviewed the manuscript. JAG and GM secured the funding.

\section*{Funding statement}
This study was funded by the Wean-to-Harvest Biosecurity Research Program, Swine Health Information Center (SHIC), and Animal and Plant Health Inspection Service (USDA-APHIS)-National Animal Disease Preparedness and Response Program (NADPRP) award number AP23VSSP0000C060.

\section*{Data Availability Statement}		
The data supporting this study's findings are not publicly available and are protected by confidential agreements; therefore, they are not available.

\bibliographystyle{elsarticle-harv}
\bibliography{mybibliography}

\end{document}


\begin{frontmatter}

\title{Descriptive and risk analysis of vehicle movements linked to porcine reproductive and respiratory syndrome (PRRS) and porcine epidemic diarrhea (PED) transmission in US commercial swine farms}

\author[1]{Jason A. Galvis} 
\author[1]{Taylor B. Parker} 
\author[2]{Cesar A. Corzo} 
\author[1]{Juliana B. Ferreira} 
\author[1]{Kelly A. Meiklejohn} 
\author[1]{Gustavo Machado\corref{cor2}}
\ead{gmachad@ncsu.edu}

\cortext[cor2]{Corresponding Author.}

 \affiliation[1]{organization={Department of Population Health and Pathobiology},
             addressline={North Carolina State University}, 
             city={Raleigh},
             state={NC},
             country={USA}}

\affiliation[2]{organization={Veterinary Population Medicine Department, College of Veterinary Medicine},
             addressline={University of  Minnesota}, 
             city={St. Paul},
             state={MN},
             country={USA}}

\end{frontmatter}     

\begin{figure}[H]
\centering
\includegraphics[width=0.9\textwidth]{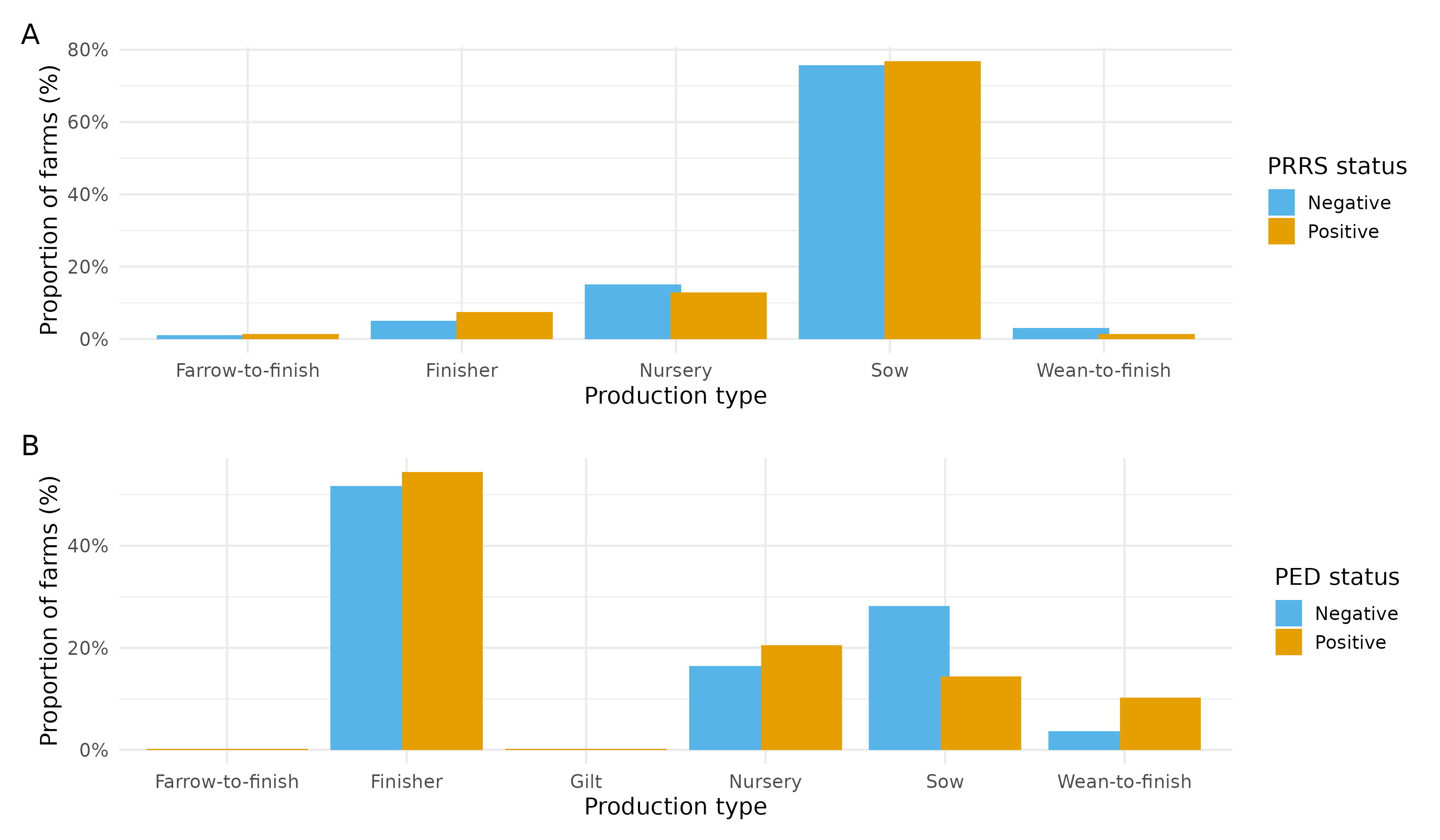}
\caption{Proportion of negative and positive farms with available vehicle movement data used in the study.}
\end{figure}

\begin{figure}[H]
\centering
\includegraphics[width=0.9\textwidth]{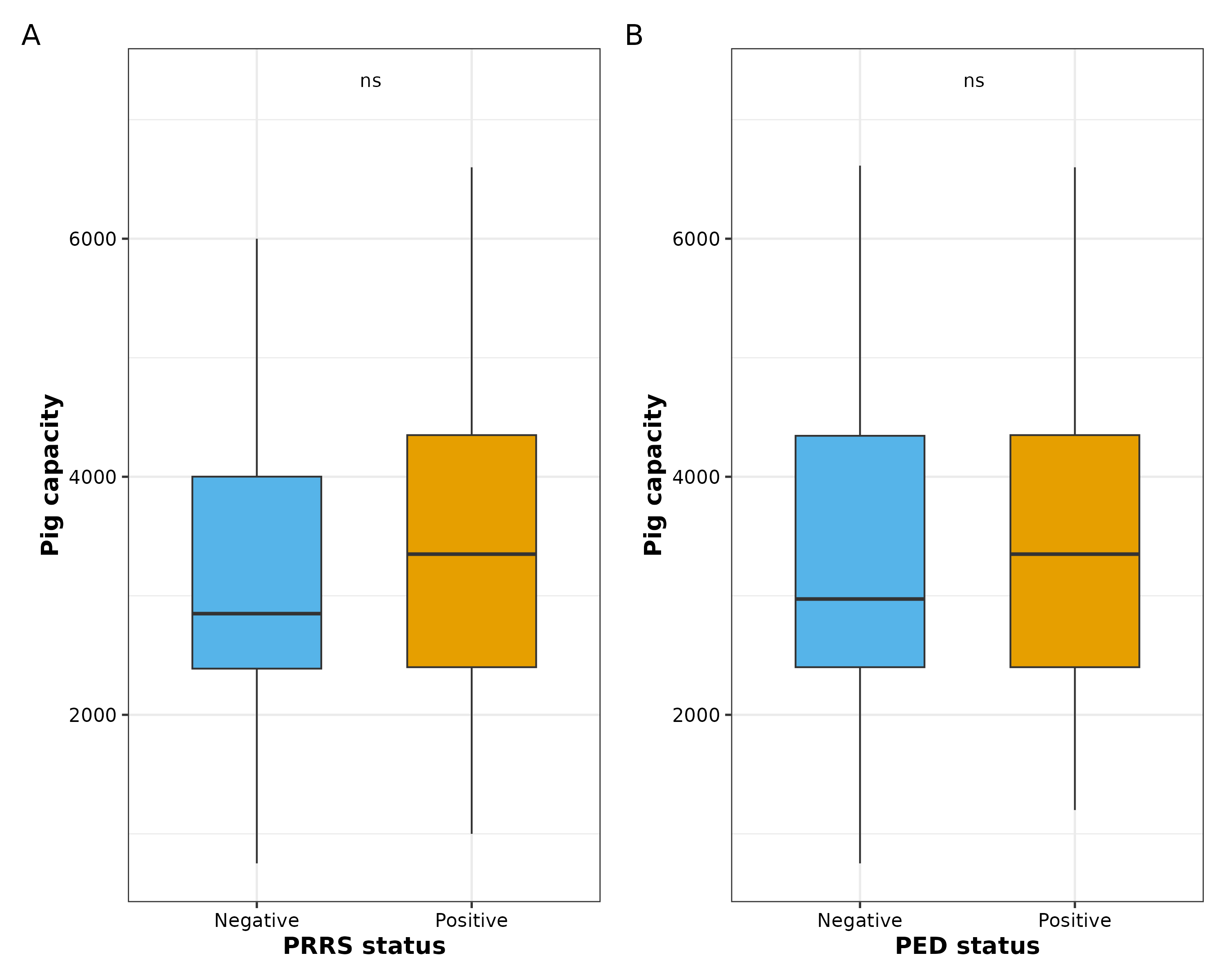}
\caption{Pig capacity}
\end{figure}

\begin{figure}[H]
\centering
\includegraphics[width=0.9\textwidth]{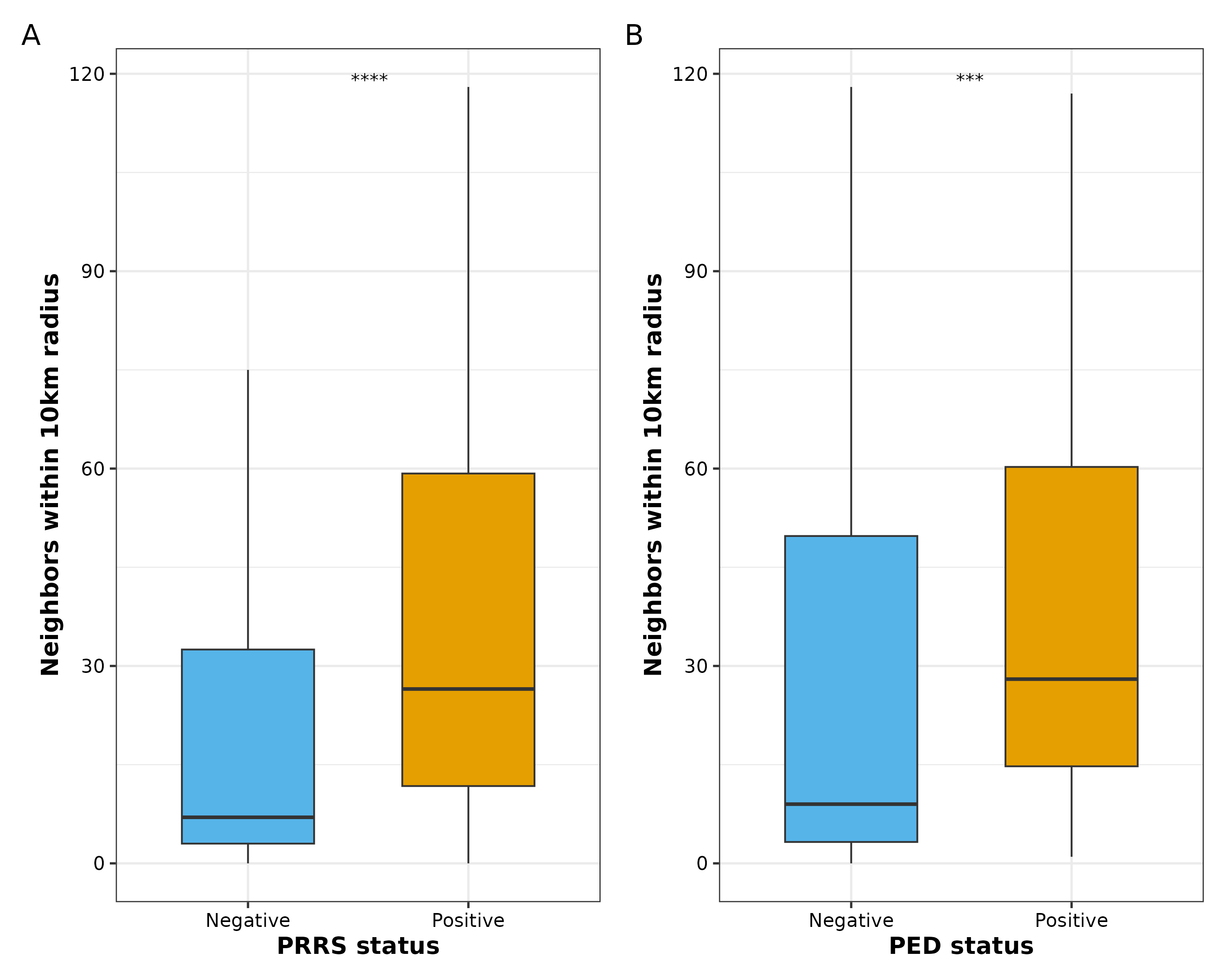}
\caption{Neighbors within 10km radius}
\end{figure}

\begin{figure}[H]
\centering
\includegraphics[width=0.9\textwidth]{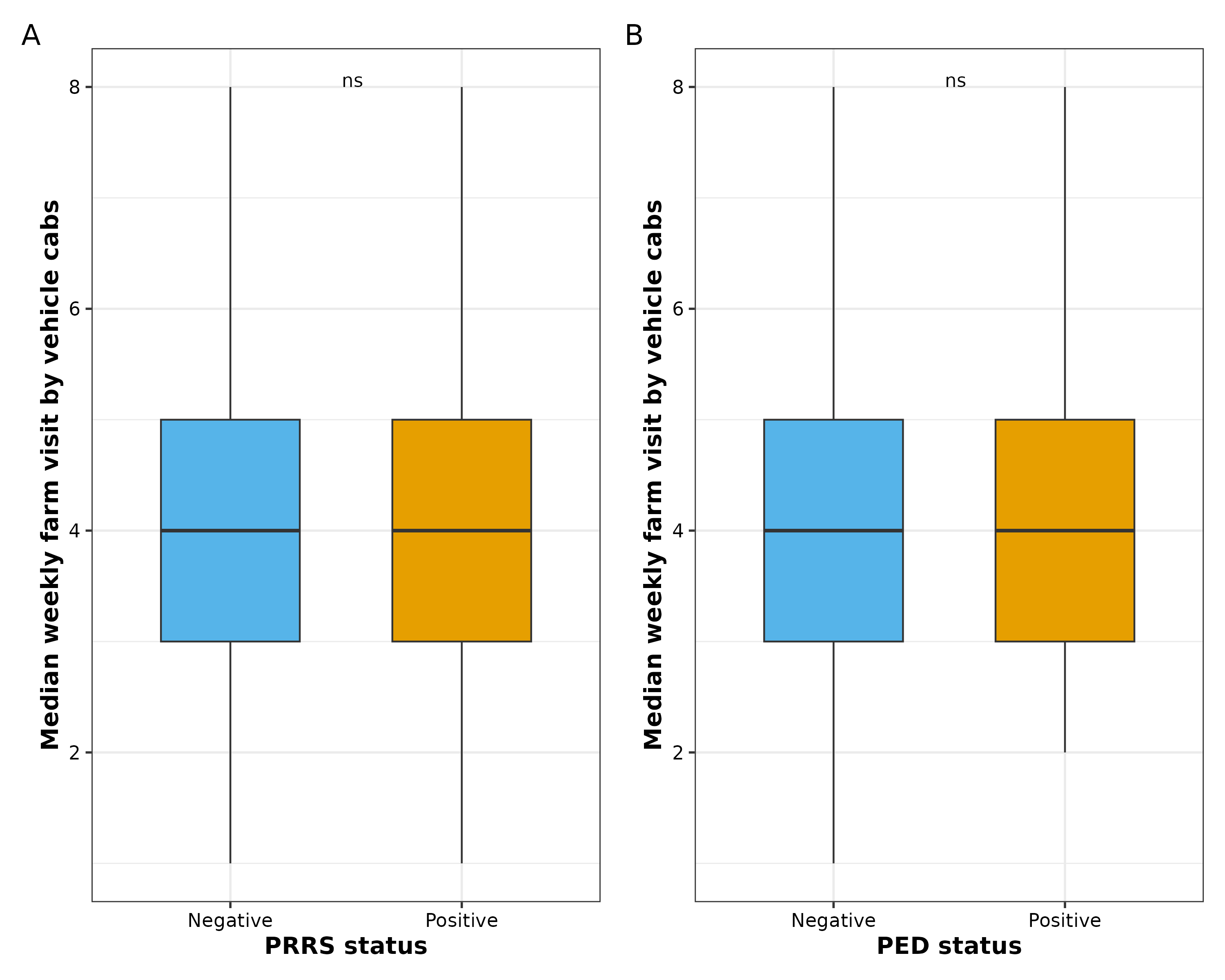}
\caption{Median weekly farm visit by vehicle cabs}
\label{fig:SM_03_disease_weekly_farm_truck_visit_plot}
\end{figure}

\begin{figure}[H]
\centering
\includegraphics[width=0.9\textwidth]{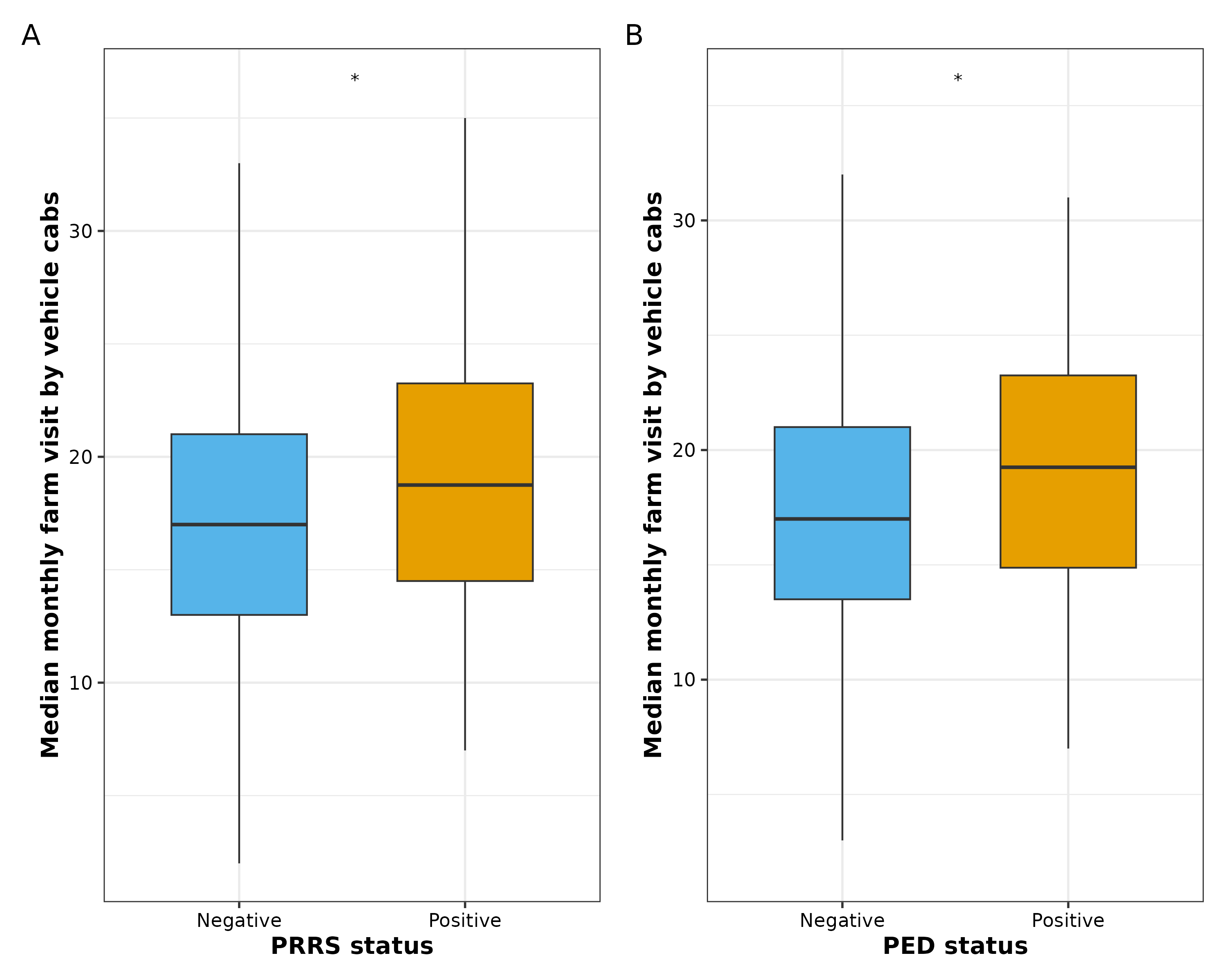}
\caption{Median monthly farm visit by vehicle cabs}
\label{fig:SM_04_disease_monthly_farm_truck_visit_plot}
\end{figure}

\begin{figure}[H]
\centering
\includegraphics[width=0.9\textwidth]{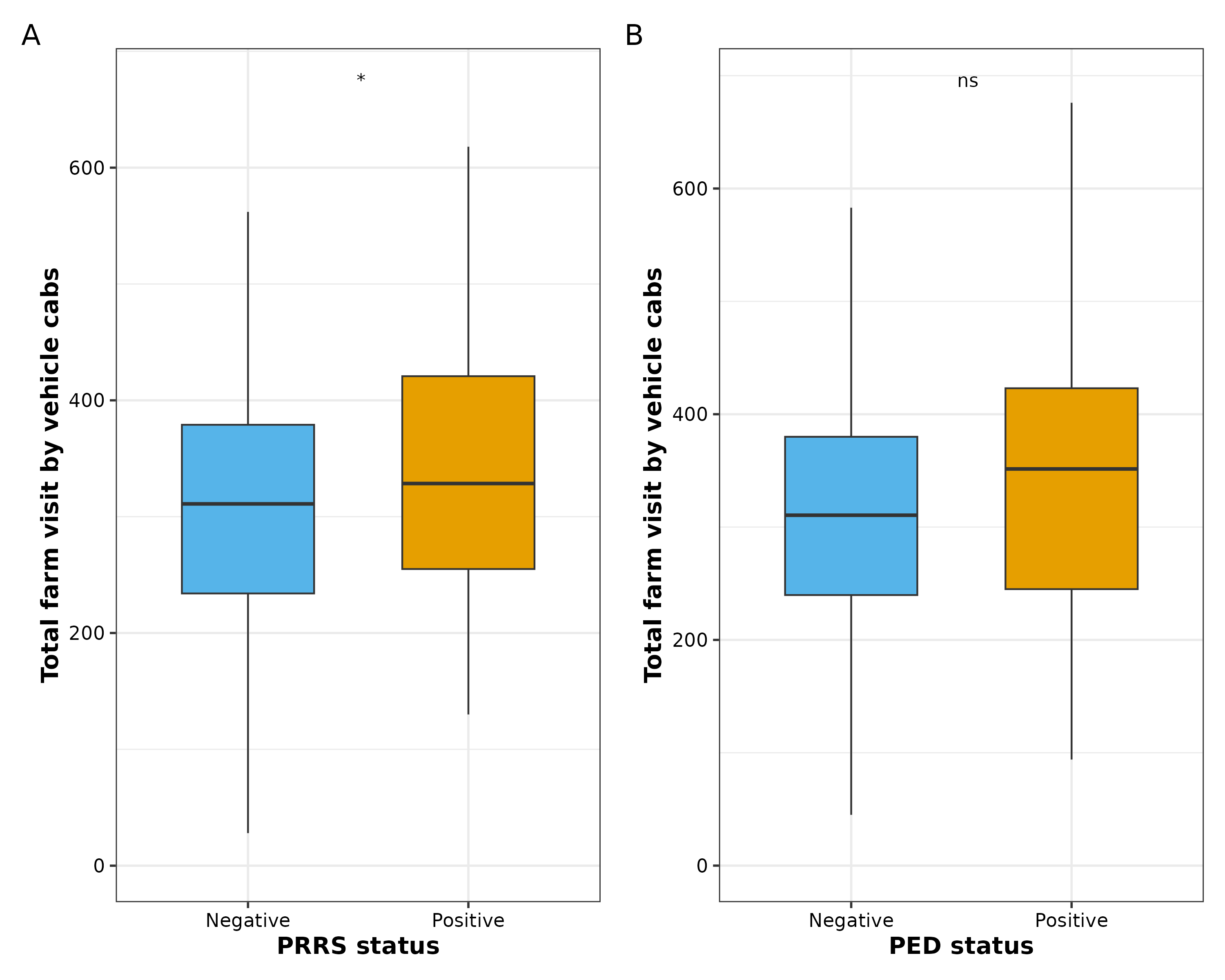}
\caption{Total farm visit by vehicle cabs}
\label{fig:SM_05_disease_total_farm_truck_visit_plot}
\end{figure}

\begin{figure}[H]
\centering
\includegraphics[width=0.9\textwidth]{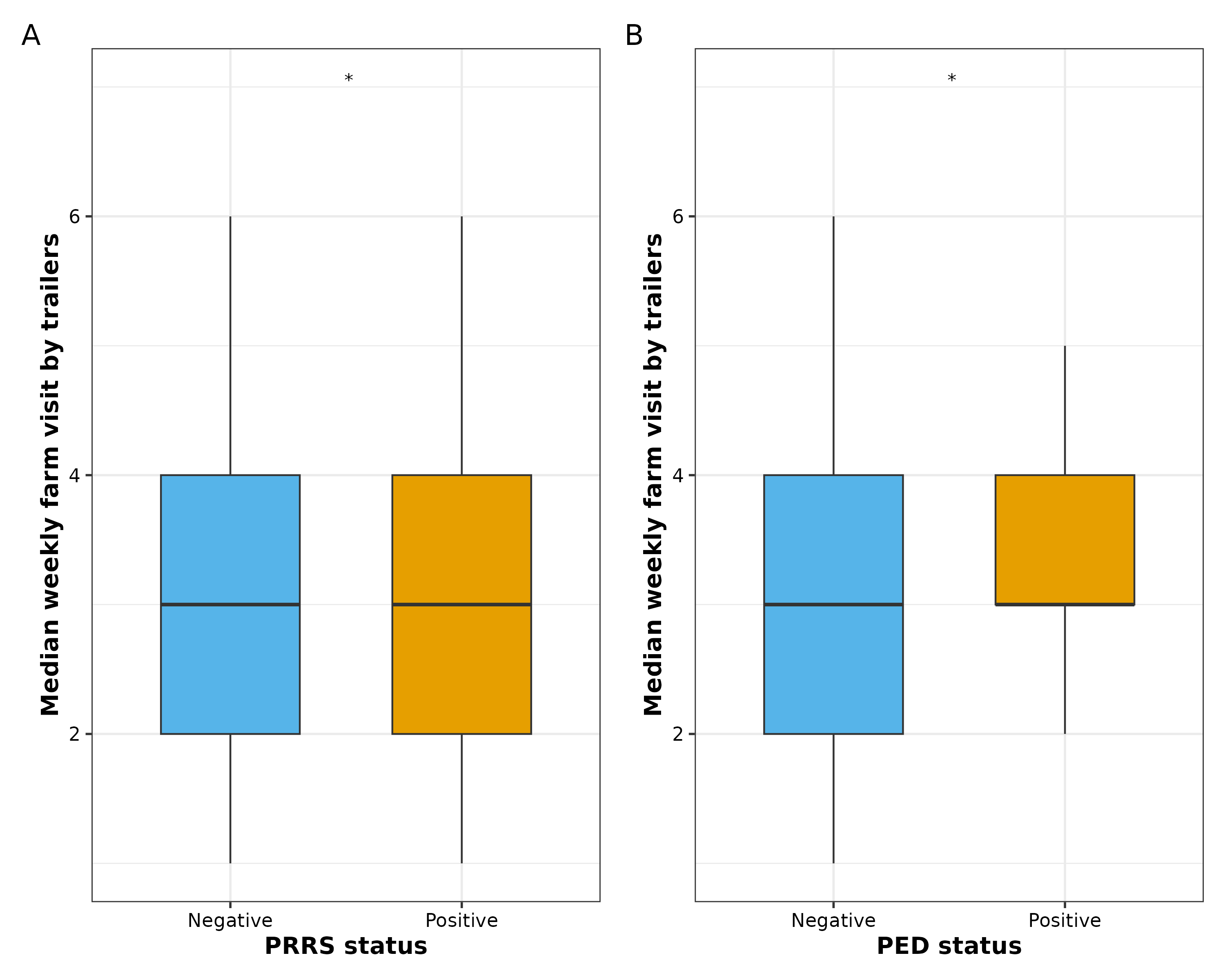}
\caption{Median weekly farm visit by trailers}
\label{fig:SM_06_disease_weekly_farm_trailers_visit_plot}
\end{figure}

\begin{figure}[H]
\centering
\includegraphics[width=0.9\textwidth]{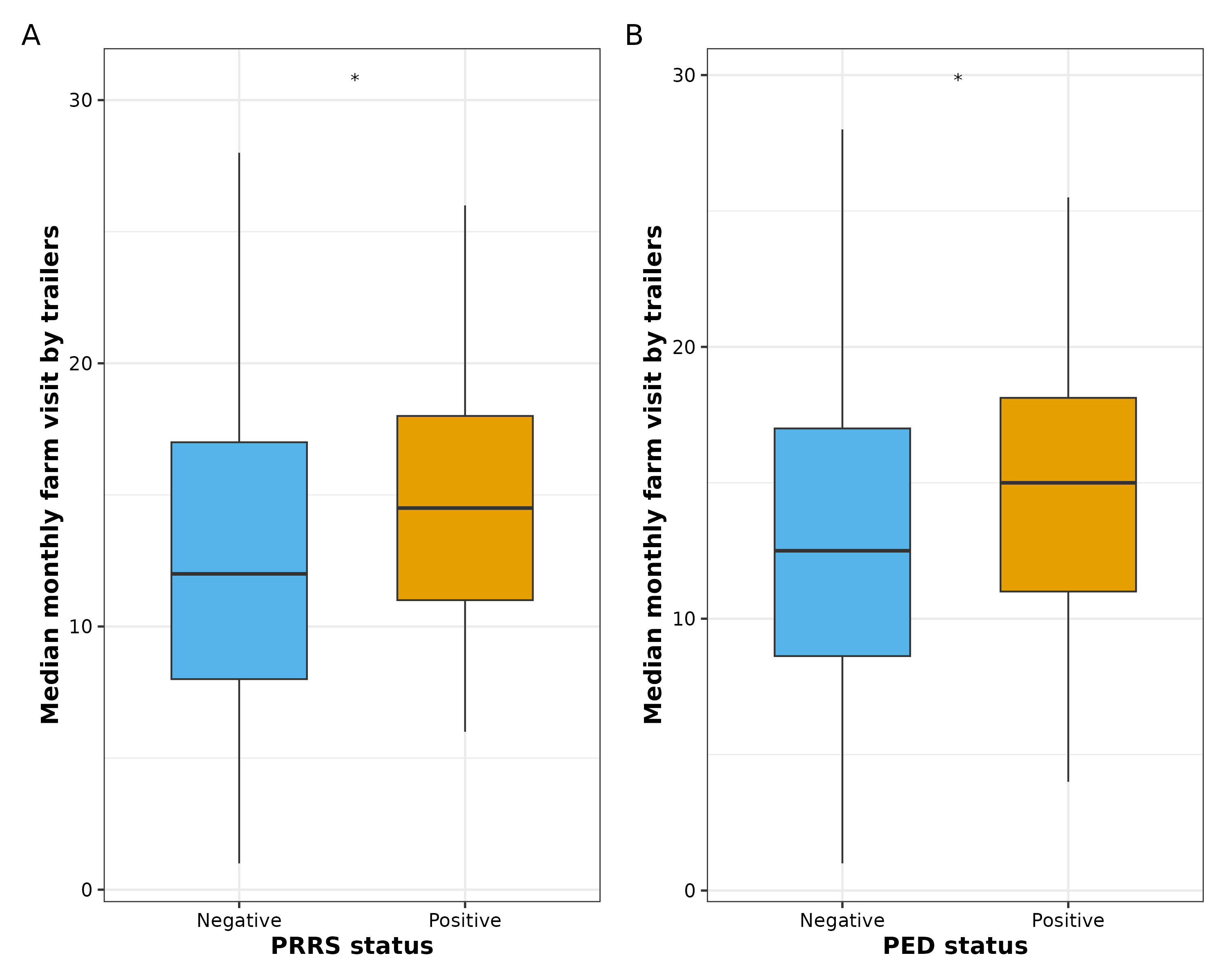}
\caption{Median monthly farm visit by trailers}
\label{fig:SM_07_disease_monthly_farm_trailers_visit_plot}
\end{figure}

\begin{figure}[H]
\centering
\includegraphics[width=0.9\textwidth]{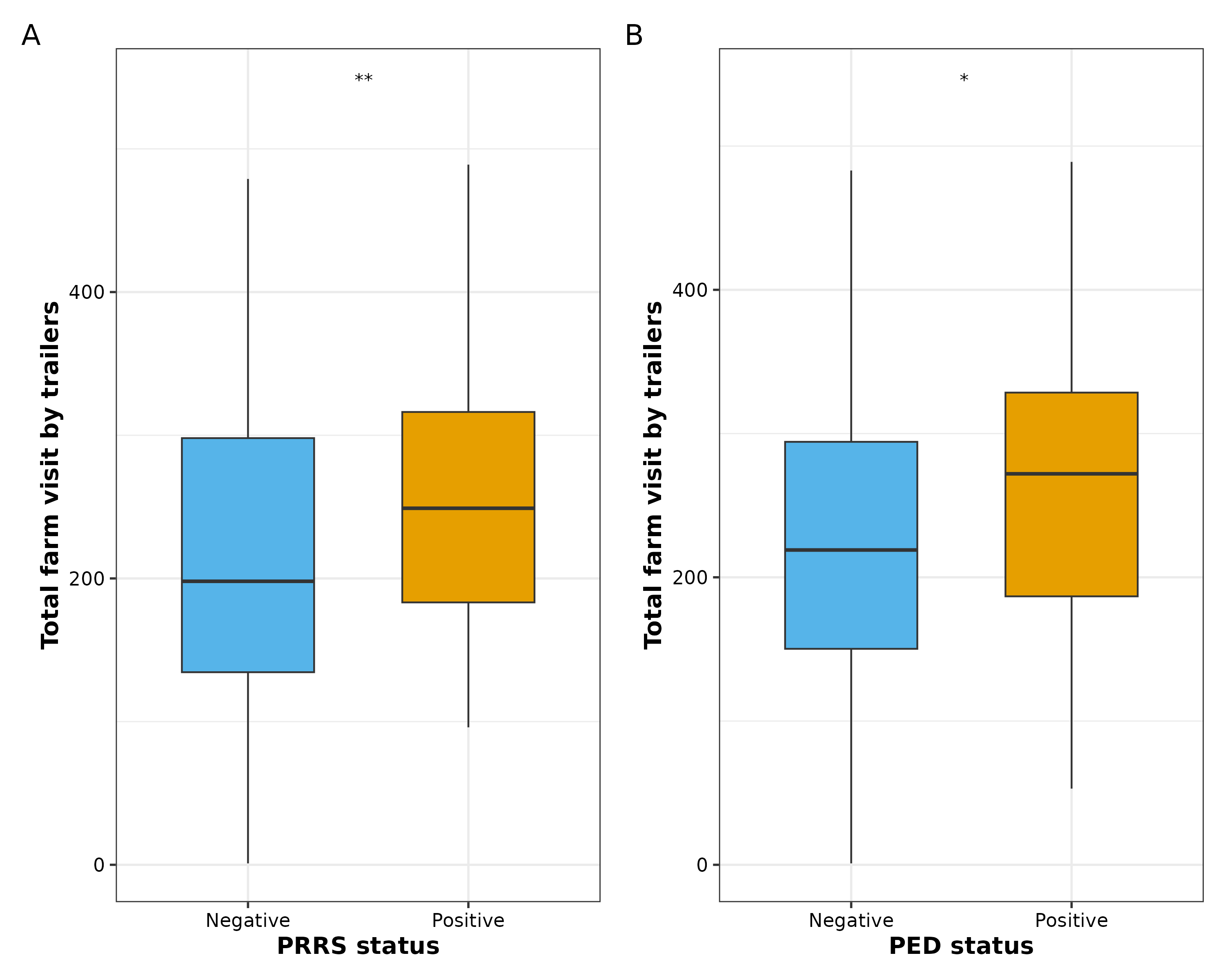}
\caption{Total farm visit by trailers}
\label{fig:SM_08_disease_total_farm_trailers_visit_plot}
\end{figure}

\begin{figure}[H]
\centering
\includegraphics[width=0.9\textwidth]{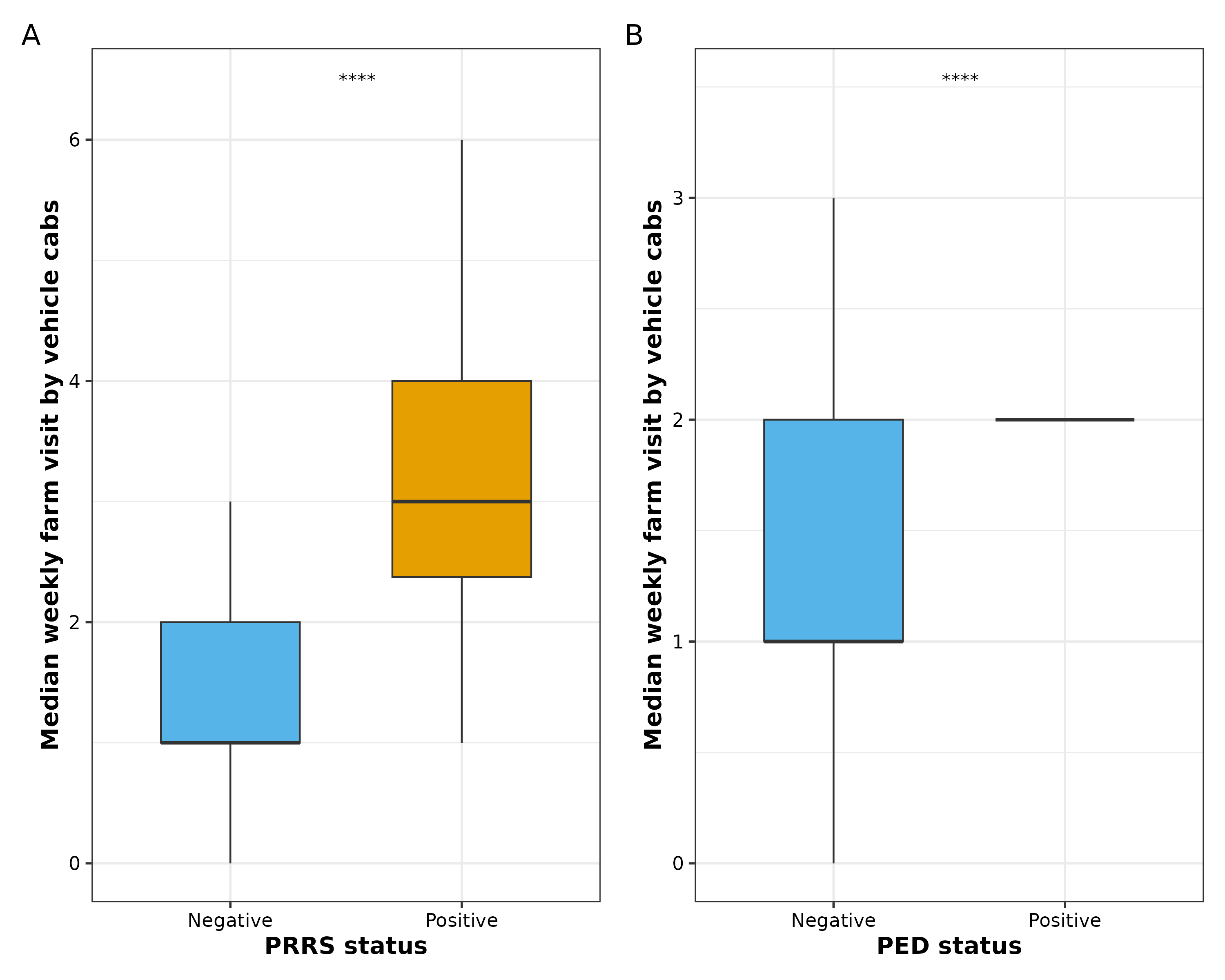}
\caption{Median weekly farm visit by vehicle cabs exposed to positive farms one day before the visit}
\label{fig:SM_09_disease_weekly_farm_contaminated_truck_visit_1dayB}
\end{figure}

\begin{figure}[H]
\centering
\includegraphics[width=0.9\textwidth]{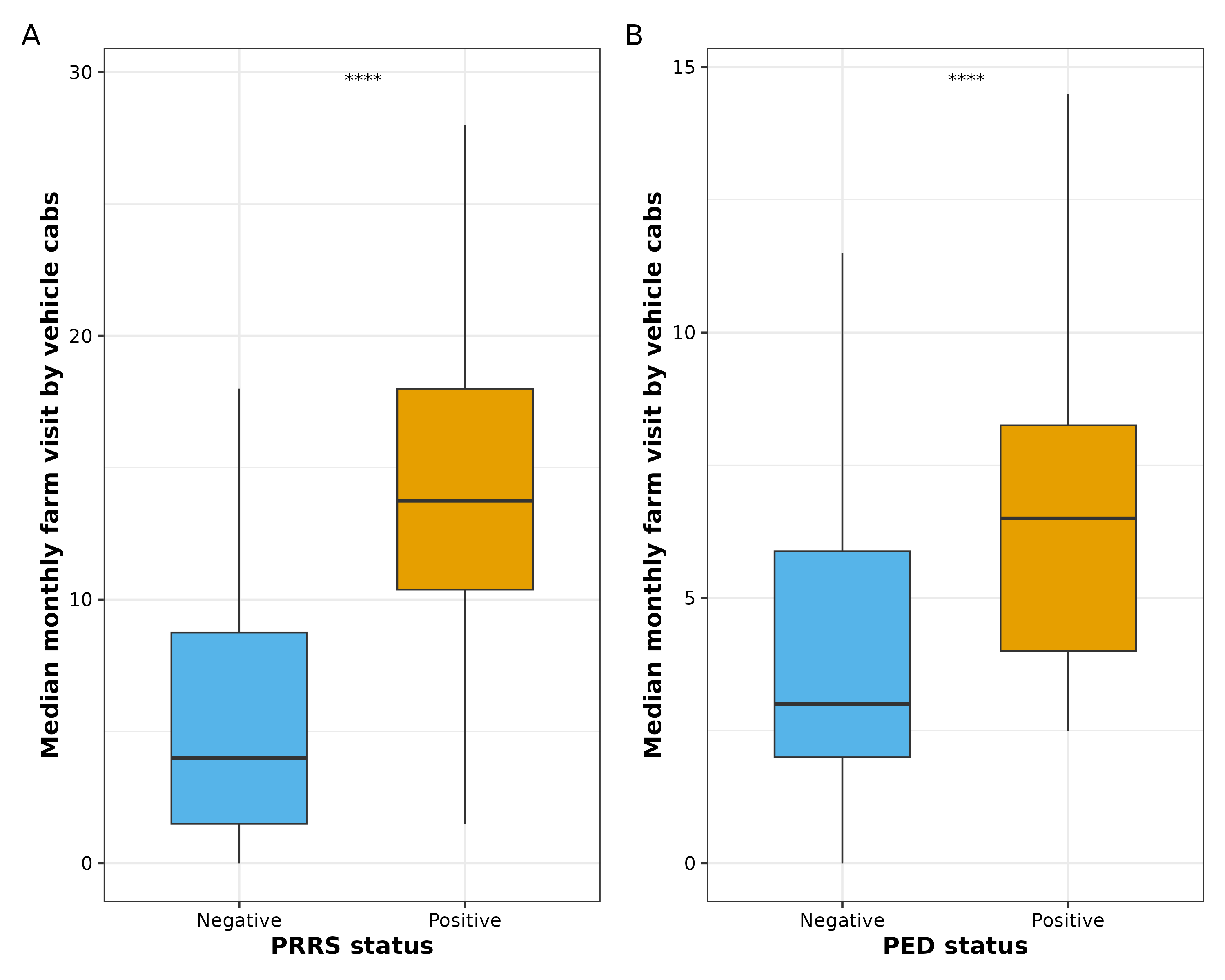}
\caption{Median monthly farm visit by vehicle cabs exposed to positive farms one day before the visit}
\label{fig:SM_10_disease_monthly_farm_contaminated_truck_visit_1dayB}
\end{figure}

\begin{figure}[H]
\centering
\includegraphics[width=0.9\textwidth]{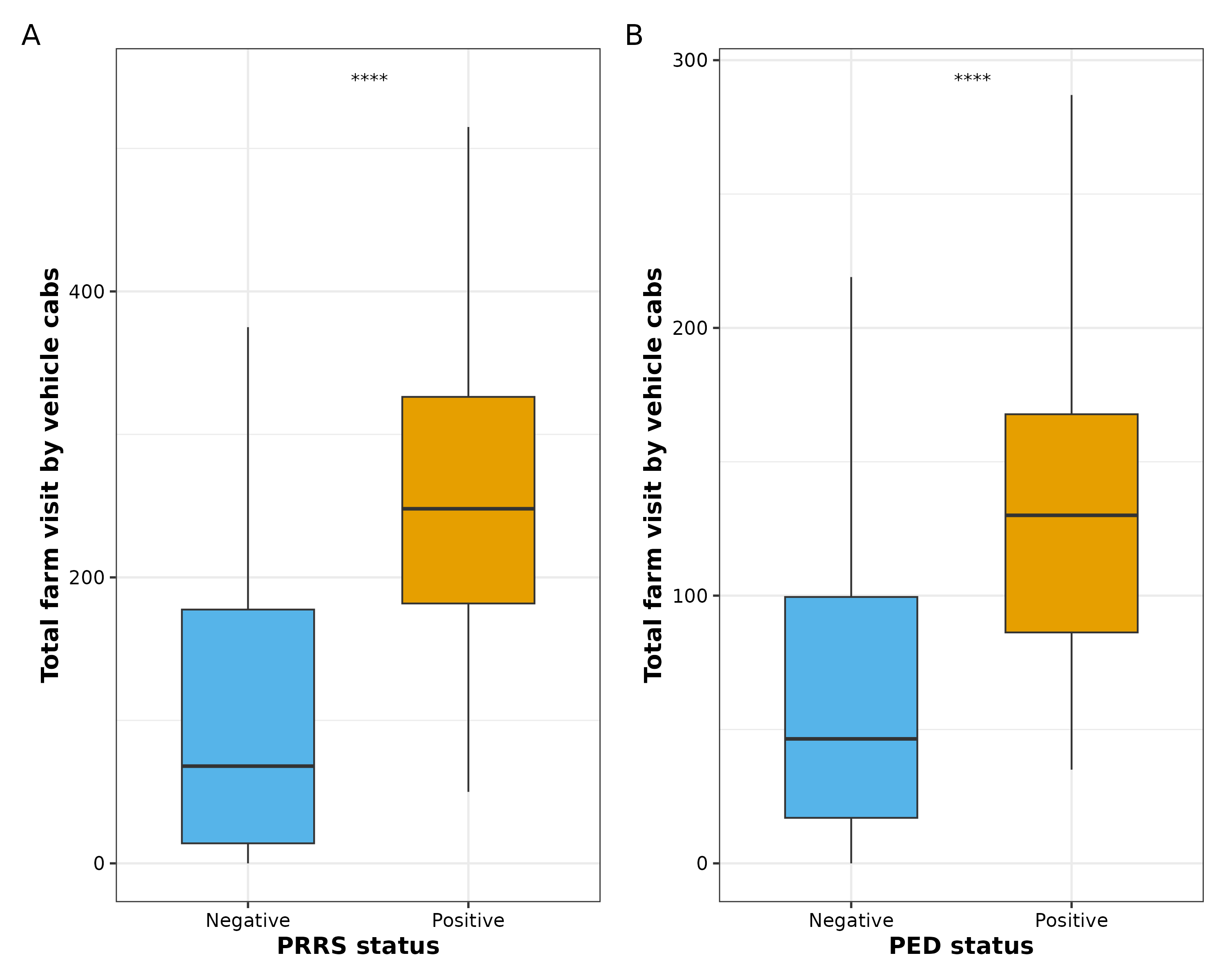}
\caption{Total farm visit by vehicle cabs exposed to positive farms one day before the visit}
\label{fig:SM_11_disease_total_farm_contaminated_truck_visit_1dayB}
\end{figure}

\begin{figure}[H]
\centering
\includegraphics[width=0.9\textwidth]{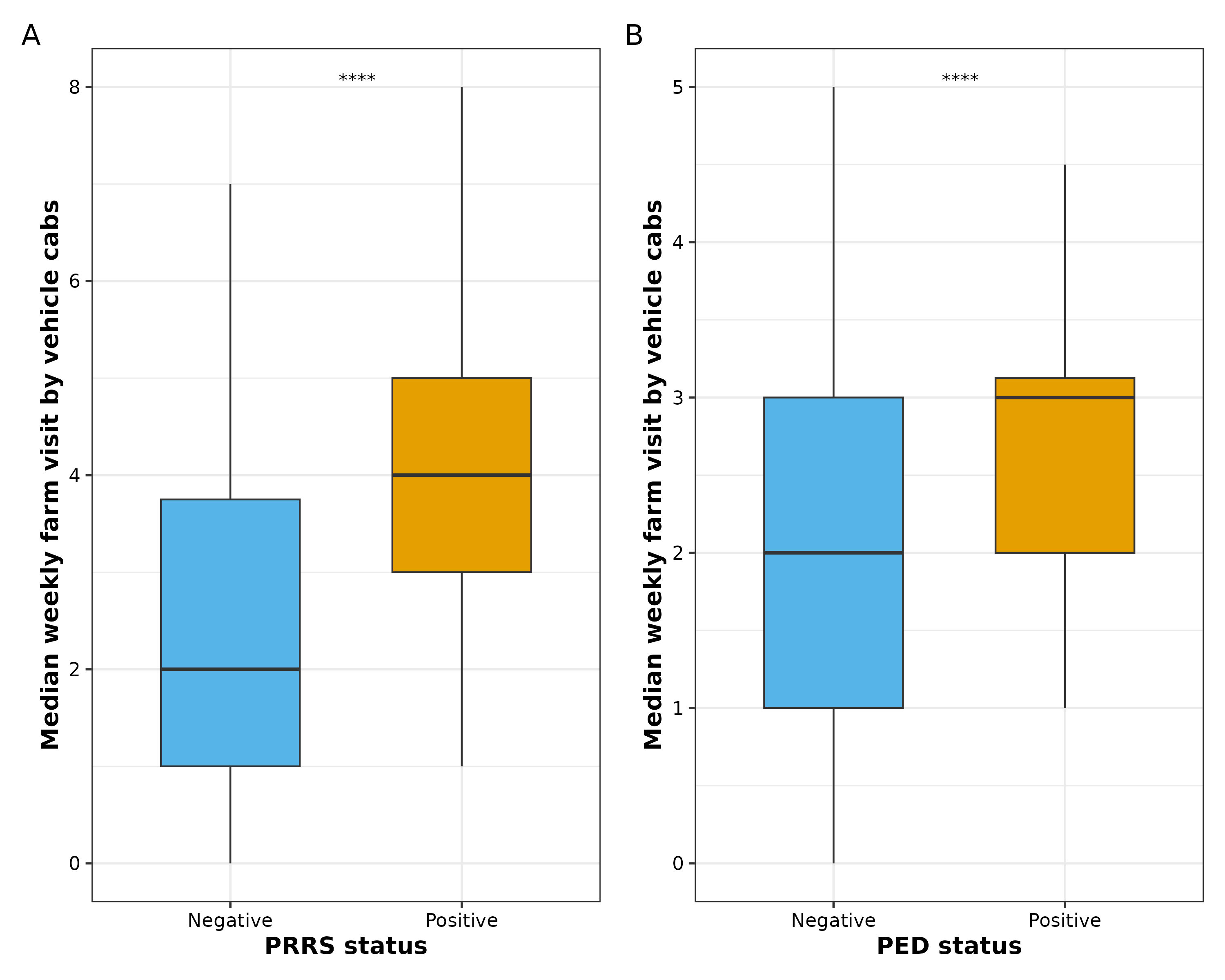}
\caption{Median weekly farm visit by vehicle cabs exposed to positive farms seven days before the visit}
\label{fig:SM_12_disease_weekly_farm_contaminated_truck_visit_7dayB}
\end{figure}

\begin{figure}[H]
\centering
\includegraphics[width=0.9\textwidth]{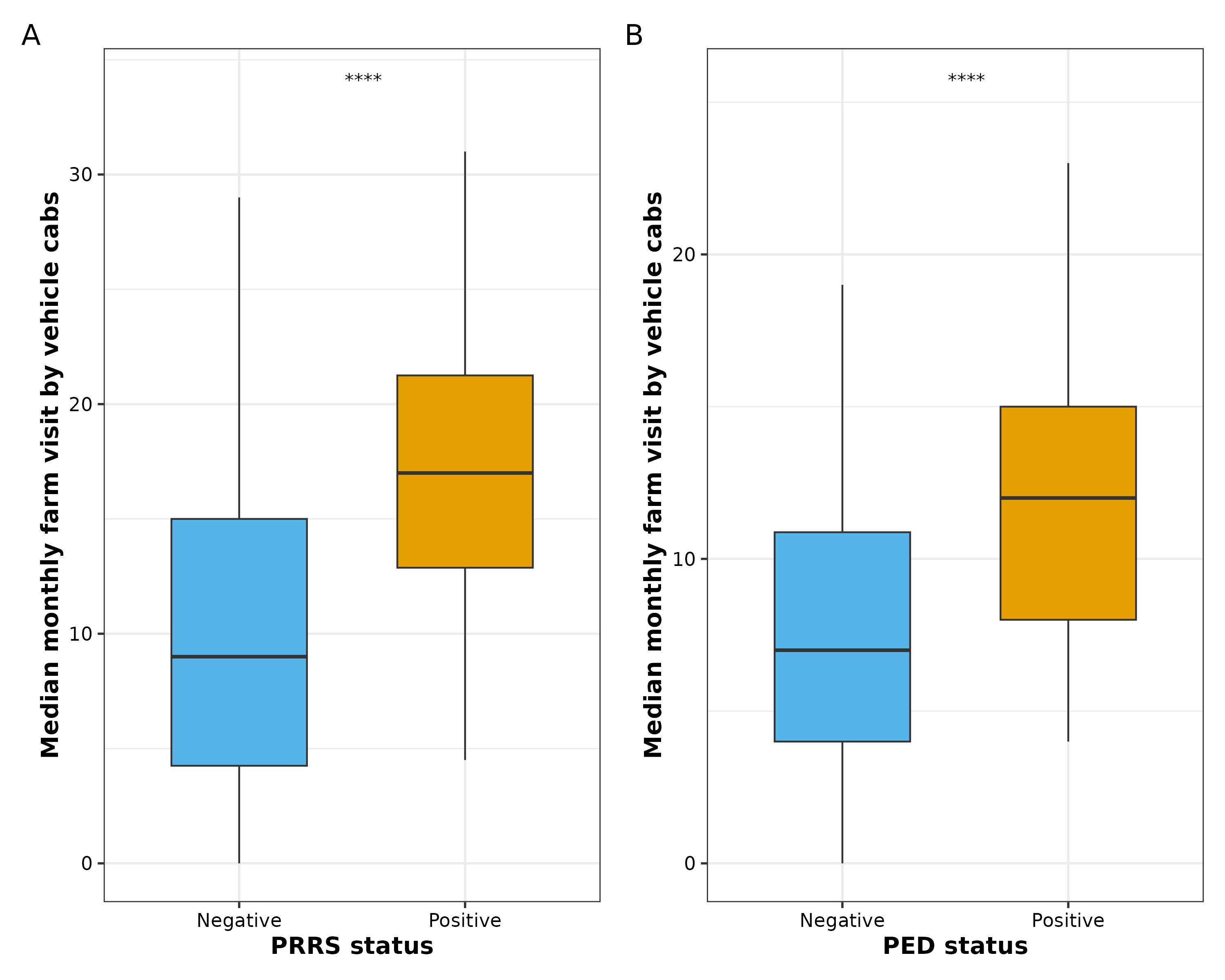}
\caption{Median monthly farm visit by vehicle cabs exposed to positive farms seven days before the visit}
\label{fig:SM_13_disease_monthly_farm_contaminated_truck_visit_7dayB}
\end{figure}

\begin{figure}[H]
\centering
\includegraphics[width=0.9\textwidth]{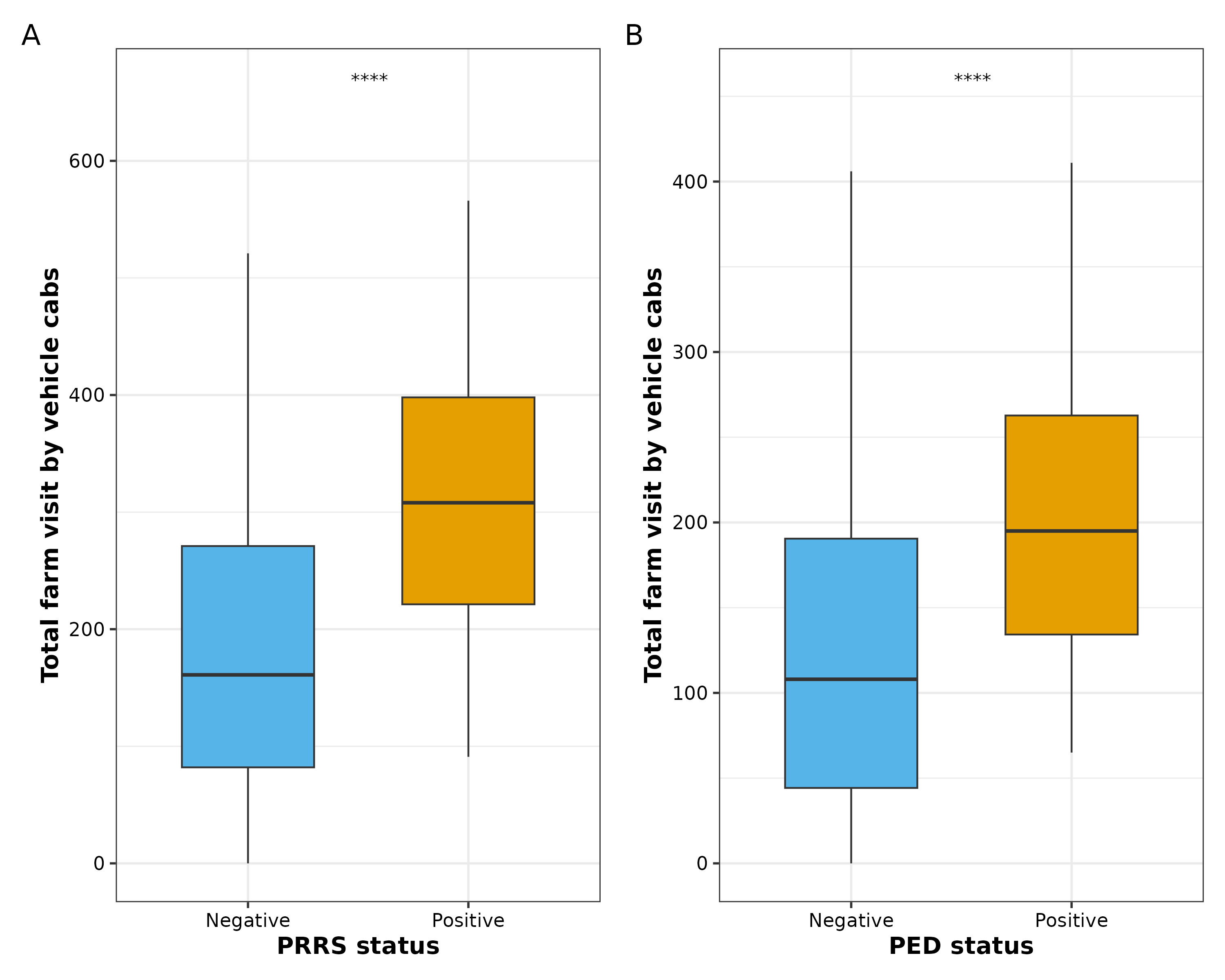}
\caption{Total farm visit by vehicle cabs exposed to positive farms seven days before the visit}
\label{fig:SM_14_disease_total_farm_contaminated_truck_visit_7dayB}
\end{figure}

\begin{figure}[H]
\centering
\includegraphics[width=0.9\textwidth]{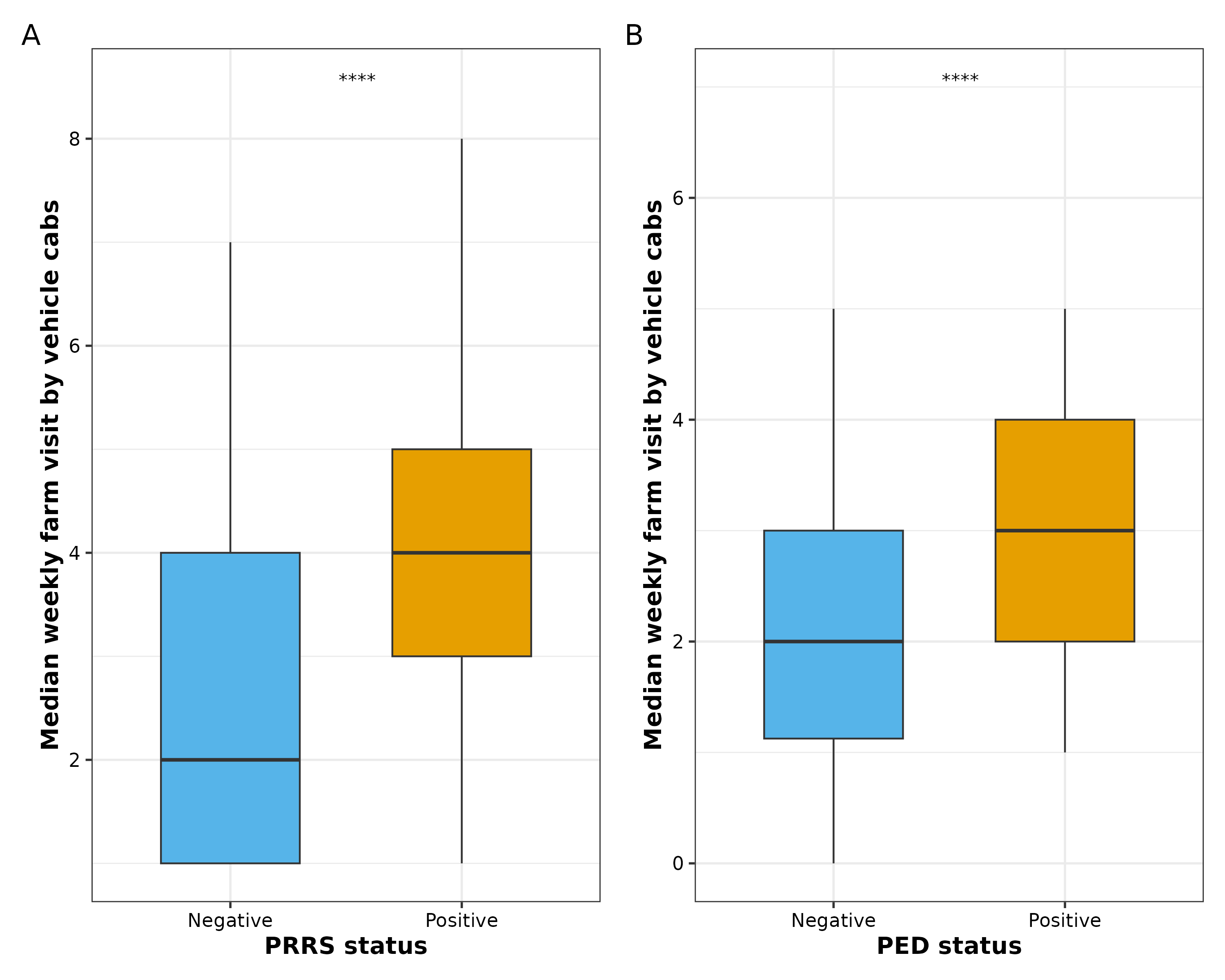}
\caption{Median weekly farm visit by vehicle cabs exposed to positive farms 11 days before the visit}
\label{fig:SM_15_disease_weekly_farm_contaminated_truck_visit_11dayB}
\end{figure}

\begin{figure}[H]
\centering
\includegraphics[width=0.9\textwidth]{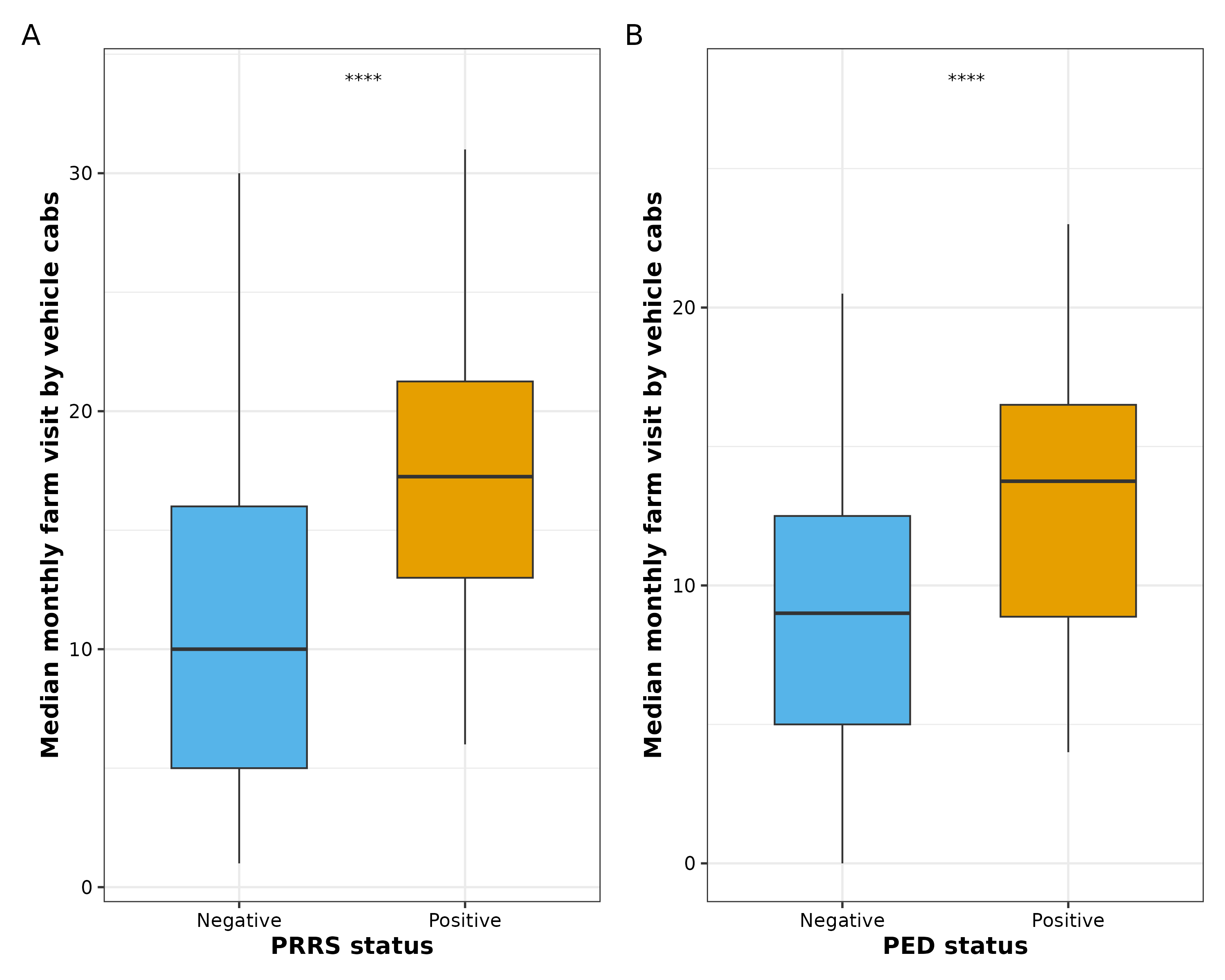}
\caption{Median monthly farm visit by vehicle cabs exposed to positive farms 11 days before the visit}
\label{fig:SM_16_disease_monthly_farm_contaminated_truck_visit_11dayB}
\end{figure}

\begin{figure}[H]
\centering
\includegraphics[width=0.9\textwidth]{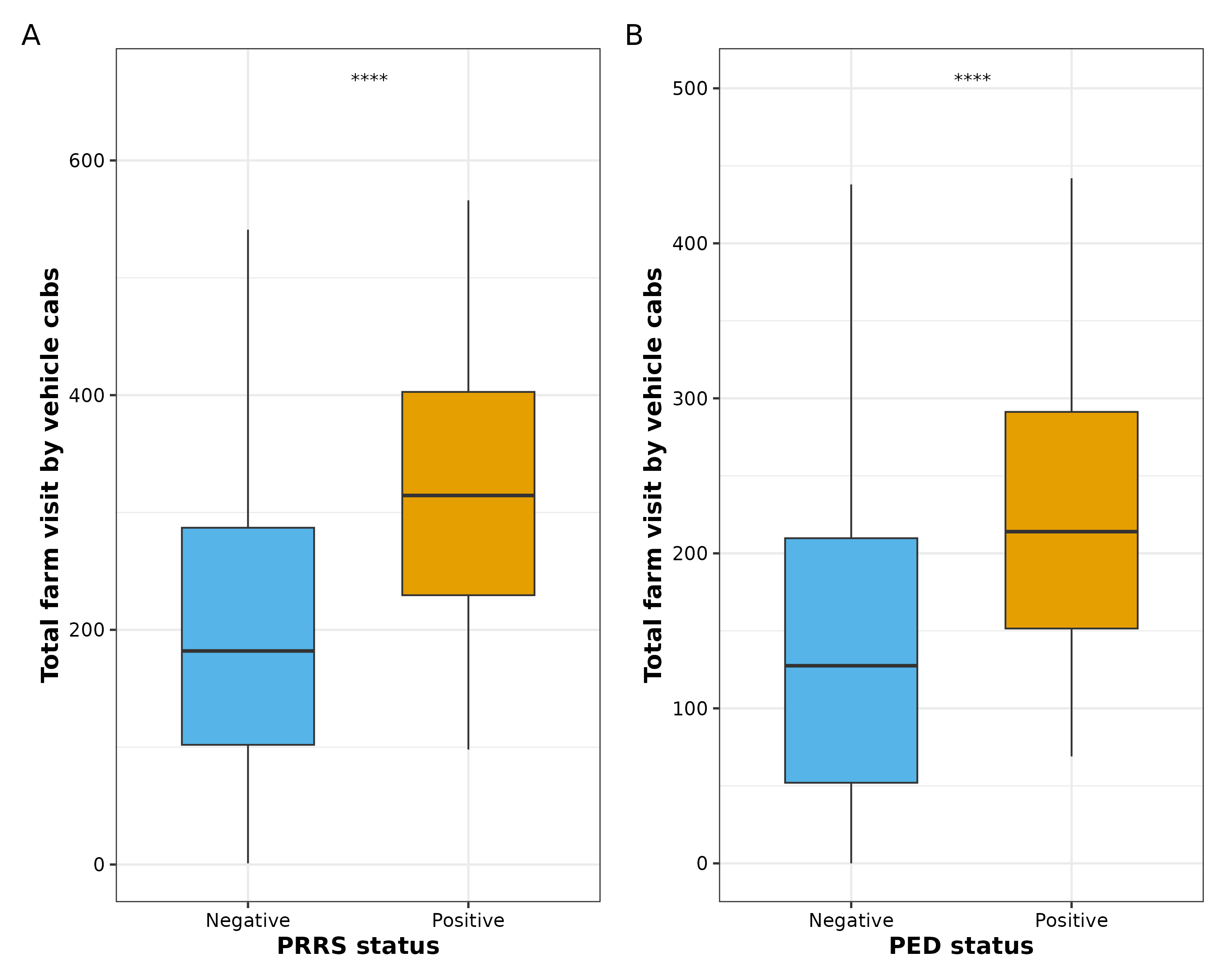}
\caption{Total farm visit by vehicle cabs exposed to positive farms 11 days before the visit}
\label{fig:SM_17_disease_total_farm_contaminated_truck_visit_11dayB}
\end{figure}

\begin{figure}[H]
\centering
\includegraphics[width=0.9\textwidth]{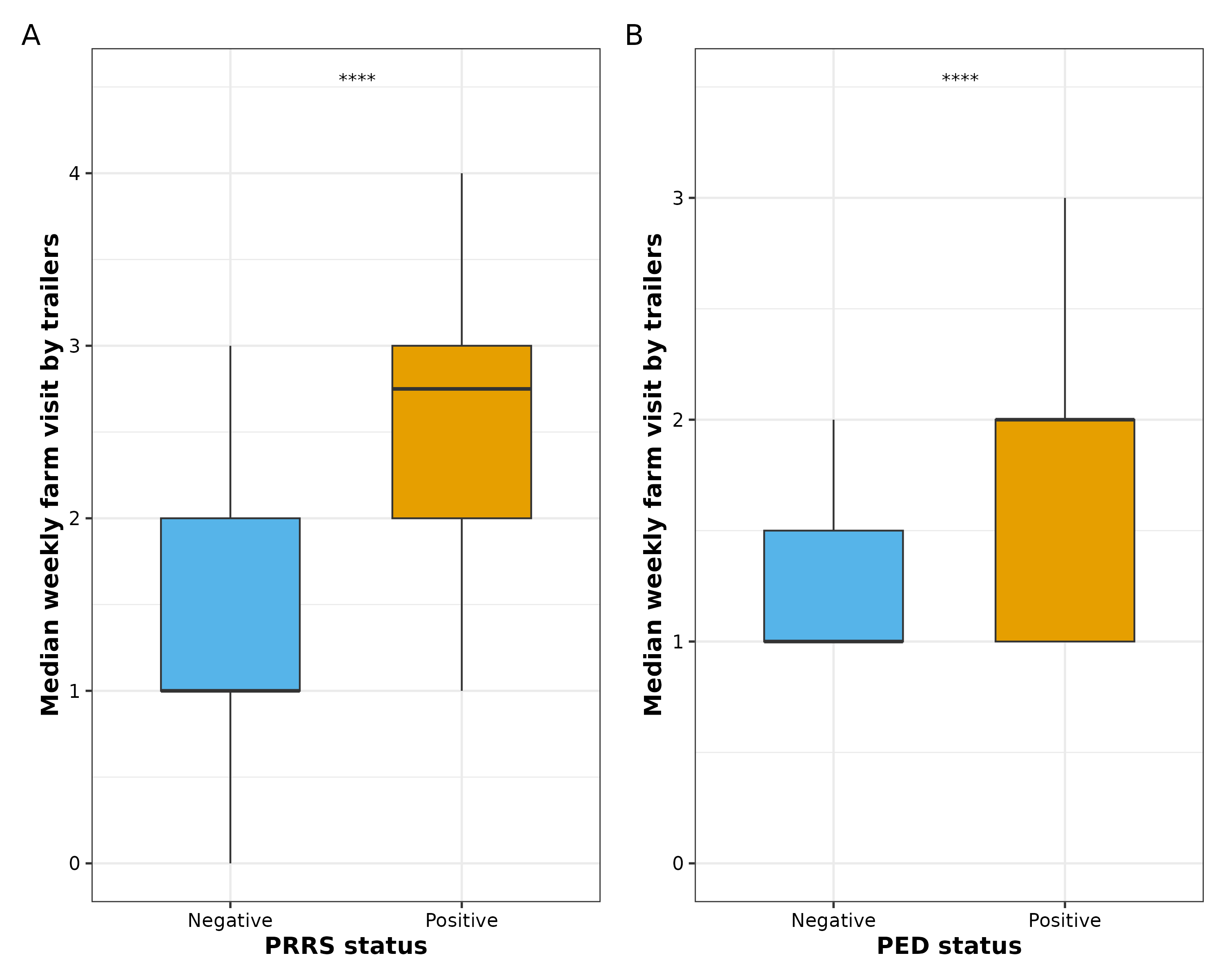}
\caption{Median weekly farm visit by trailers exposed to positive farms one day before the visit}
\label{fig:SM_18_disease_weekly_farm_contaminated_trailer_visit_1dayB}
\end{figure}

\begin{figure}[H]
\centering
\includegraphics[width=0.9\textwidth]{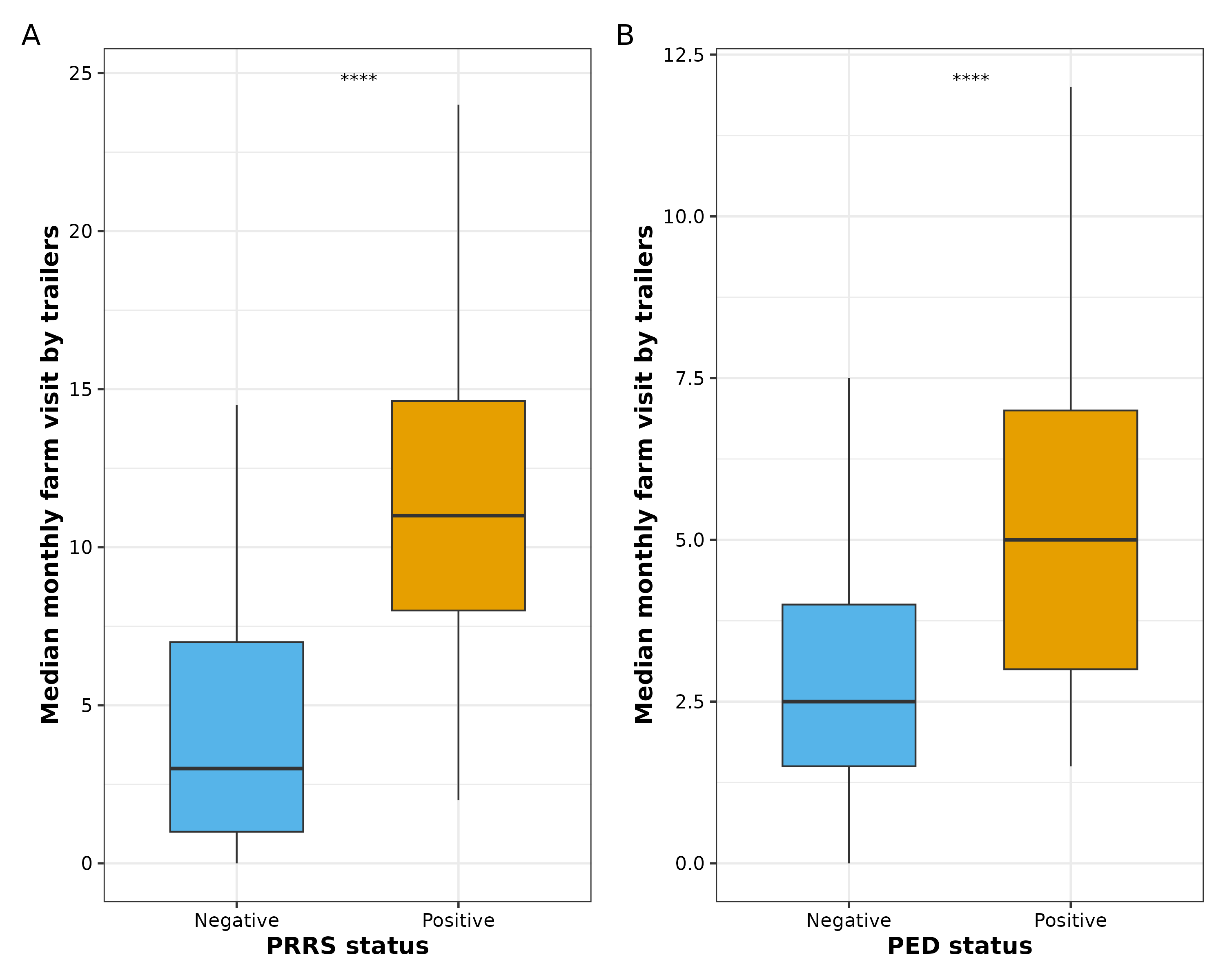}
\caption{Median monthly farm visit by trailers exposed to positive farms one day before the visit}
\label{fig:SM_19_disease_monthly_farm_contaminated_trailer_visit_1dayB}
\end{figure}

\begin{figure}[H]
\centering
\includegraphics[width=0.9\textwidth]{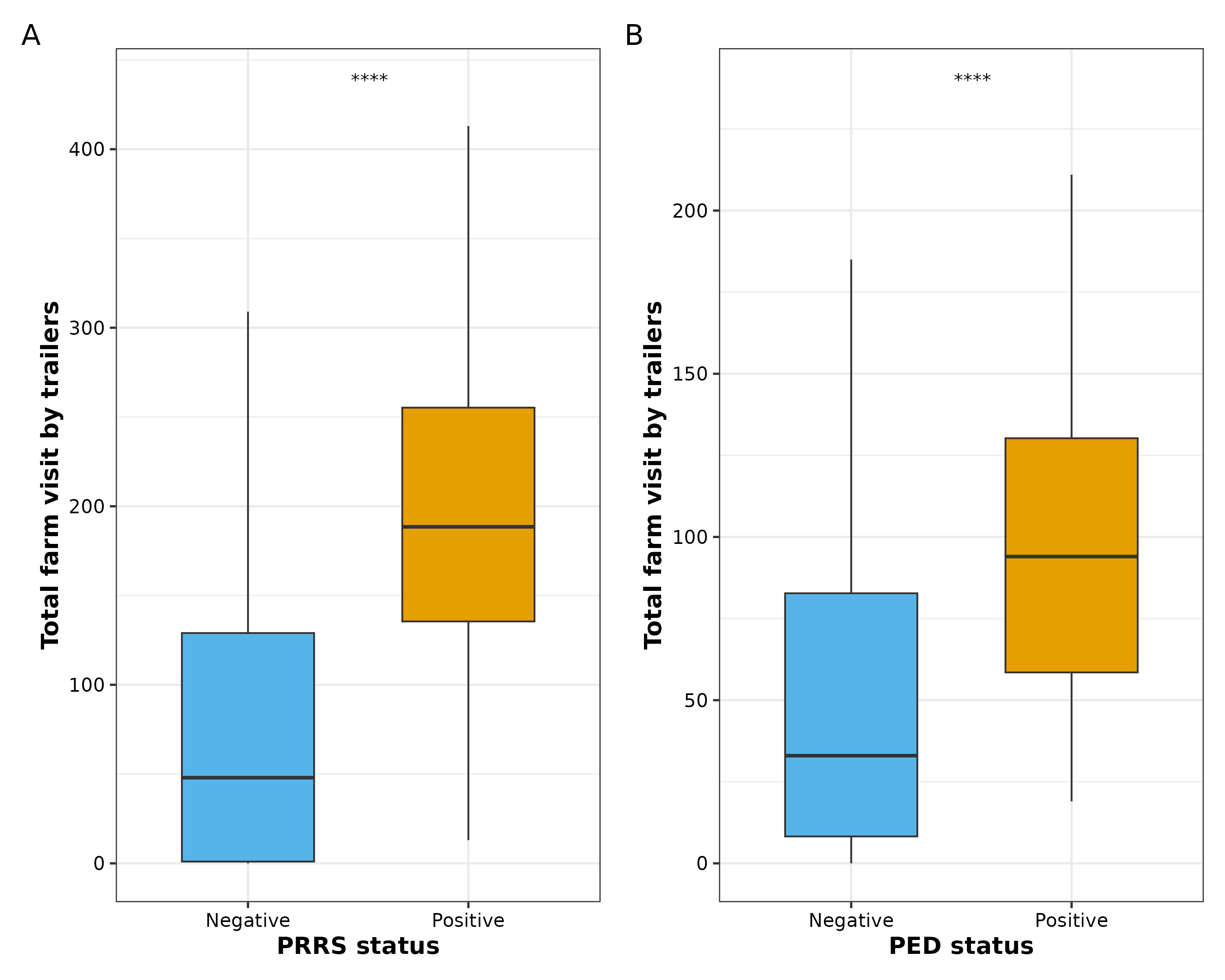}
\caption{Total farm visit by trailers exposed to positive farms one day before the visit}
\label{fig:SM_20_disease_total_farm_contaminated_trailer_visit_1dayB}
\end{figure}

\begin{figure}[H]
\centering
\includegraphics[width=0.9\textwidth]{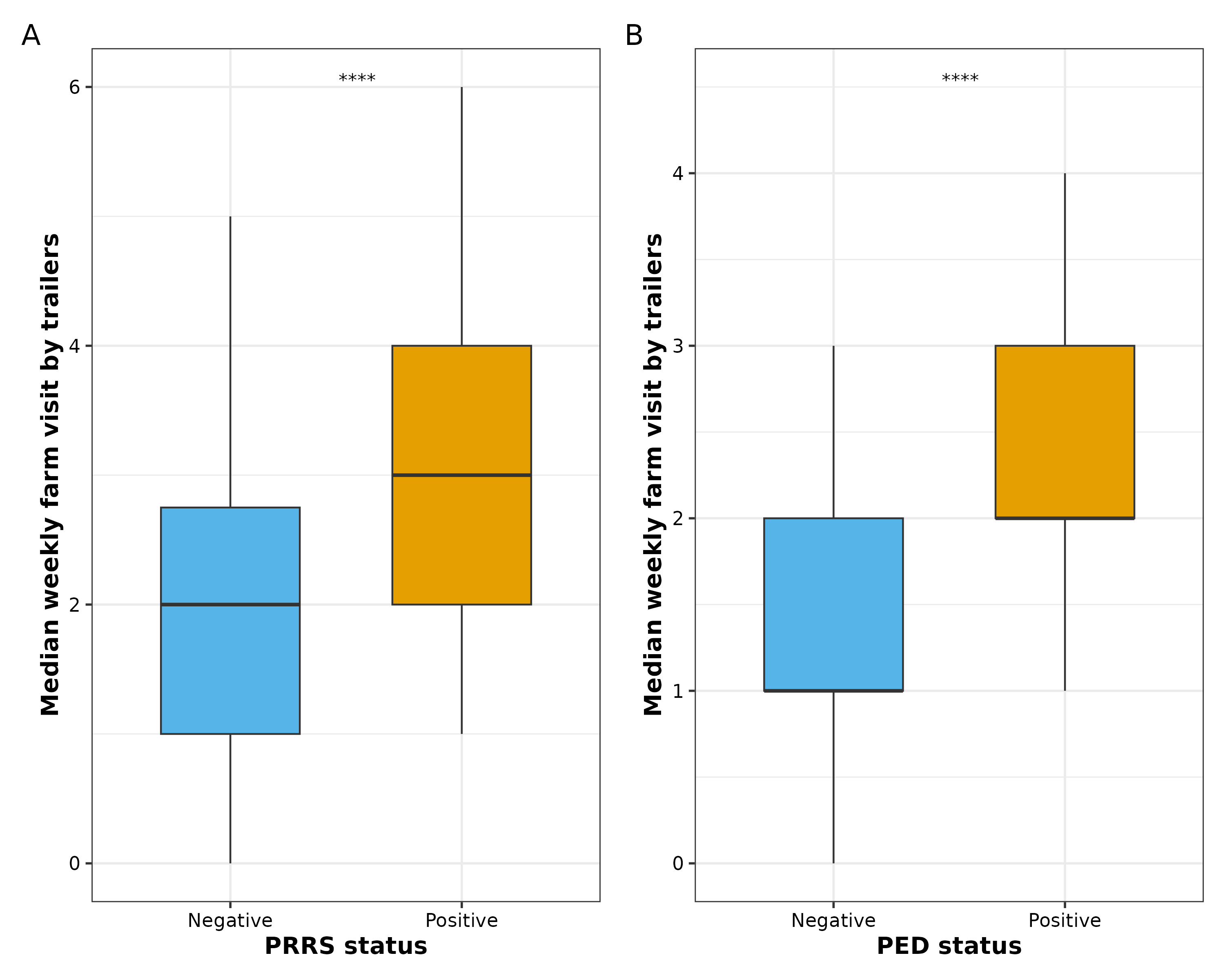}
\caption{Median weekly farm visit by trailers exposed to positive farms seven days before the visit}
\label{fig:SM_21_disease_weekly_farm_contaminated_trailer_visit_7dayB}
\end{figure}

\begin{figure}[H]
\centering
\includegraphics[width=0.9\textwidth]{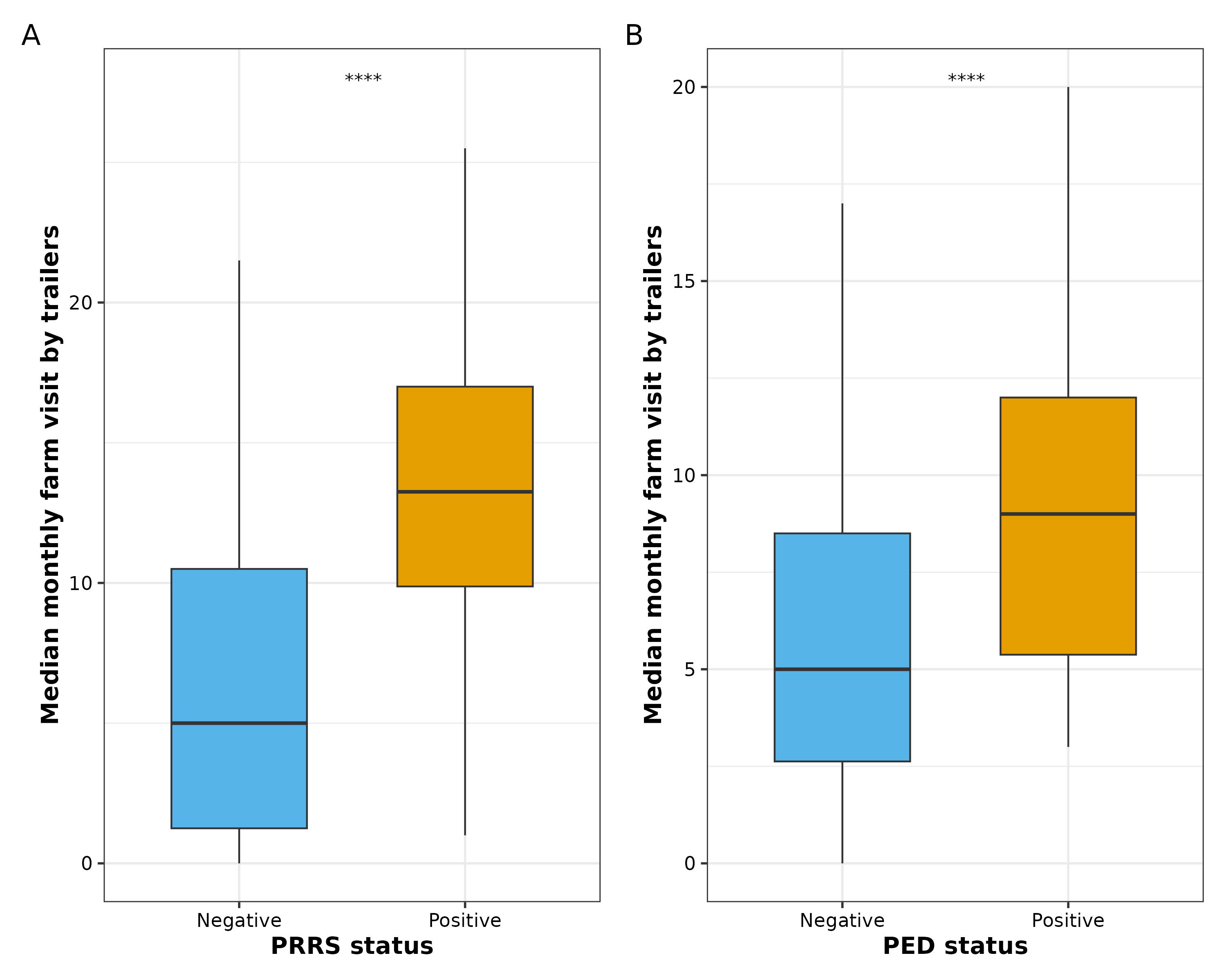}
\caption{Median monthly farm visit by trailers exposed to positive farms seven days before the visit}
\label{fig:SM_22_disease_monthly_farm_contaminated_trailer_visit_7dayB}
\end{figure}

\begin{figure}[H]
\centering
\includegraphics[width=0.9\textwidth]{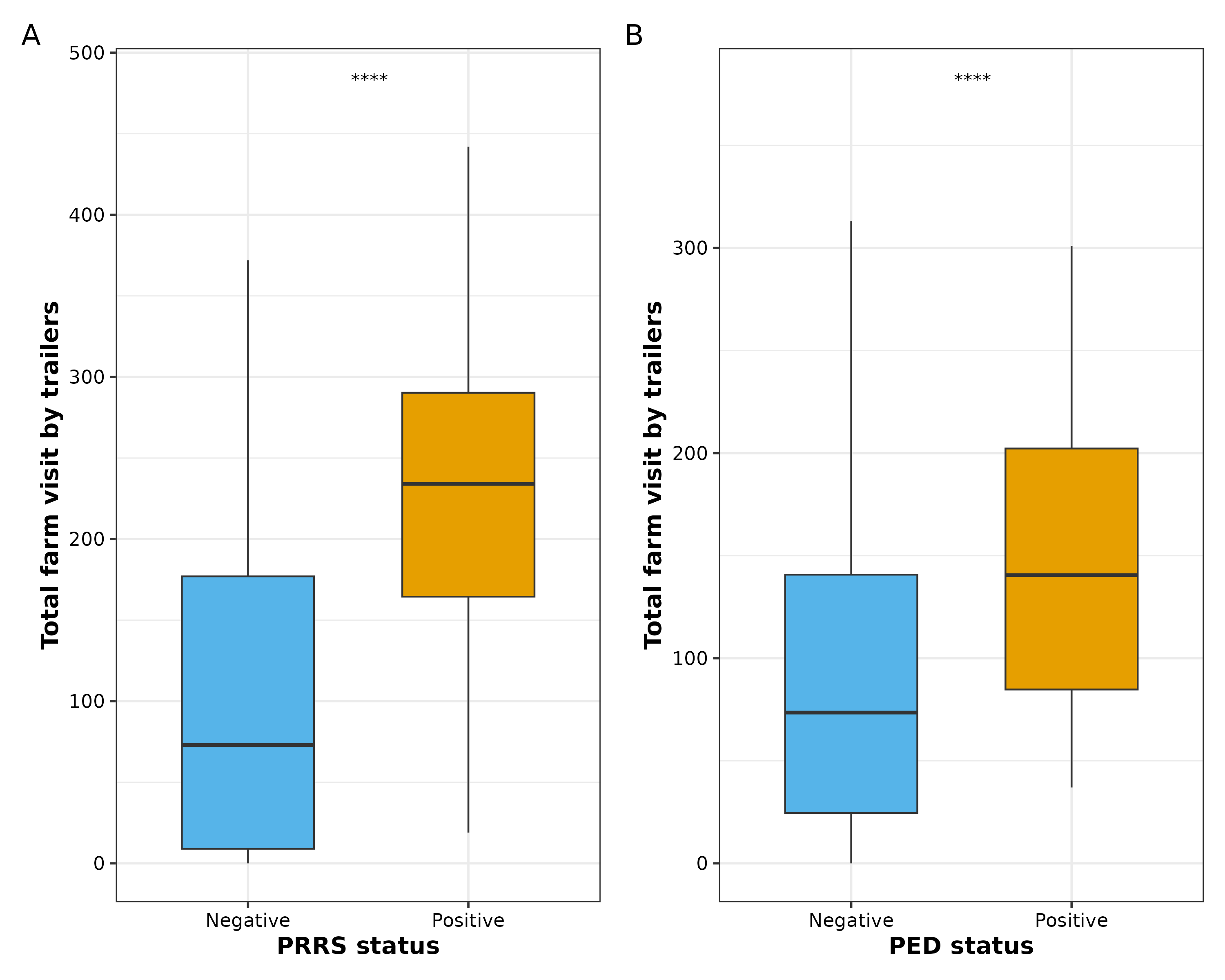}
\caption{Total farm visit by trailers exposed to positive farms seven days before the visit}
\label{fig:SM_23_disease_total_farm_contaminated_trailer_visit_7dayB}
\end{figure}

\begin{figure}[H]
\centering
\includegraphics[width=0.9\textwidth]{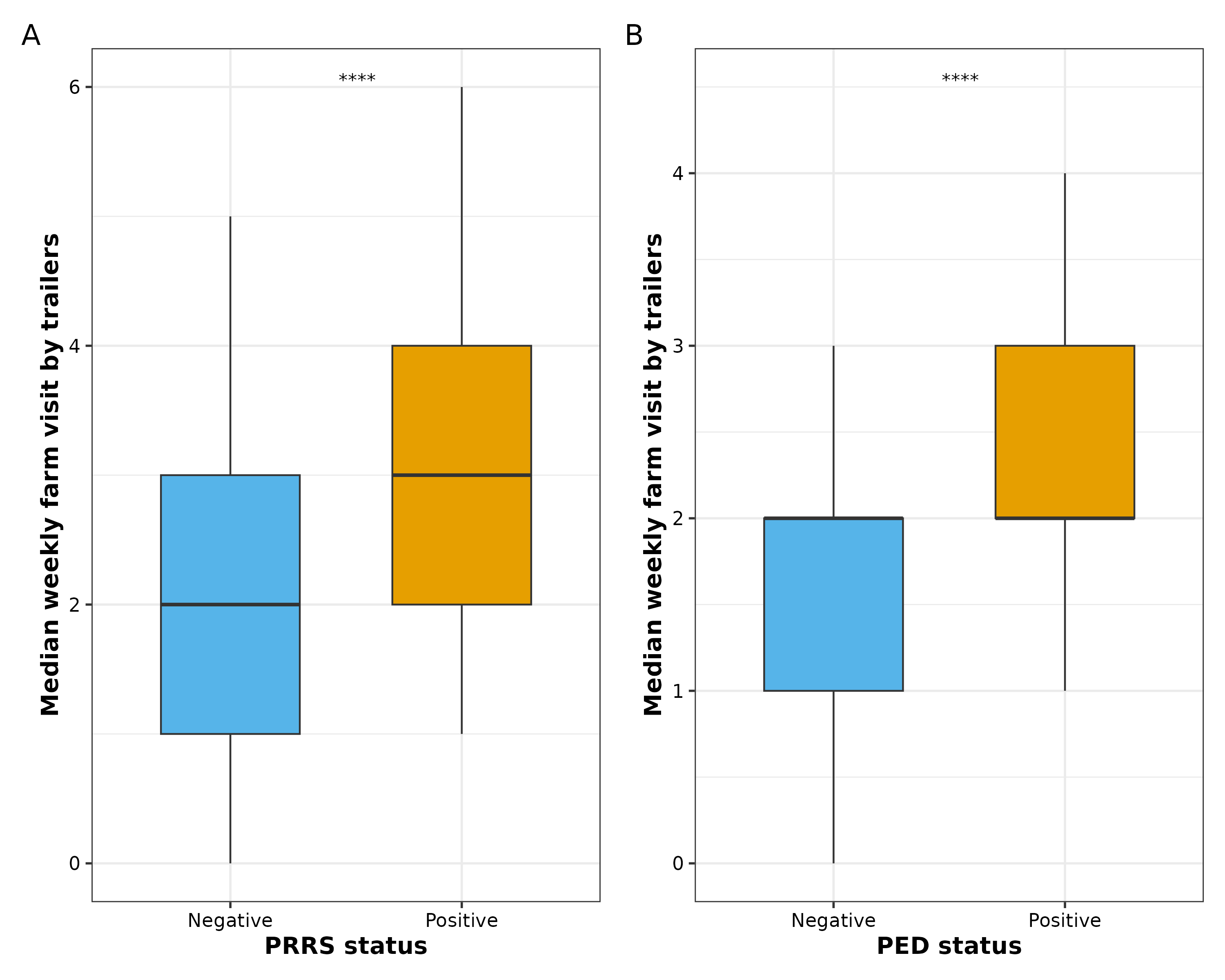}
\caption{Median weekly farm visit by trailers exposed to positive farms 11 days before the visit}
\label{fig:SM_24_disease_weekly_farm_contaminated_trailer_visit_11dayB}
\end{figure}

\begin{figure}[H]
\centering
\includegraphics[width=0.9\textwidth]{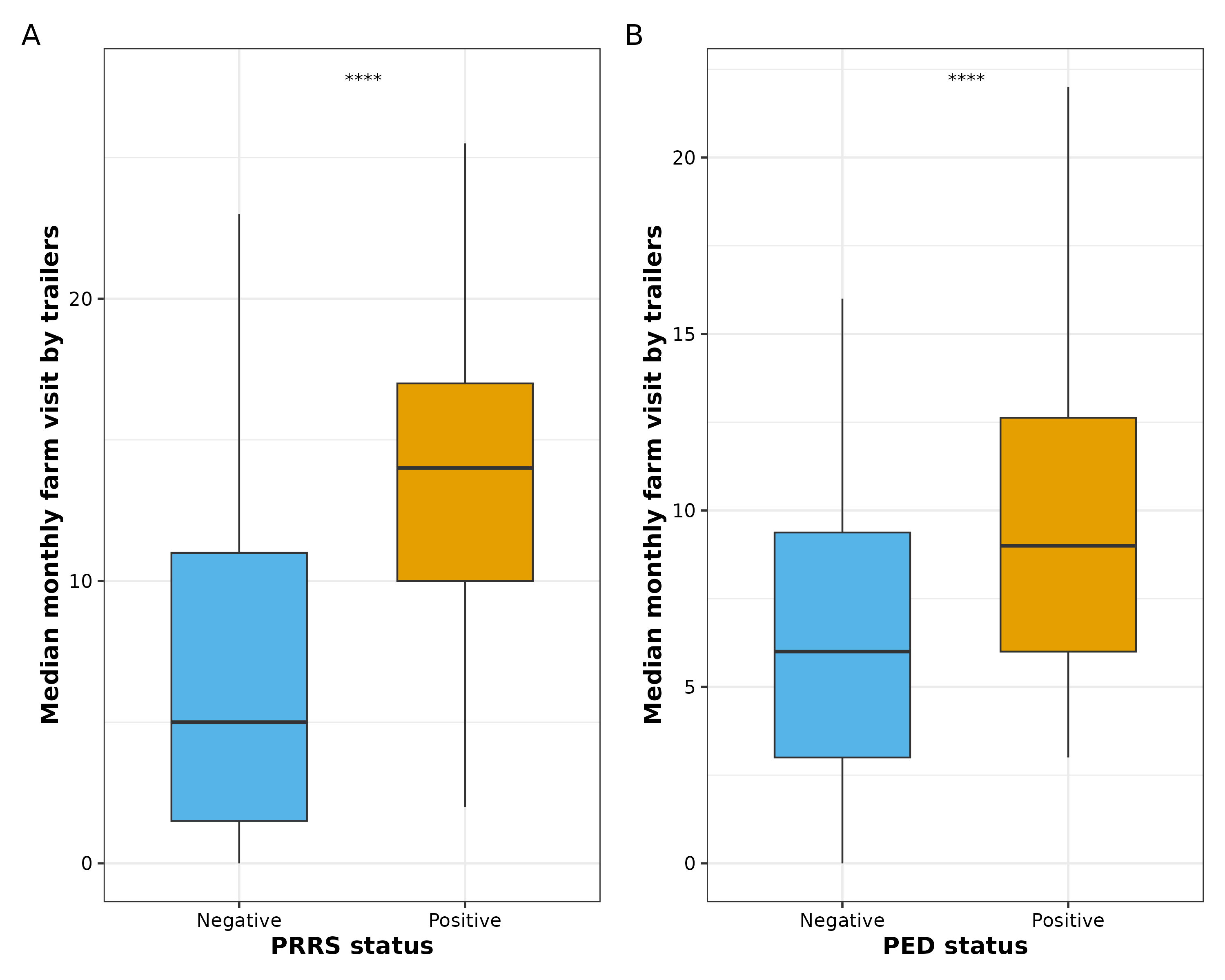}
\caption{Median monthly farm visit by trailers exposed to positive farms 11 days before the visit}
\label{fig:SM_25_disease_monthly_farm_contaminated_trailer_visit_11dayB}
\end{figure}

\begin{figure}[H]
\centering
\includegraphics[width=0.9\textwidth]{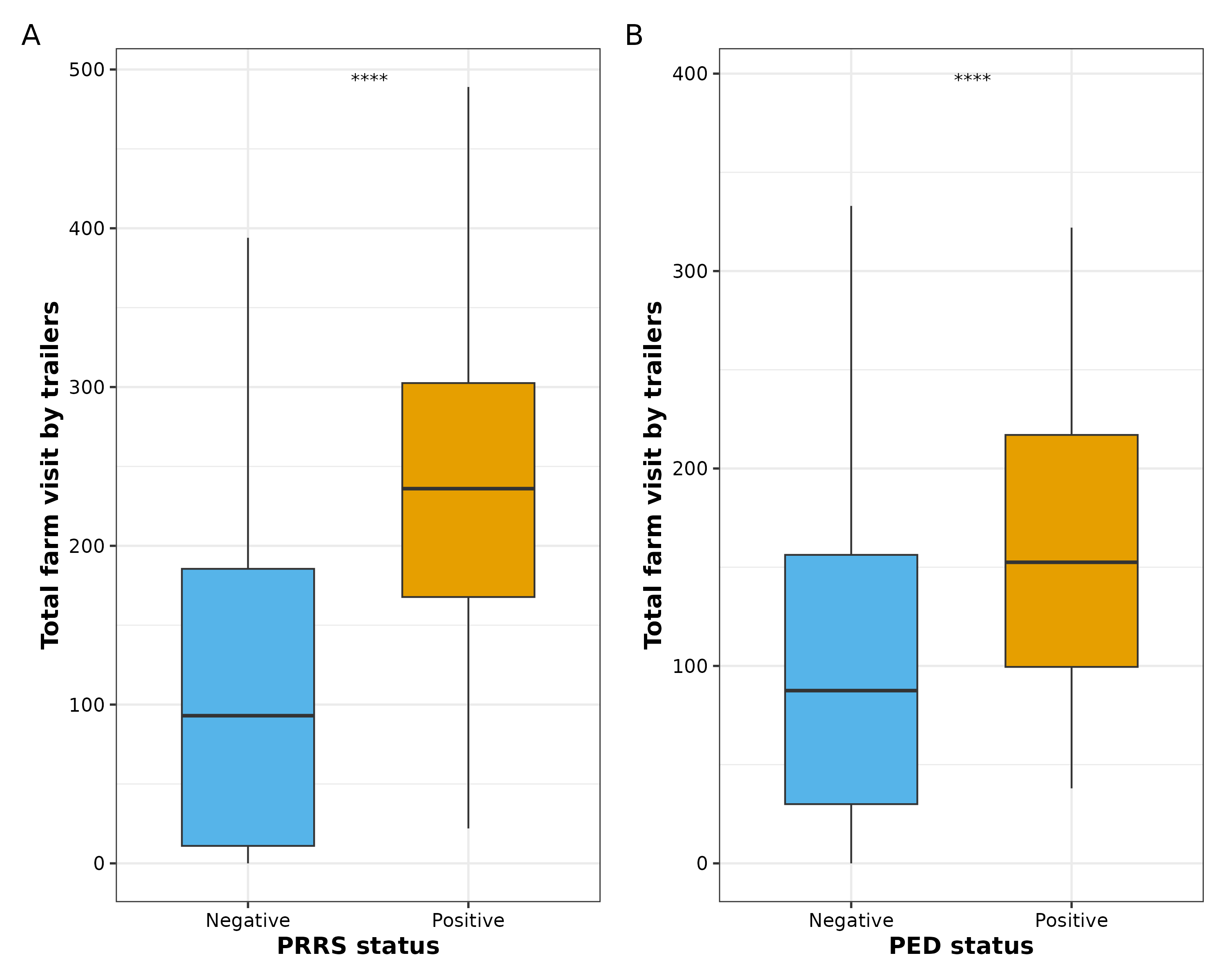}
\caption{Total farm visit by trailers exposed to positive farms 11 days before the visit}
\label{fig:SM_26_disease_total_farm_contaminated_trailer_visit_11dayB}
\end{figure}

\begin{figure}[H]
\centering
\includegraphics[width=0.9\textwidth]{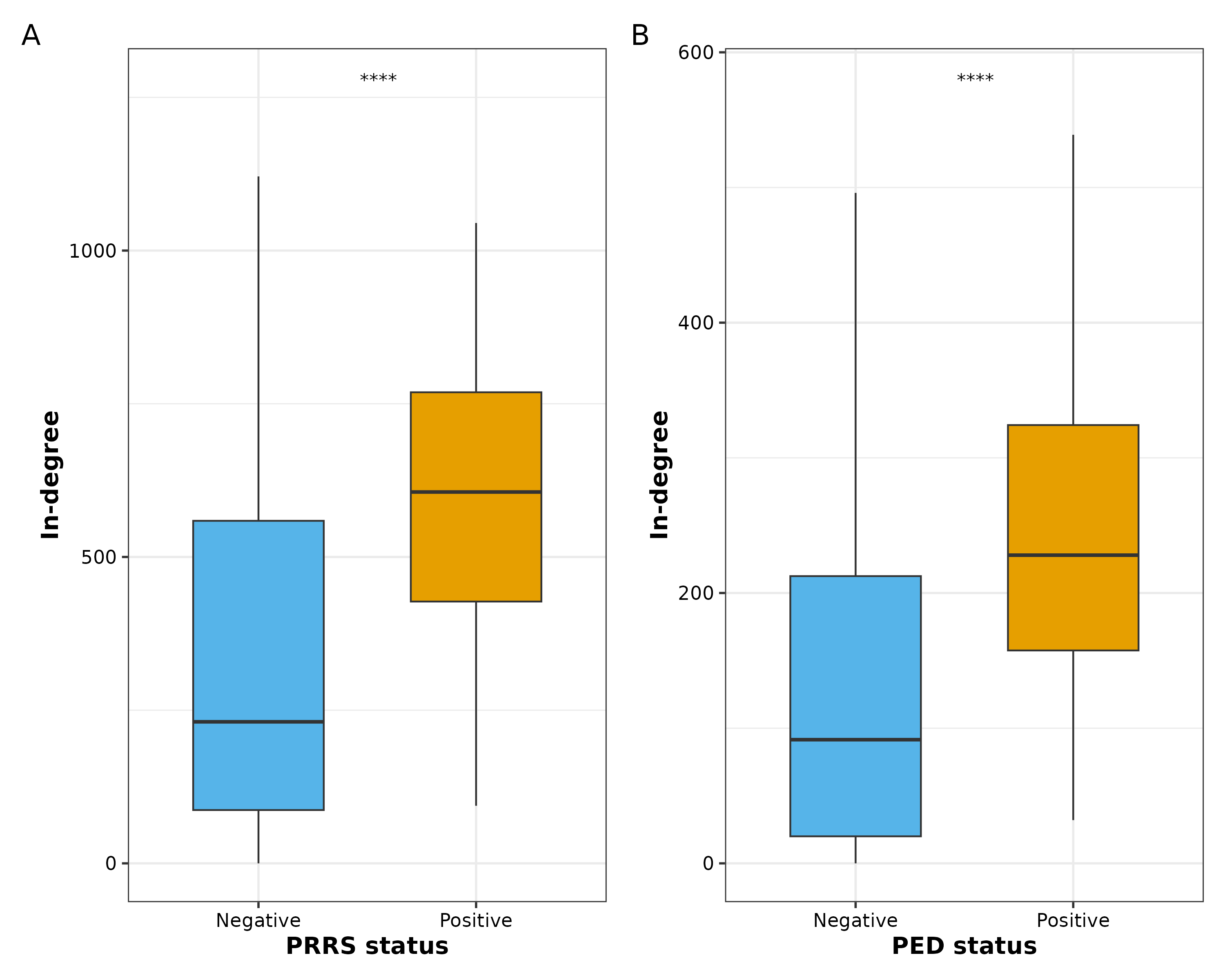}
\caption{Vehicle cab network in-degree of positive farms}
\label{fig:SM_27_disease_infected_indegree_truck}
\end{figure}

\begin{figure}[H]
\centering
\includegraphics[width=0.9\textwidth]{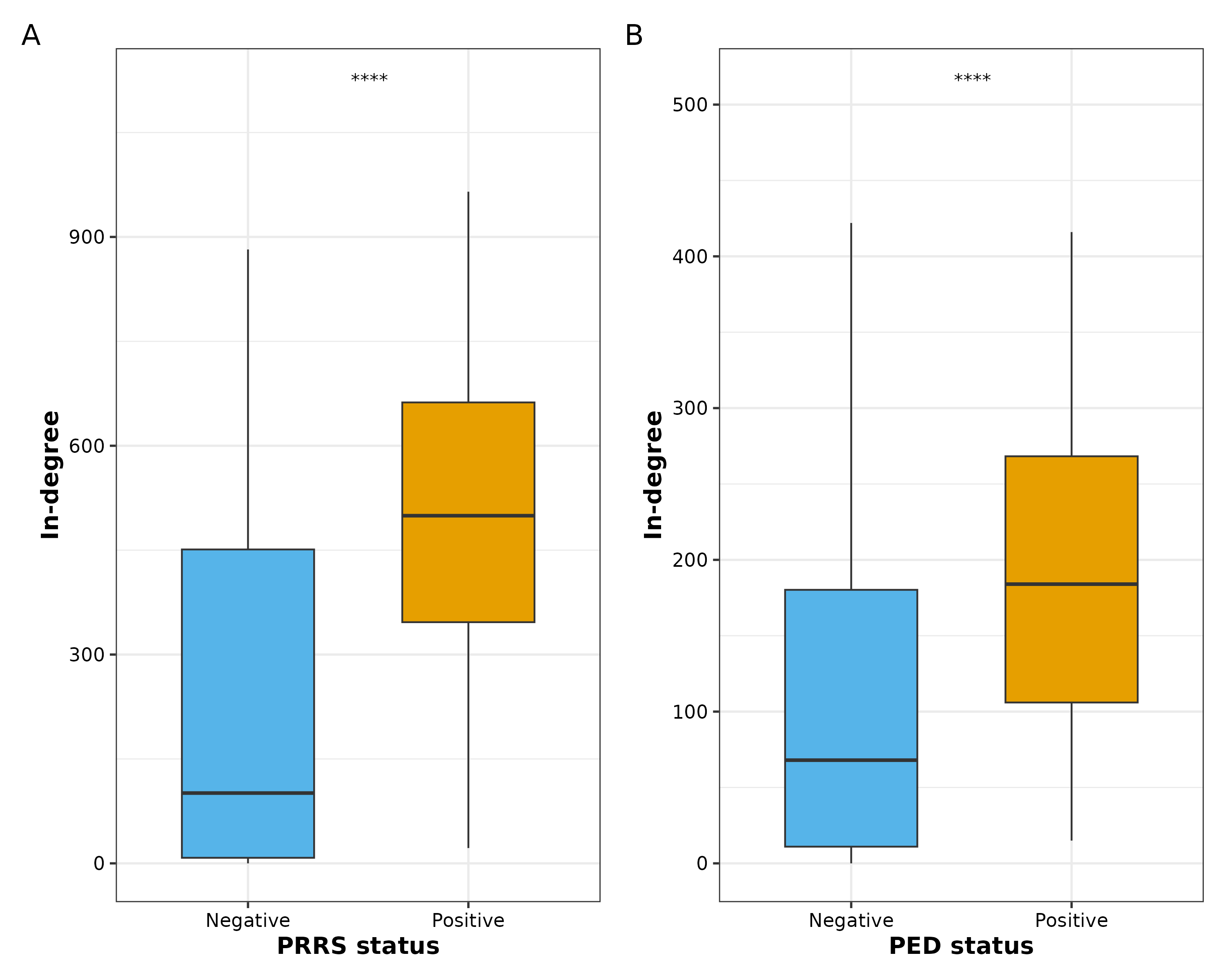}
\caption{Trailer network in-degree of positive farms}
\label{fig:SM_28_disease_infected_indegree_trailer}
\end{figure}

\begin{figure}[H]
\centering
\includegraphics[width=0.9\textwidth]{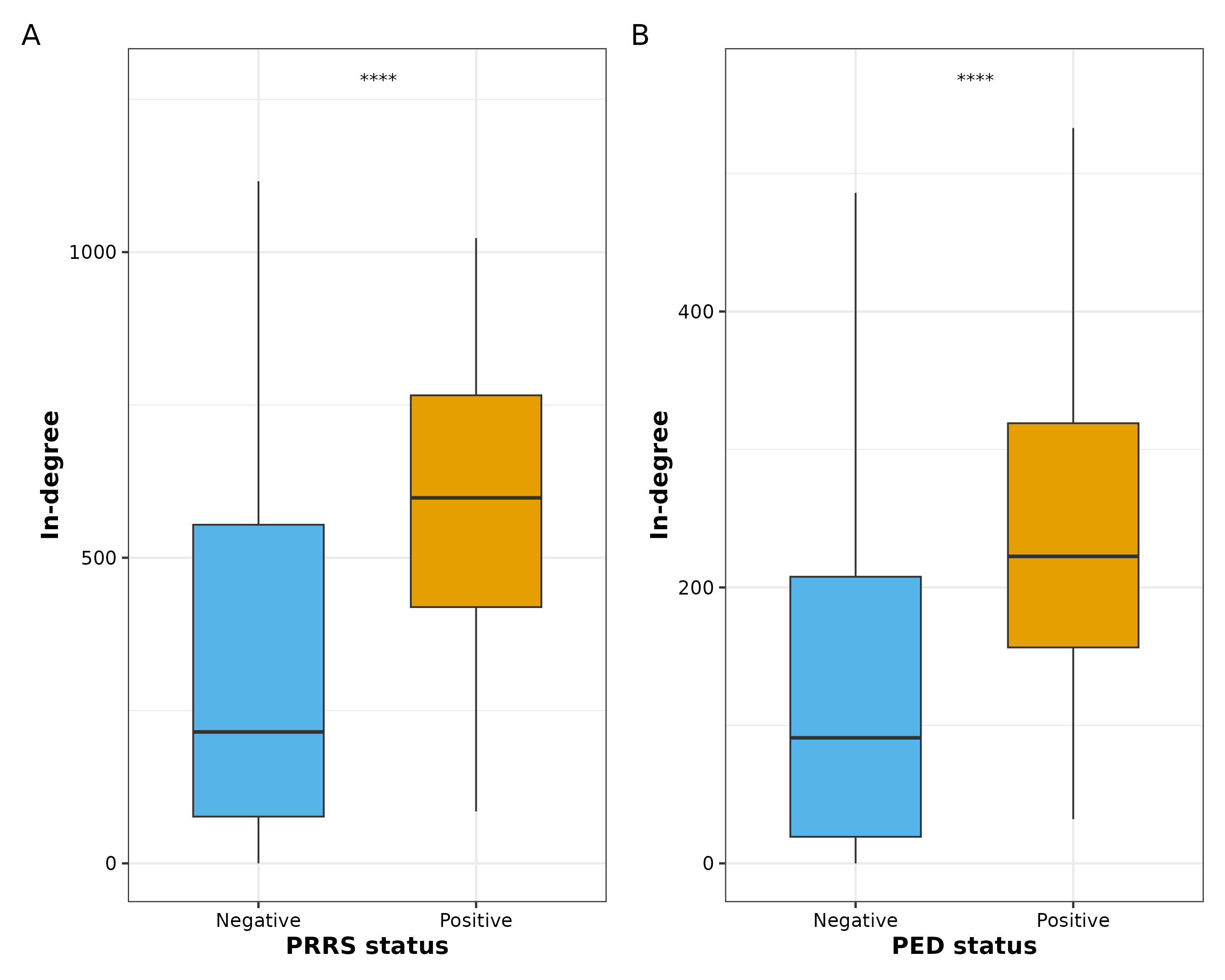}
\caption{Vehicle cab network in-degree of positive farms with C\&D}
\label{fig:SM_29_disease_infected_indegree_truck_cd}
\end{figure}

\begin{figure}[H]
\centering
\includegraphics[width=0.9\textwidth]{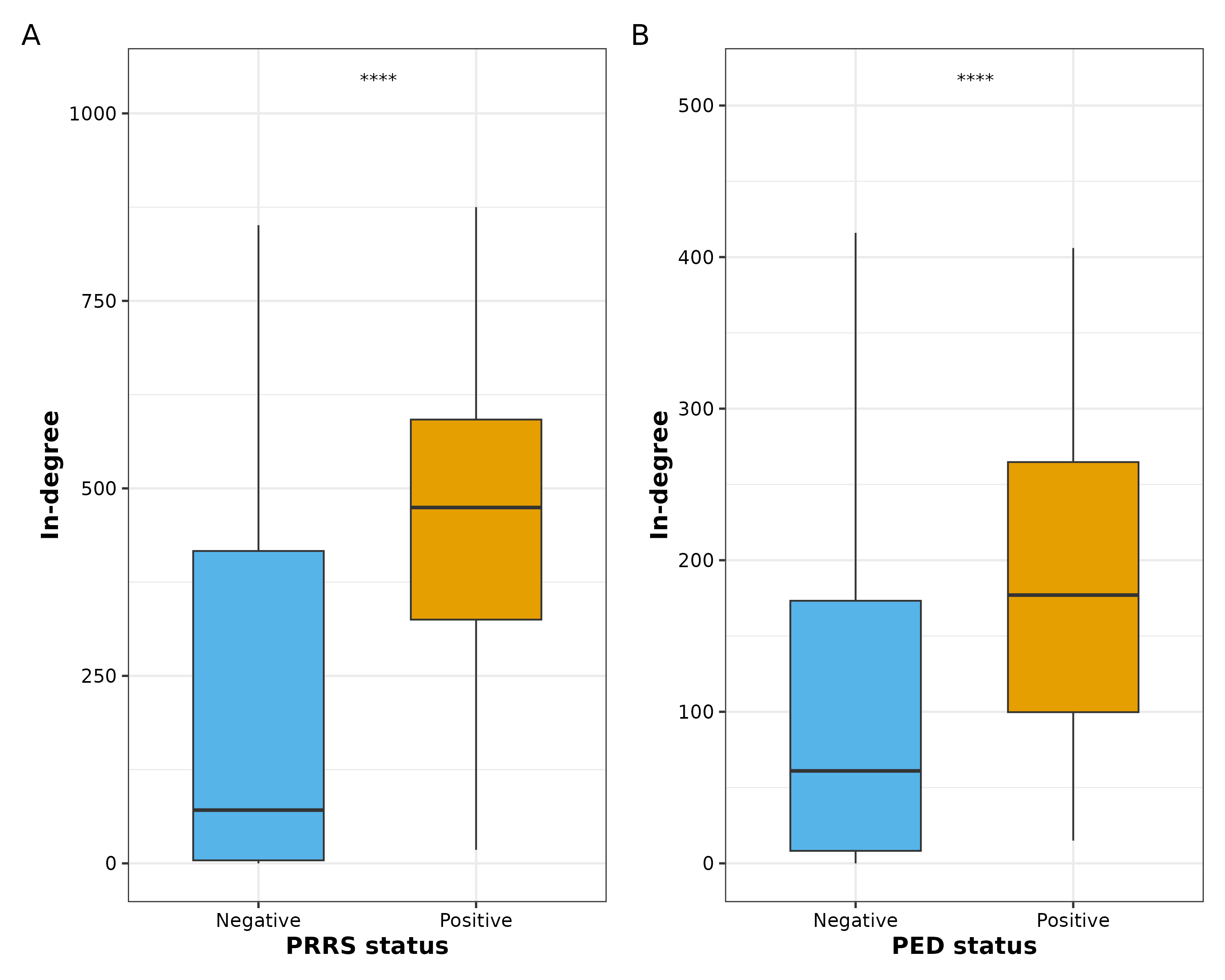}
\caption{Trailer network in-degree of positive farms  with C\&D}
\label{fig:SM_30_disease_infected_indegree_trailer_cd}
\end{figure}

\begin{figure}[H]
\centering
\includegraphics[width=0.9\textwidth]{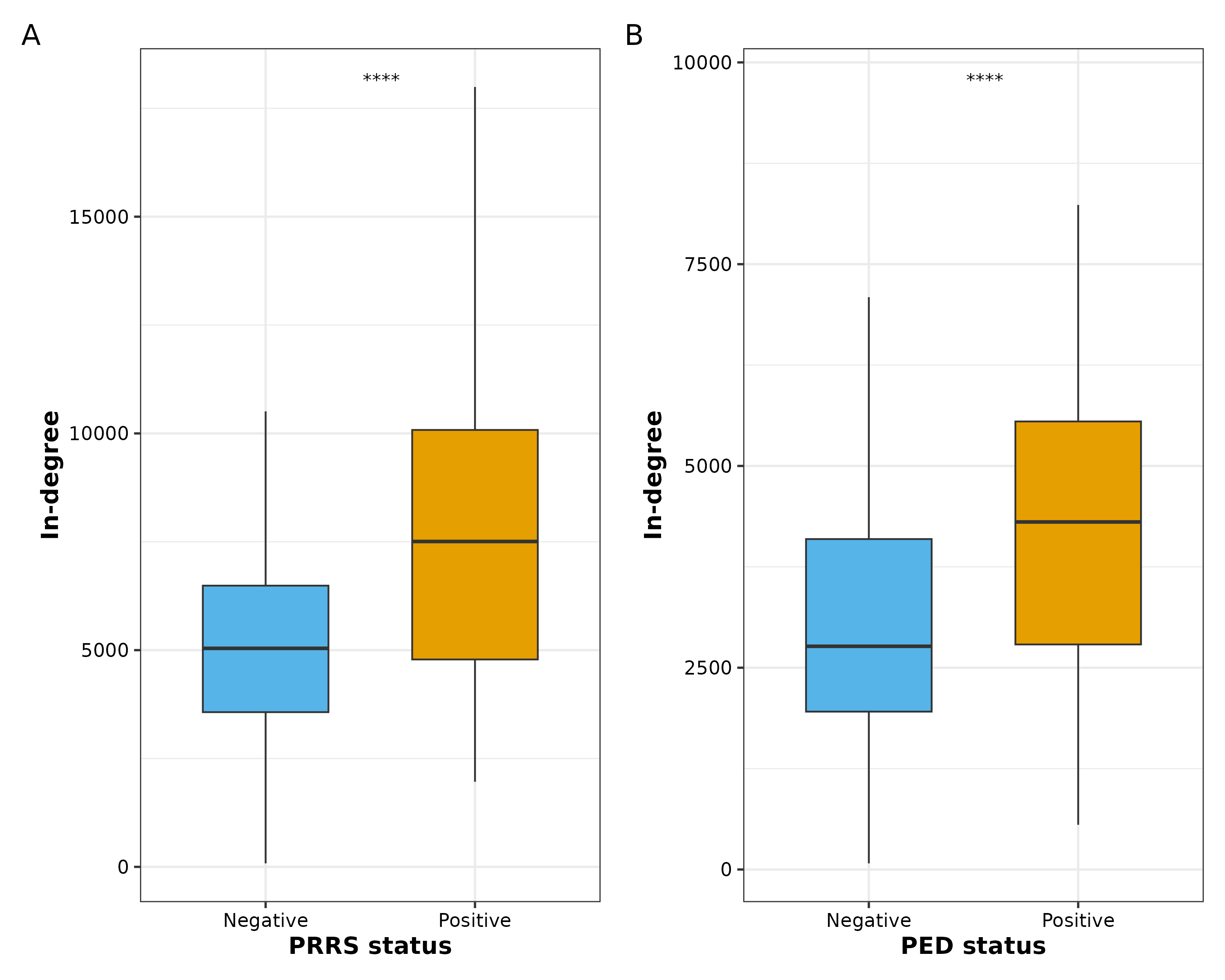}
\caption{Vehicle cab network in-degree}
\label{fig:SM_31_disease_indegree_truck}
\end{figure}

\begin{figure}[H]
\centering
\includegraphics[width=0.9\textwidth]{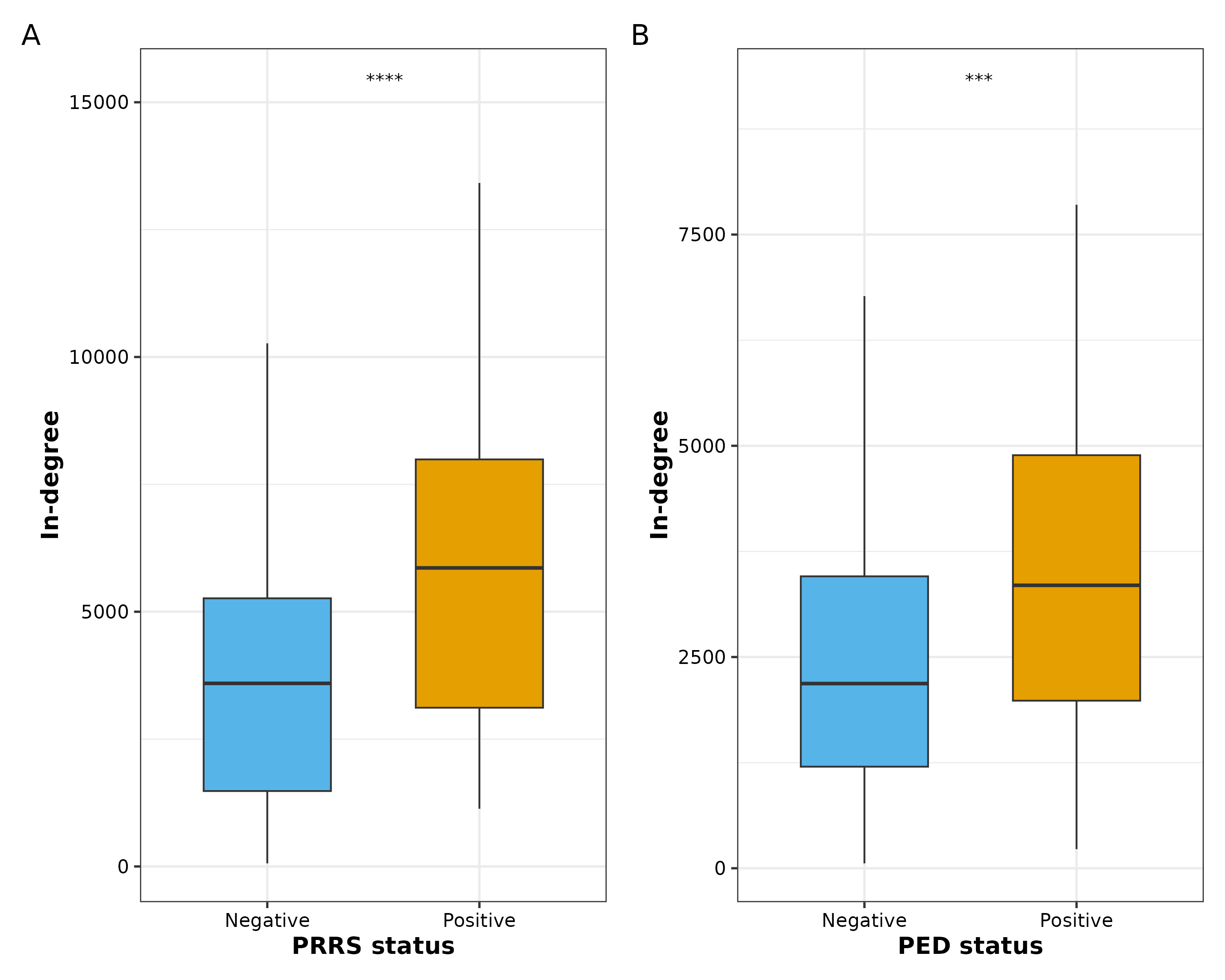}
\caption{Trailer network in-degree}
\label{fig:SM_32_disease_indegree_trailer}
\end{figure}

\begin{figure}[H]
\centering
\includegraphics[width=0.9\textwidth]{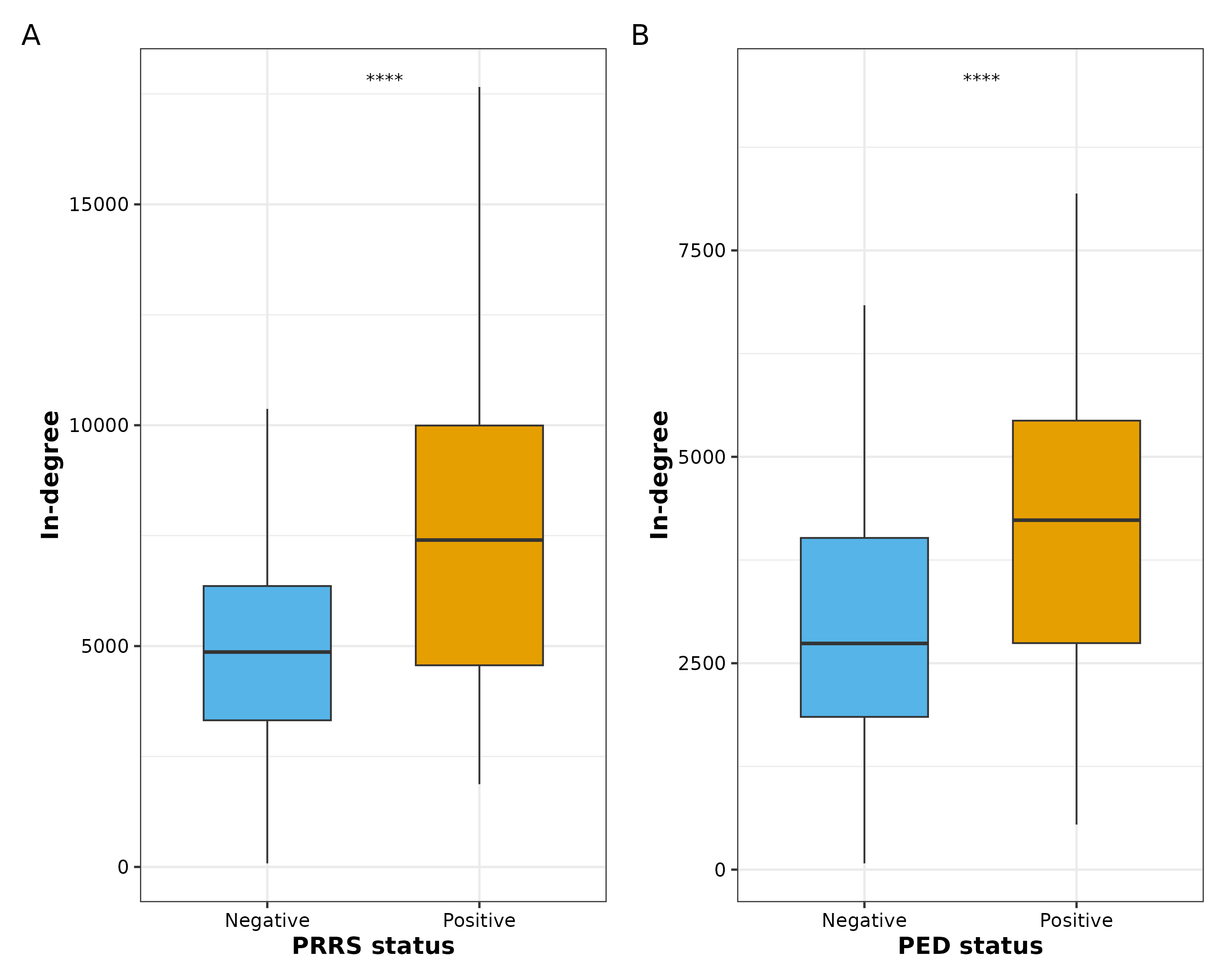}
\caption{Vehicle cab network in-degree with C\&D}
\label{fig:SM_33_disease_indegree_truck_cd}
\end{figure}

\begin{figure}[H]
\centering
\includegraphics[width=0.9\textwidth]{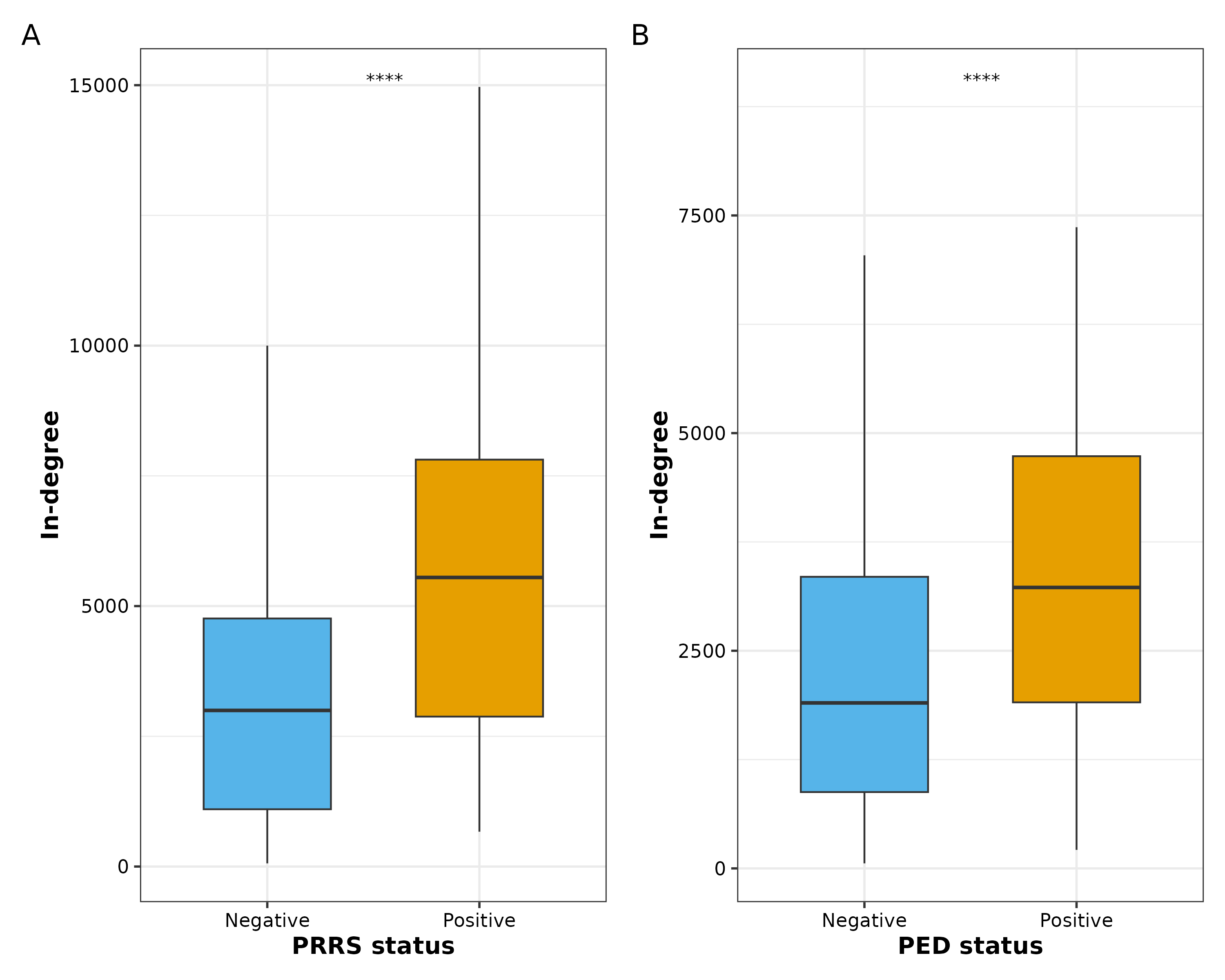}
\caption{Trailer network in-degree with C\&D}
\label{fig:SM_34_disease_indegree_trailer_cd}
\end{figure}

\begin{figure}[H]
\centering
\includegraphics[width=0.9\textwidth]{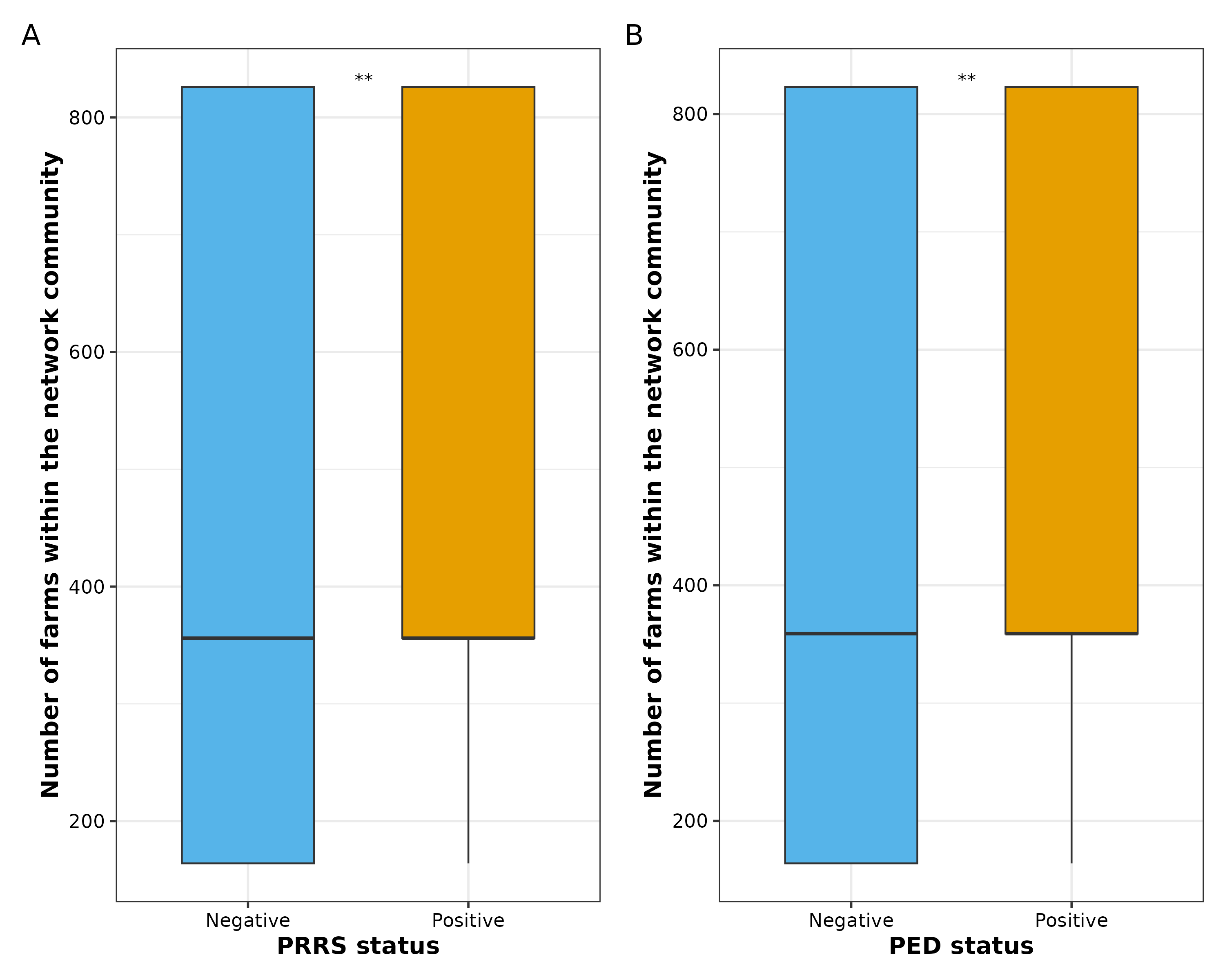}
\caption{Number of farms within the vehicle cab network community}
\label{fig:SM_35_disease_community_size_truck}
\end{figure}

\begin{figure}[H]
\centering
\includegraphics[width=0.9\textwidth]{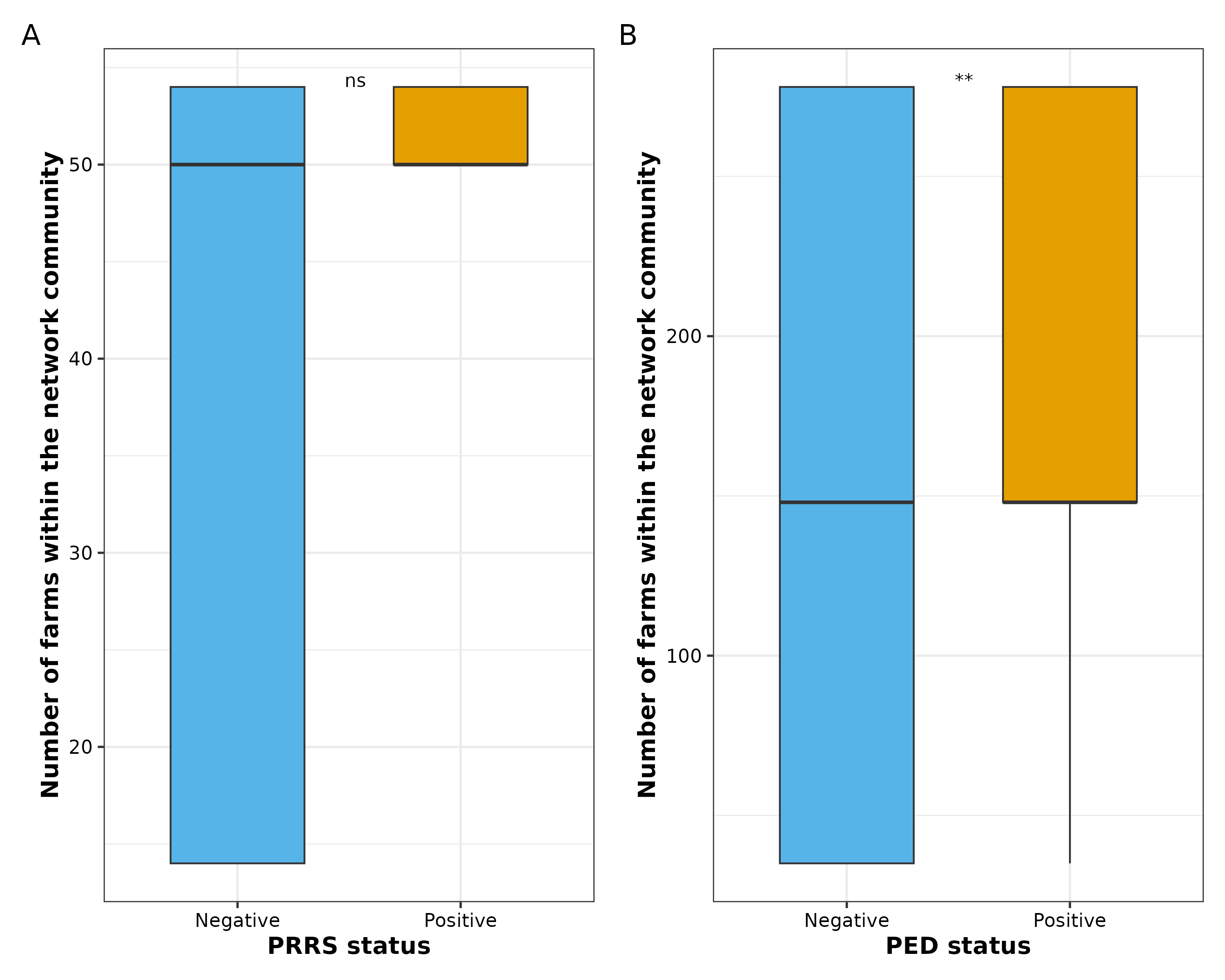}
\caption{Number of positive farms within the vehicle cab network community}
\label{fig:SM_36_disease_infected_community_size_truck}
\end{figure}

\begin{figure}[H]
\centering
\includegraphics[width=0.9\textwidth]{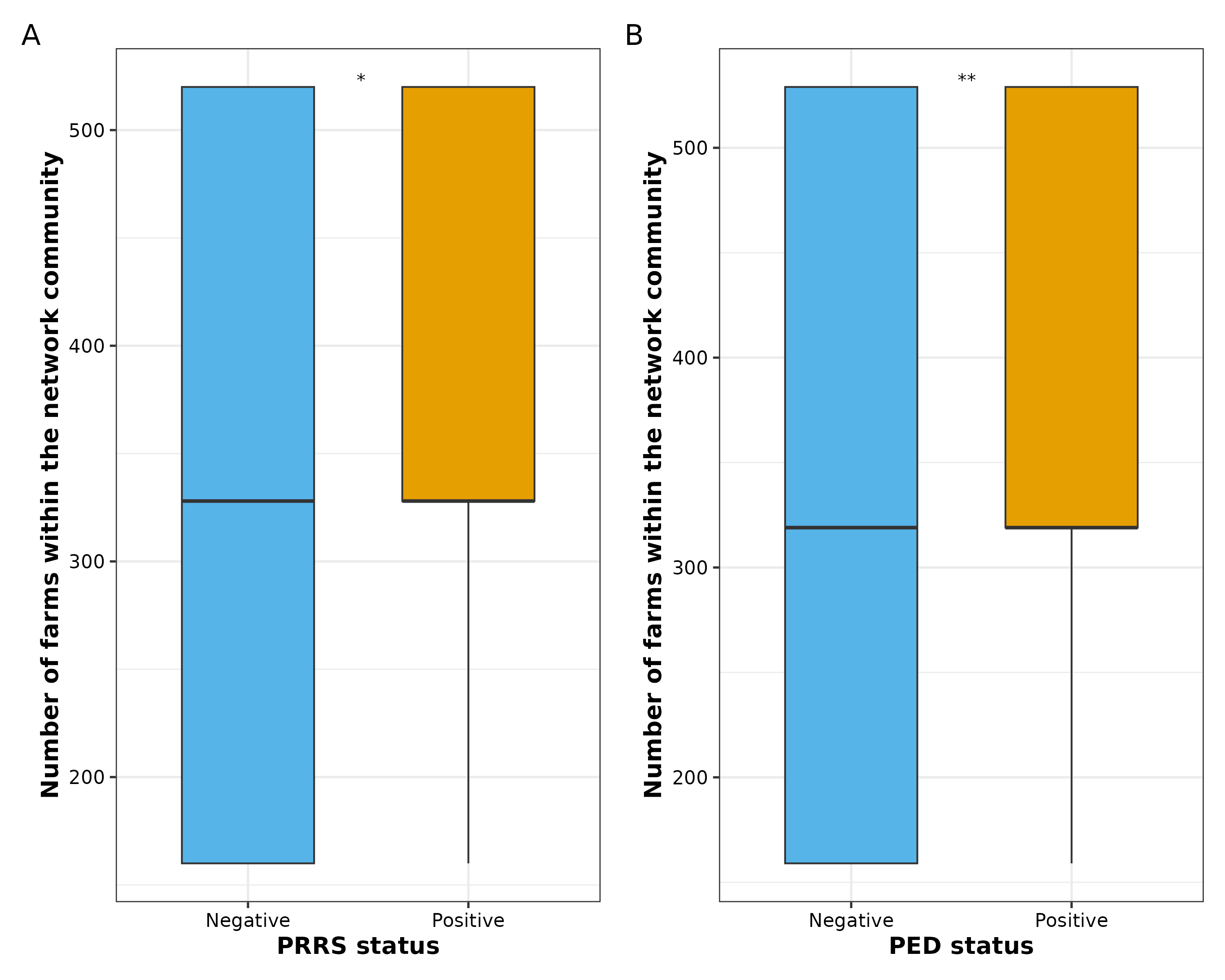}
\caption{Number of farms within the trailer network community}
\label{fig:SM_37_disease_community_size_trailer}
\end{figure}

\begin{figure}[H]
\centering
\includegraphics[width=0.9\textwidth]{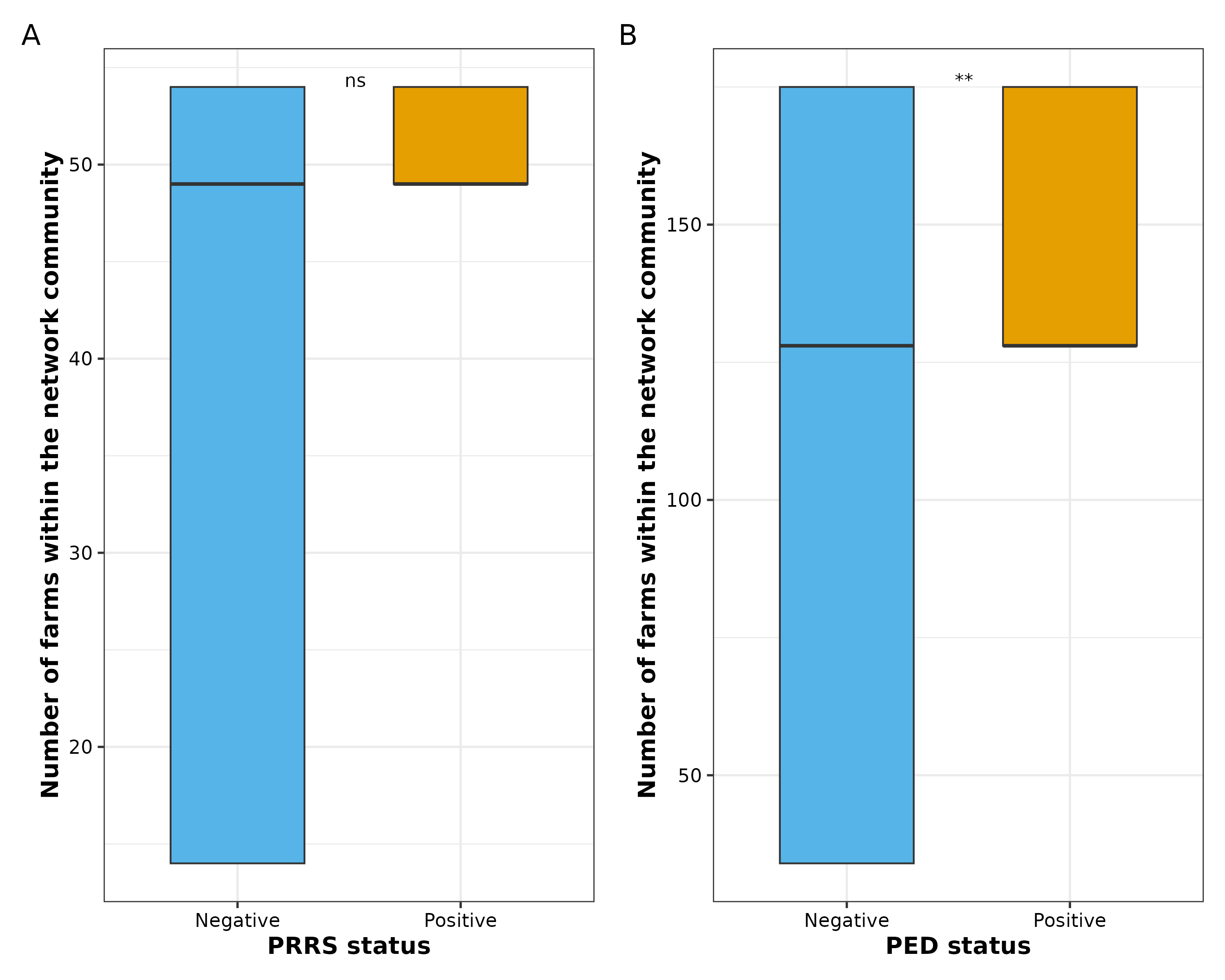}
\caption{Number of positive farms within the trailer network community}
\label{fig:SM_38_disease_infected_community_size_trailer}
\end{figure}

\begin{figure}[H]
\centering
\includegraphics[width=0.9\textwidth]{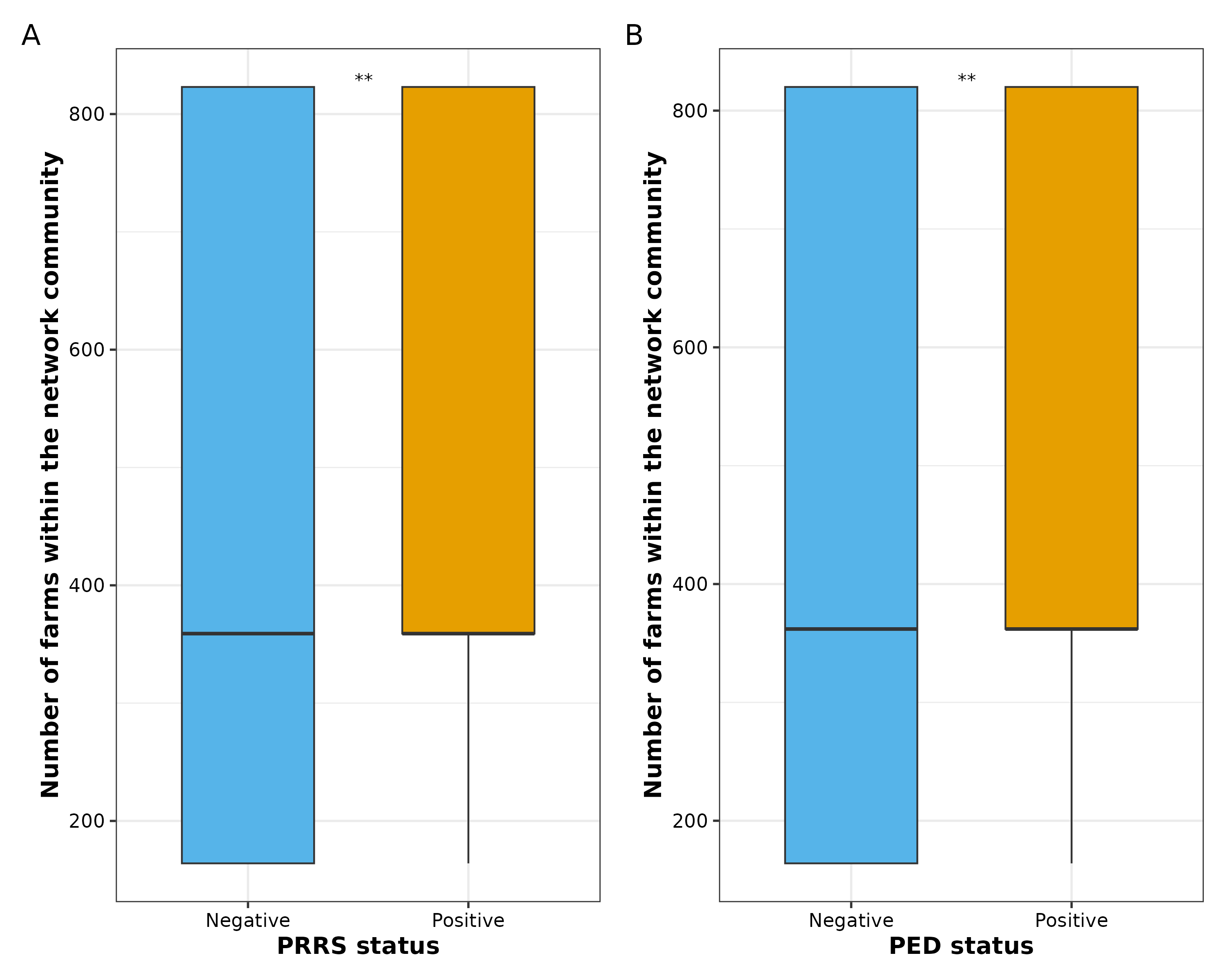}
\caption{Number of farms within the vehicle cab network community with C\&D}
\label{fig:SM_39_disease_community_size_truck_cd}
\end{figure}

\begin{figure}[H]
\centering
\includegraphics[width=0.9\textwidth]{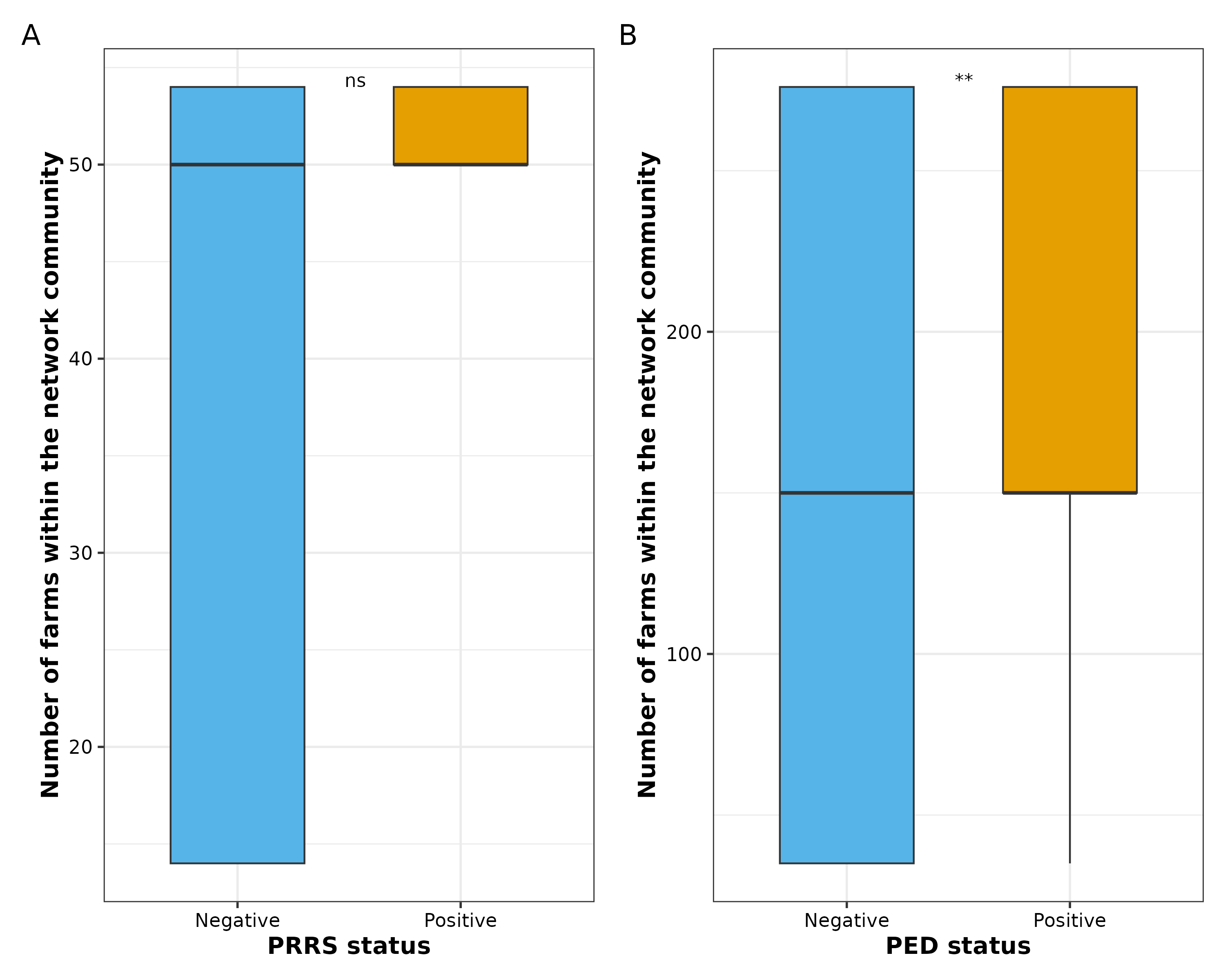}
\caption{Number of positive farms within the vehicle cab network community with C\&D}
\label{fig:SM_40_disease_infected_community_size_truck_cd}
\end{figure}

\begin{figure}[H]
\centering
\includegraphics[width=0.9\textwidth]{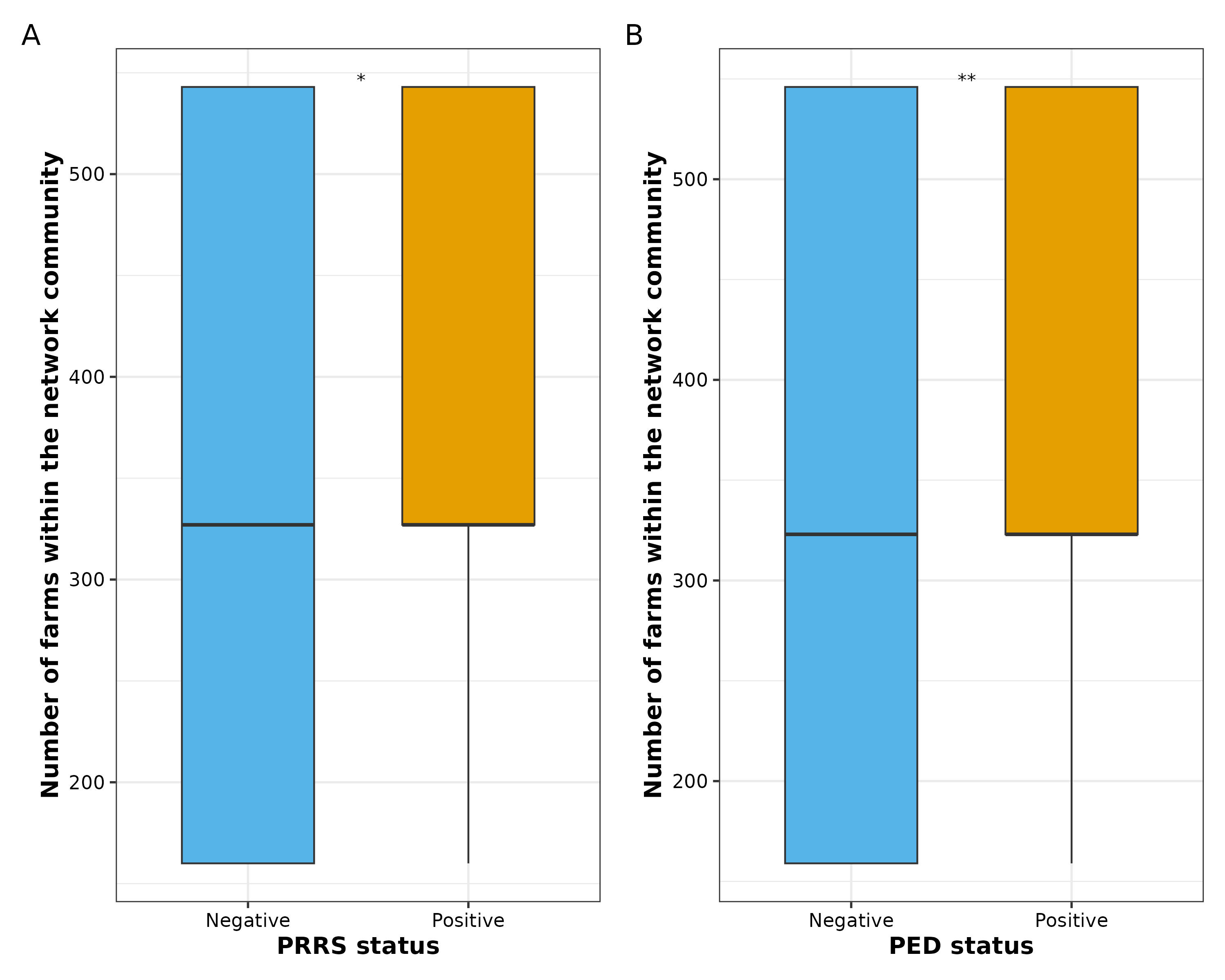}
\caption{Number of farms within the trailer network community with C\&D}
\label{fig:SM_41_disease_community_size_trailer_cd}
\end{figure}

\begin{figure}[H]
\centering
\includegraphics[width=0.9\textwidth]{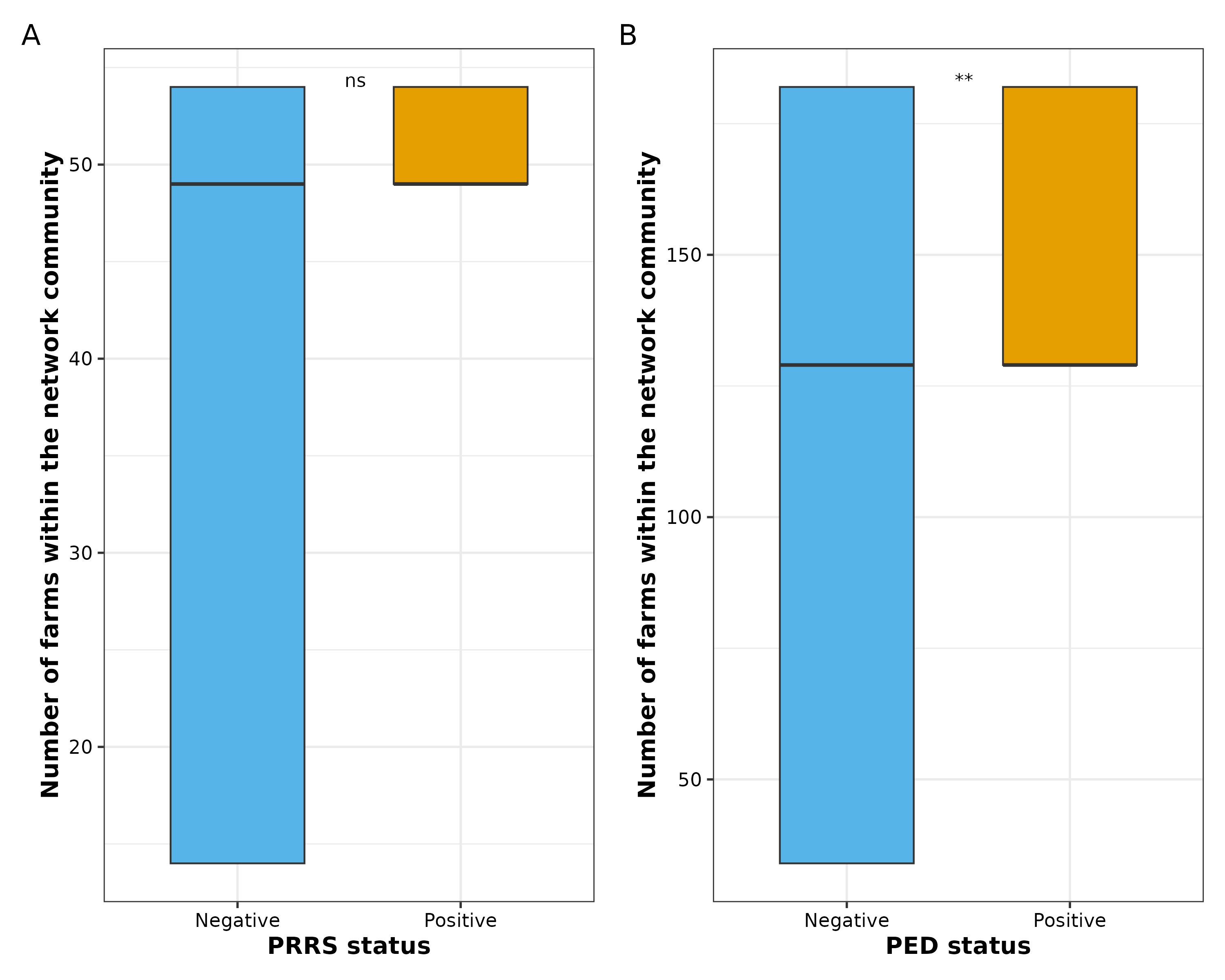}
\caption{Number of positive farms within the trailer network community with C\&D}
\label{fig:SM_42_disease_infected_community_size_trailer_cd}
\end{figure}

\begin{figure}[H]
\centering
\includegraphics[width=0.9\textwidth]{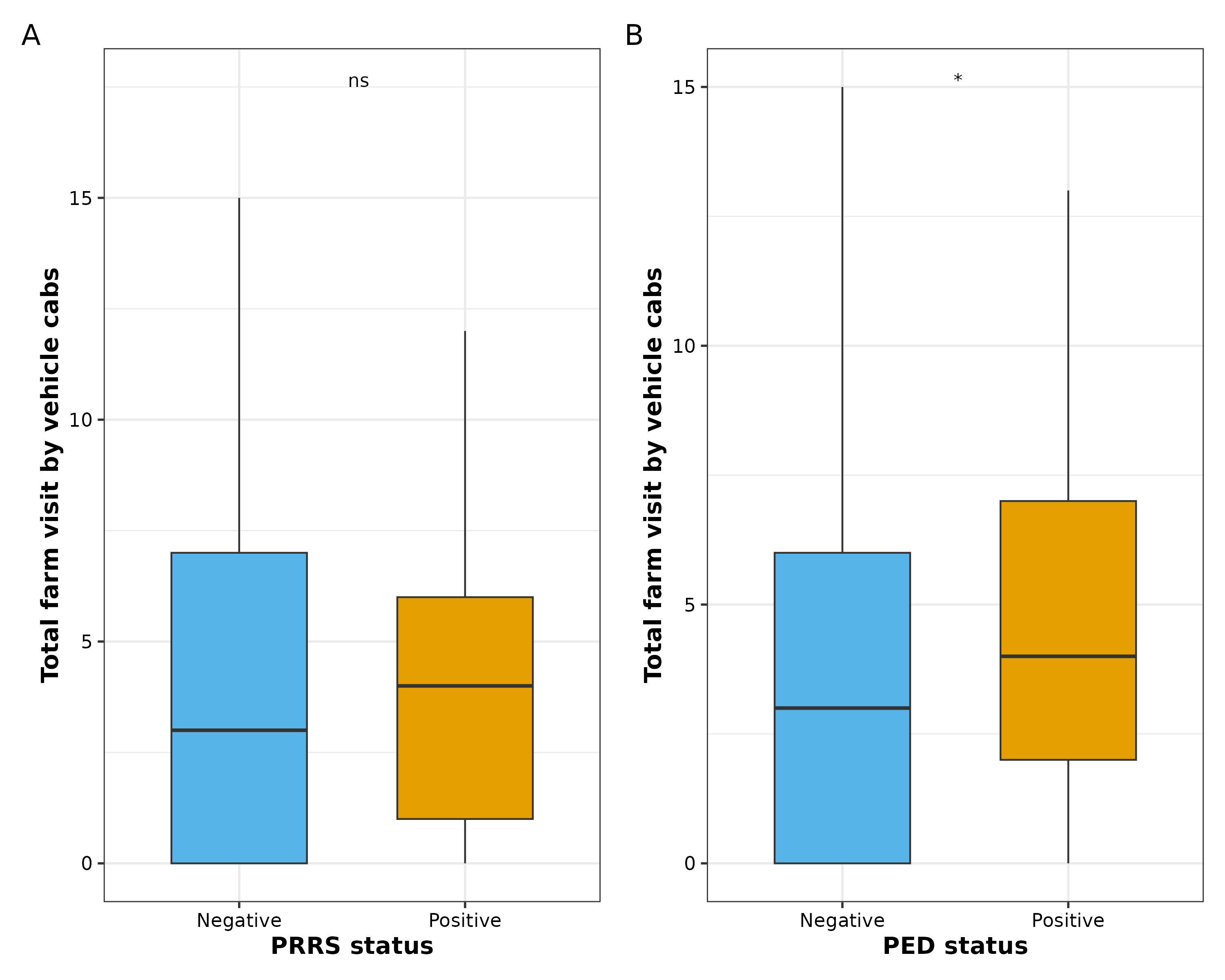}
\caption{Total farm visit by vehicle cabs exposed to slaughterhouses one day before the visit}
\label{fig:SM_43_disease_total_farm_market_truck_visit_1dayB}
\end{figure}

\begin{figure}[H]
\centering
\includegraphics[width=0.9\textwidth]{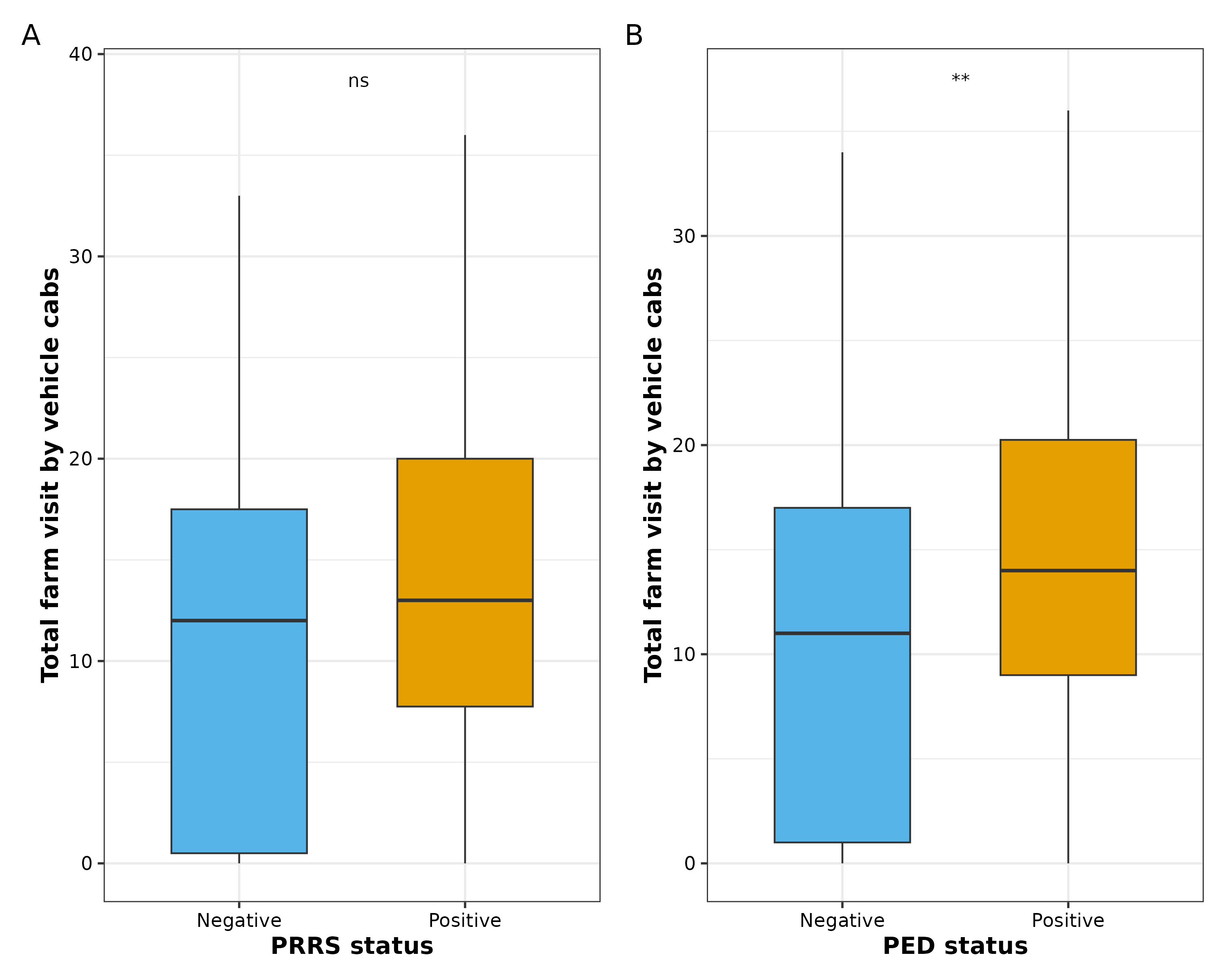}
\caption{Total farm visit by vehicle cabs exposed to slaughterhouses seven days before the visit}
\label{fig:SM_44_disease_total_farm_market_truck_visit_7dayB}
\end{figure}

\begin{figure}[H]
\centering
\includegraphics[width=0.9\textwidth]{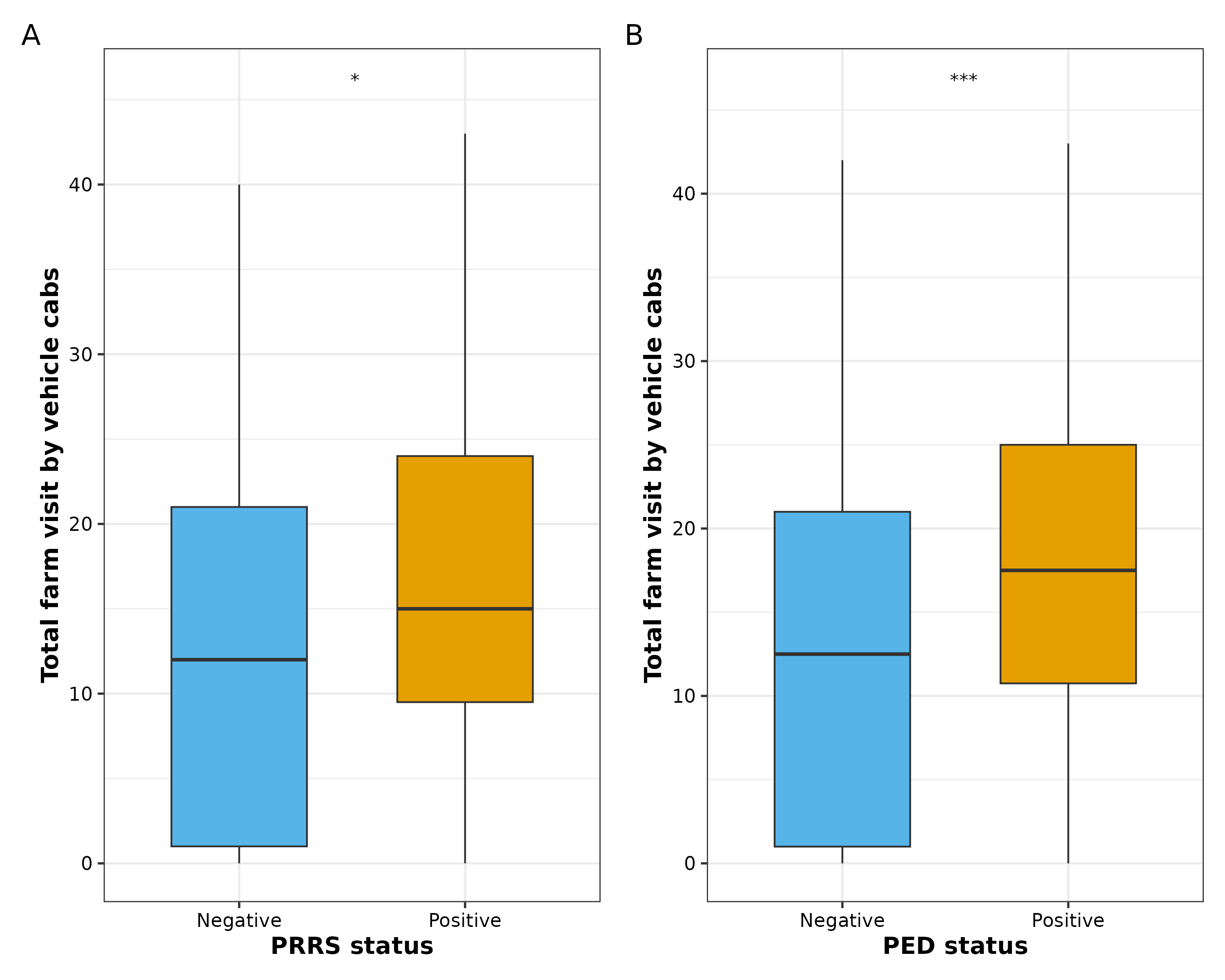}
\caption{Total farm visit by vehicle cabs exposed to slaughterhouses 11 days before the visit}
\label{fig:SM_45_disease_total_farm_market_truck_visit_11dayB}
\end{figure}

\begin{figure}[H]
\centering
\includegraphics[width=0.9\textwidth]{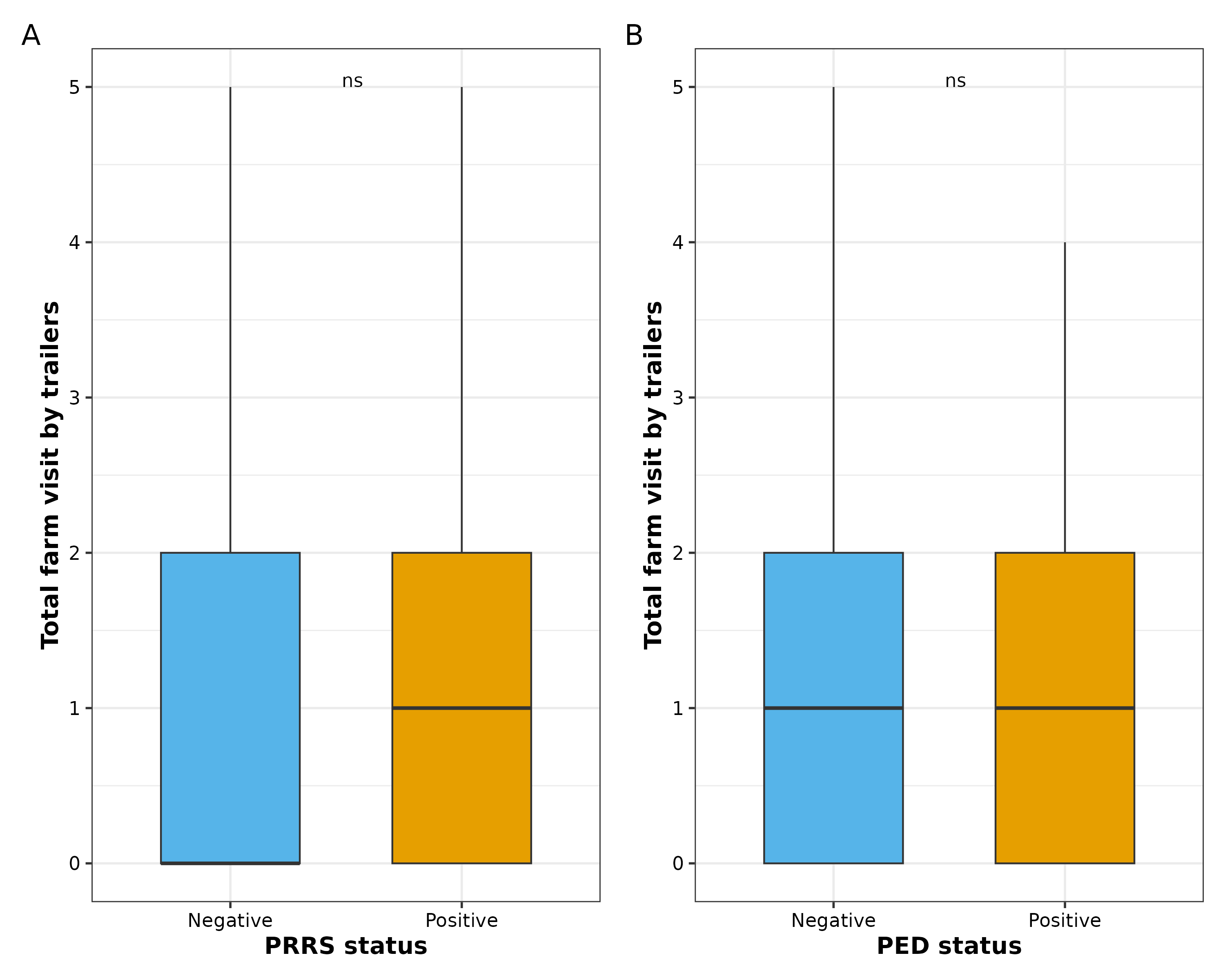}
\caption{Total farm visit by trailers exposed to slaughterhouses one day before the visit}
\label{fig:SM_46_disease_total_farm_market_trailer_visit_1dayB}
\end{figure}

\begin{figure}[H]
\centering
\includegraphics[width=0.9\textwidth]{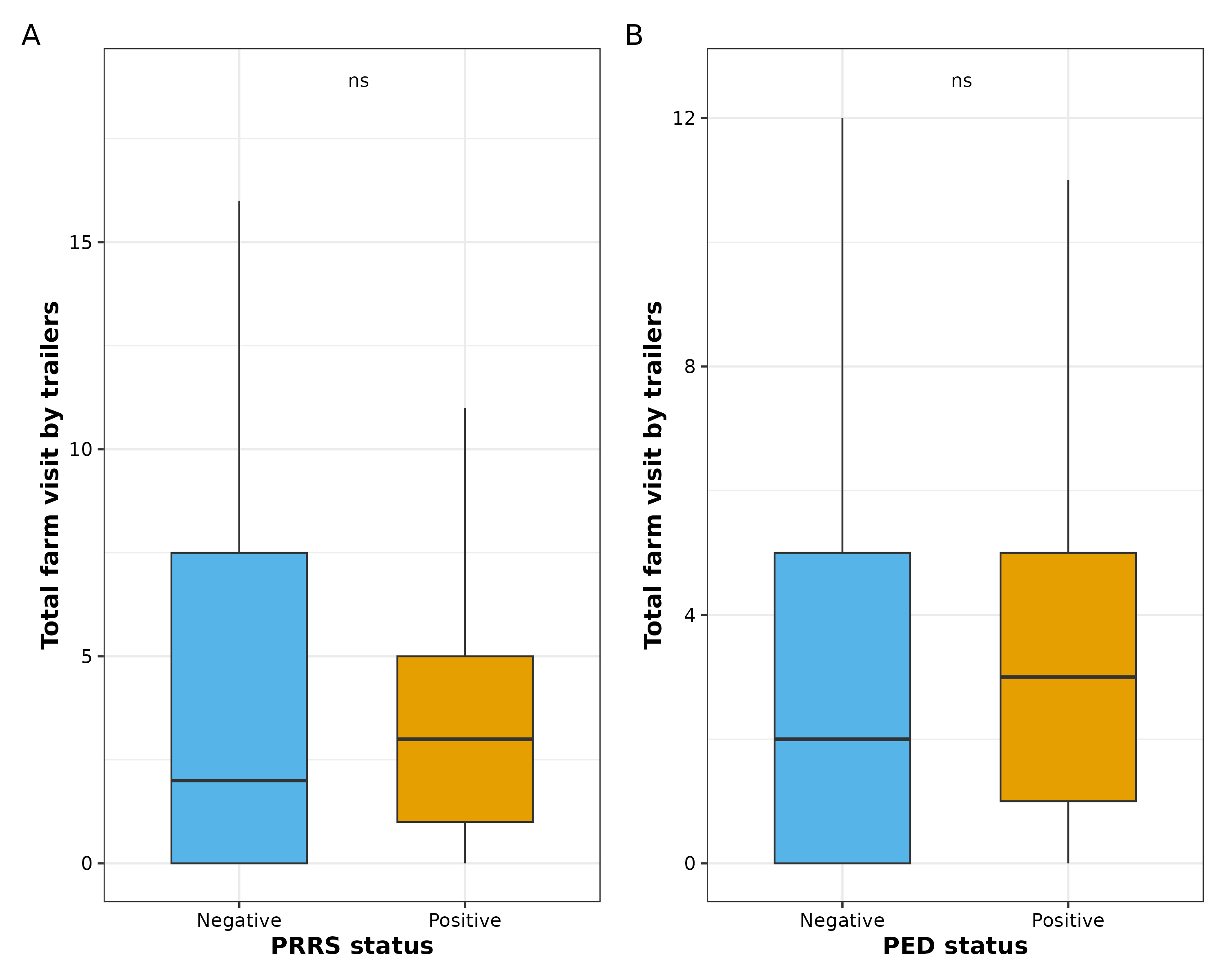}
\caption{Total farm visit by trailers exposed to slaughterhouses seven days before the visit}
\label{fig:SM_47_disease_total_farm_market_trailer_visit_7dayB}
\end{figure}

\begin{figure}[H]
\centering
\includegraphics[width=0.9\textwidth]{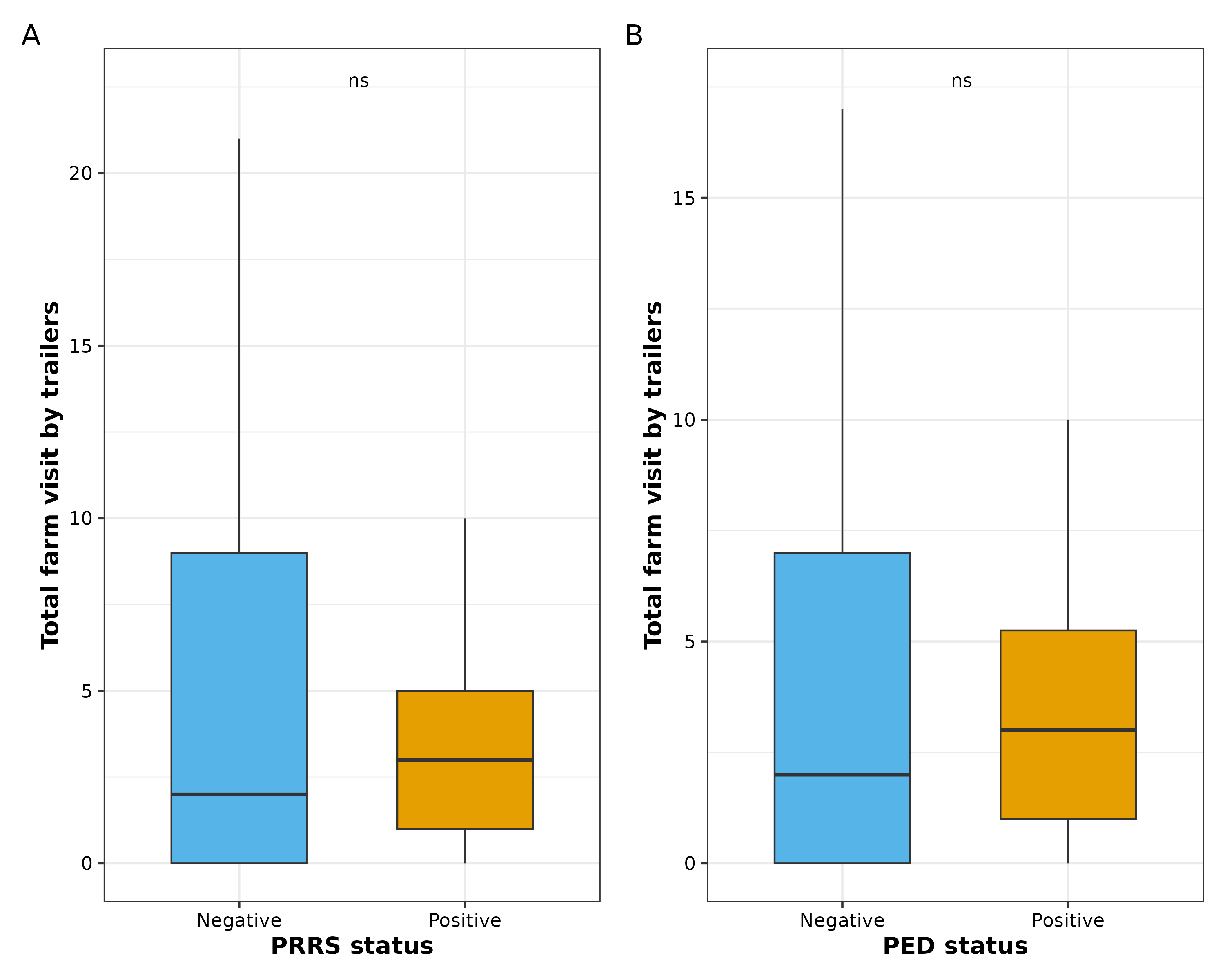}
\caption{Total farm visit by trailers exposed to slaughterhouses 11 days before the visit}
\label{fig:SM_48_disease_total_farm_market_trailer_visit_11dayB}
\end{figure}

\begin{figure}[H]
\centering
\includegraphics[width=0.9\textwidth]{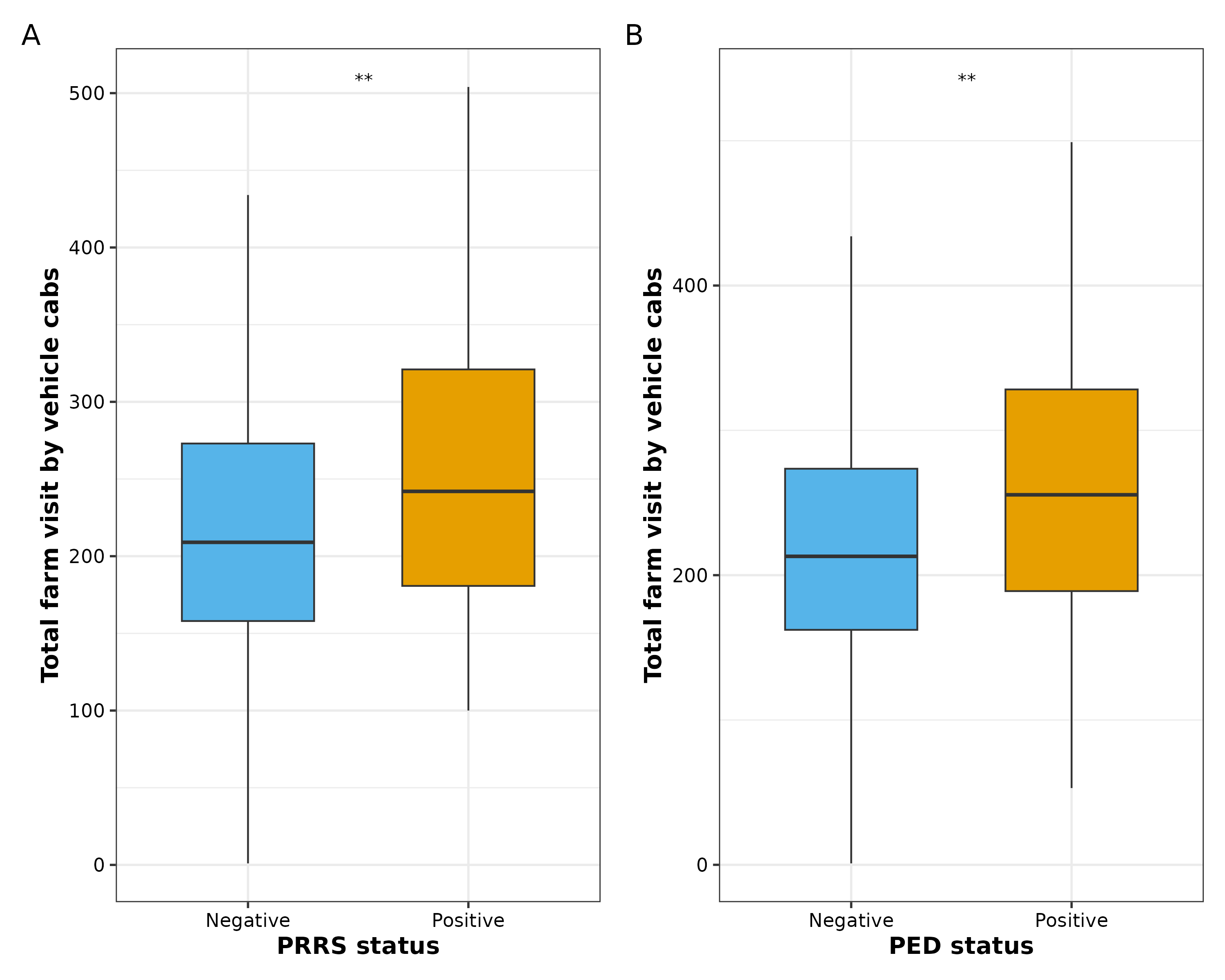}
\caption{Total farm visit by vehicle cabs exposed to finisher farms one day before the visit}
\label{fig:SM_49_disease_total_farm_finisher_truck_visit_1dayB}
\end{figure}

\begin{figure}[H]
\centering
\includegraphics[width=0.9\textwidth]{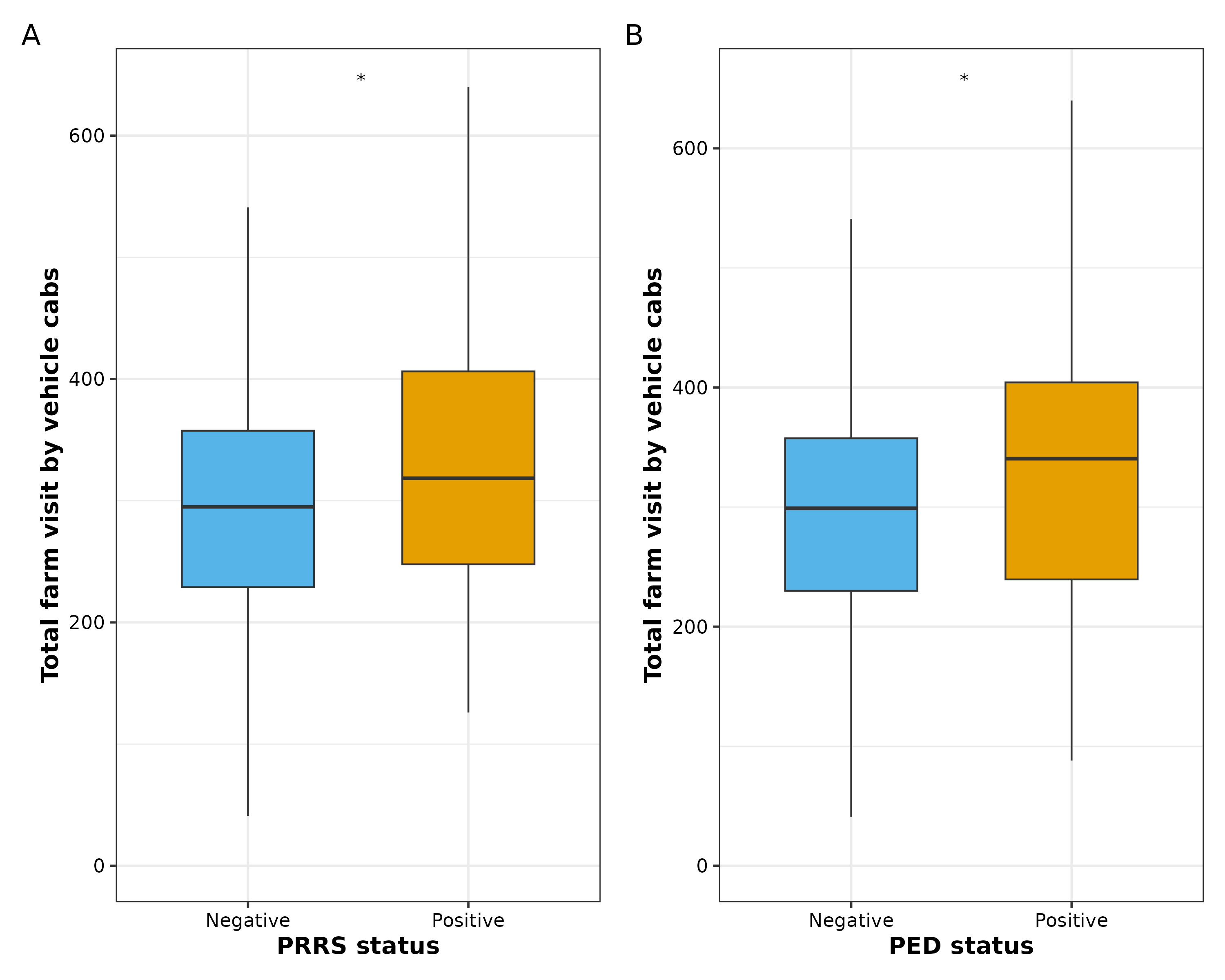}
\caption{Total farm visit by vehicle cabs exposed to finisher farms seven days before the visit}
\label{fig:SM_50_disease_total_farm_finisher_truck_visit_7dayB}
\end{figure}

\begin{figure}[H]
\centering
\includegraphics[width=0.9\textwidth]{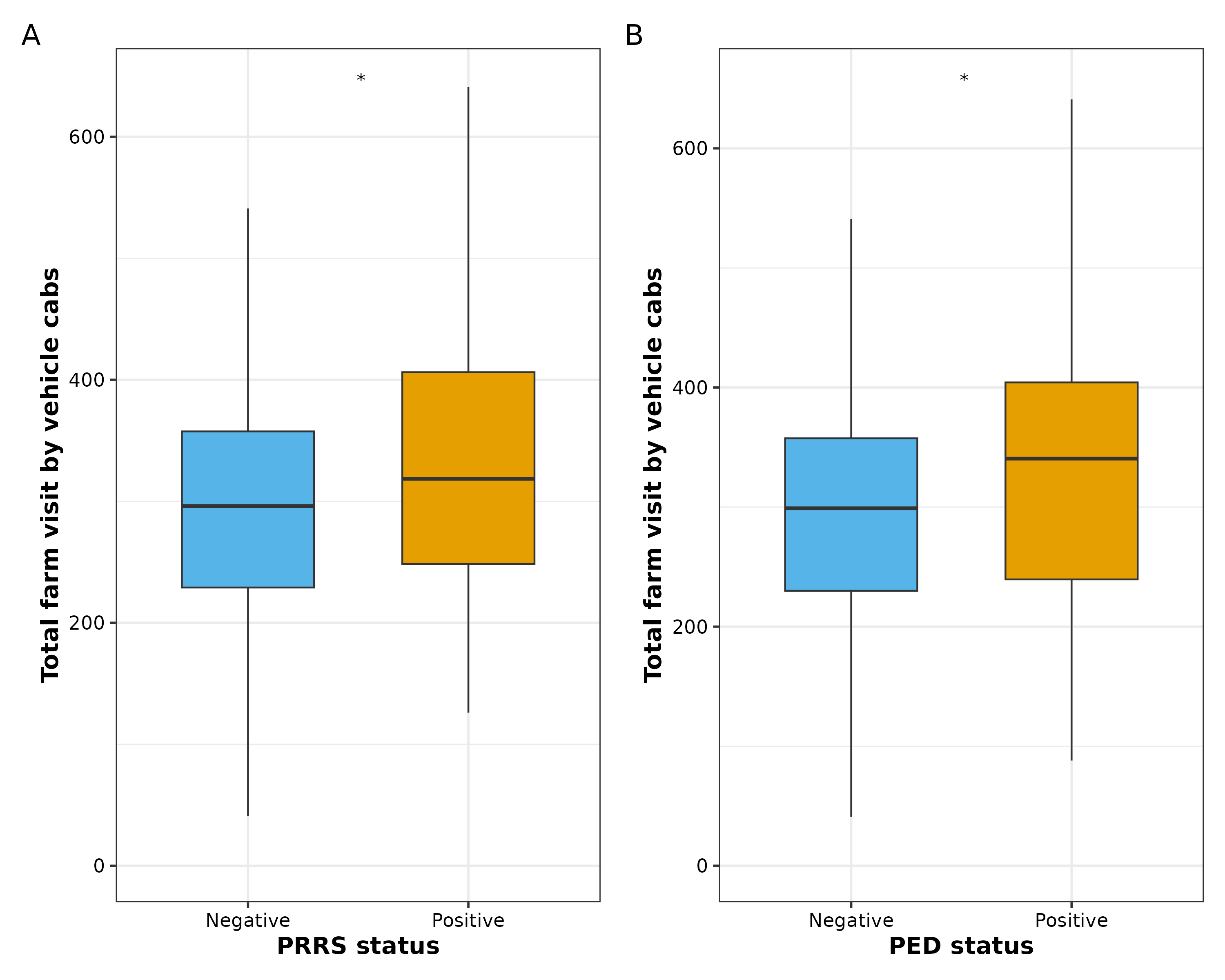}
\caption{Total farm visit by vehicle cabs exposed to finisher farms 11 days before the visit}
\label{fig:SM_51_disease_total_farm_finisher_truck_visit_11dayB}
\end{figure}

\begin{figure}[H]
\centering
\includegraphics[width=0.9\textwidth]{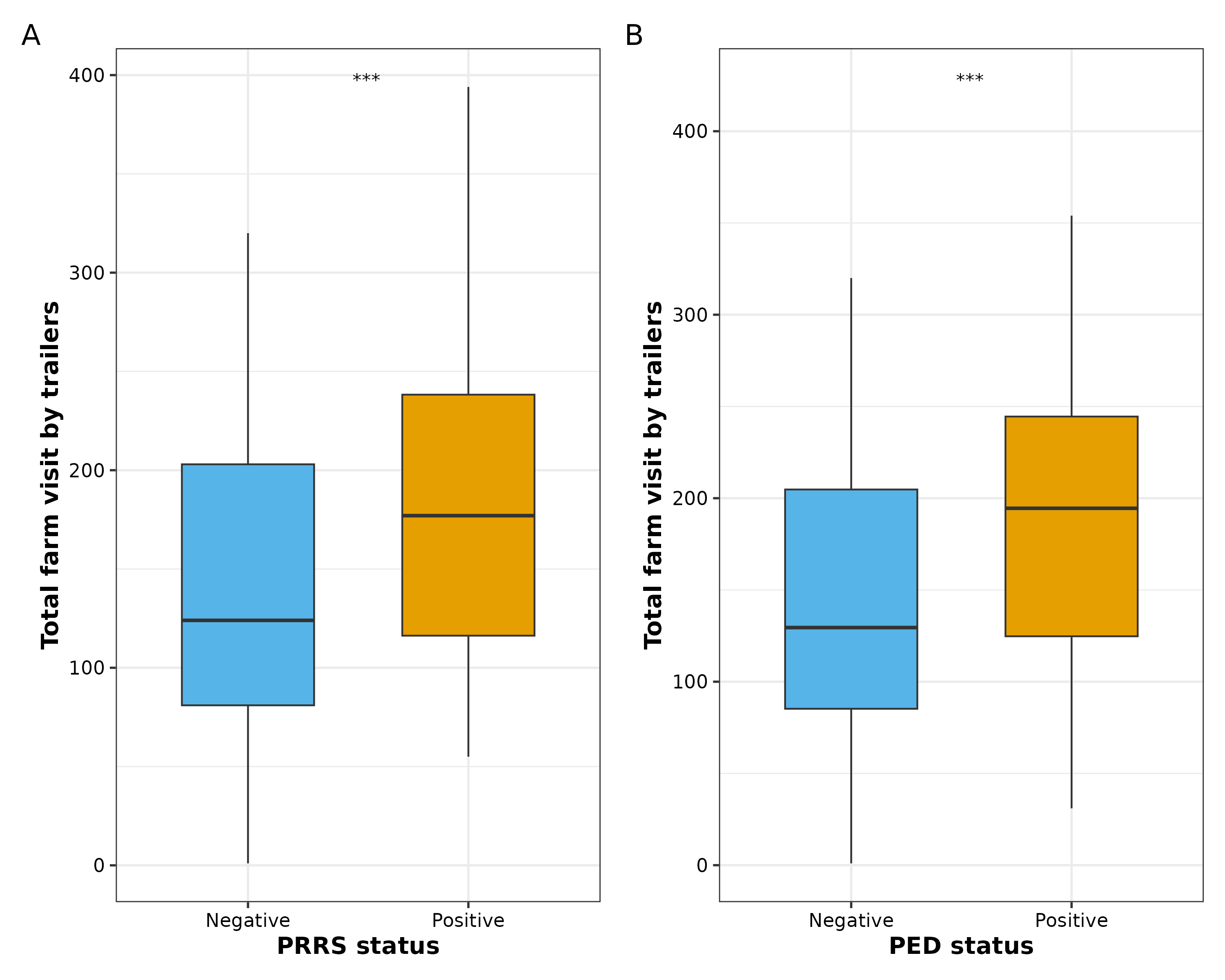}
\caption{Total farm visit by trailers exposed to finisher farms one day before the visit}
\label{fig:SM_52_disease_total_farm_finisher_trailer_visit_1dayB}
\end{figure}

\begin{figure}[H]
\centering
\includegraphics[width=0.9\textwidth]{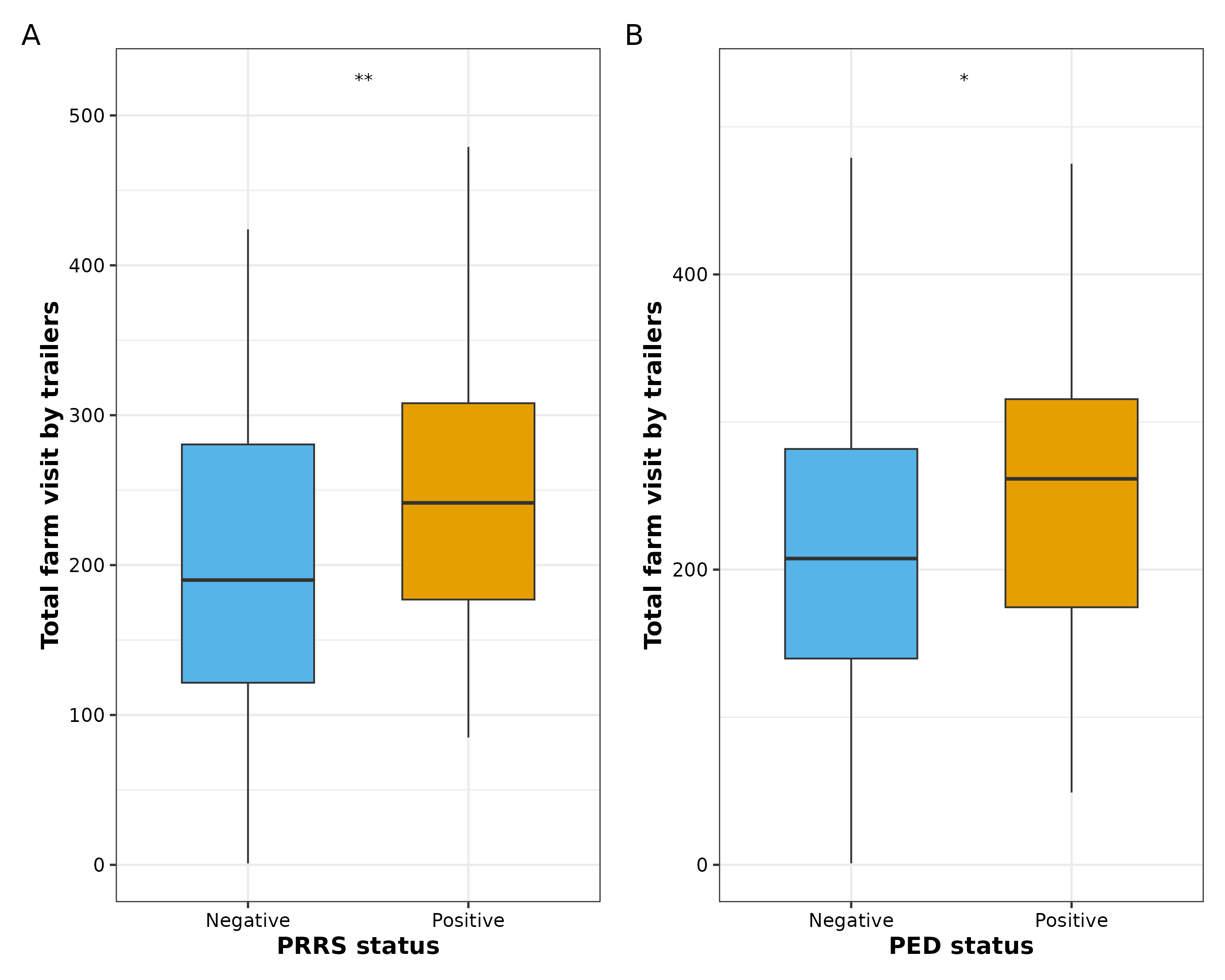}
\caption{Total farm visit by trailers exposed to finisher farms seven days before the visit}
\label{fig:SM_53_disease_total_farm_finisher_trailer_visit_7dayB}
\end{figure}

\begin{figure}[H]
\centering
\includegraphics[width=0.9\textwidth]{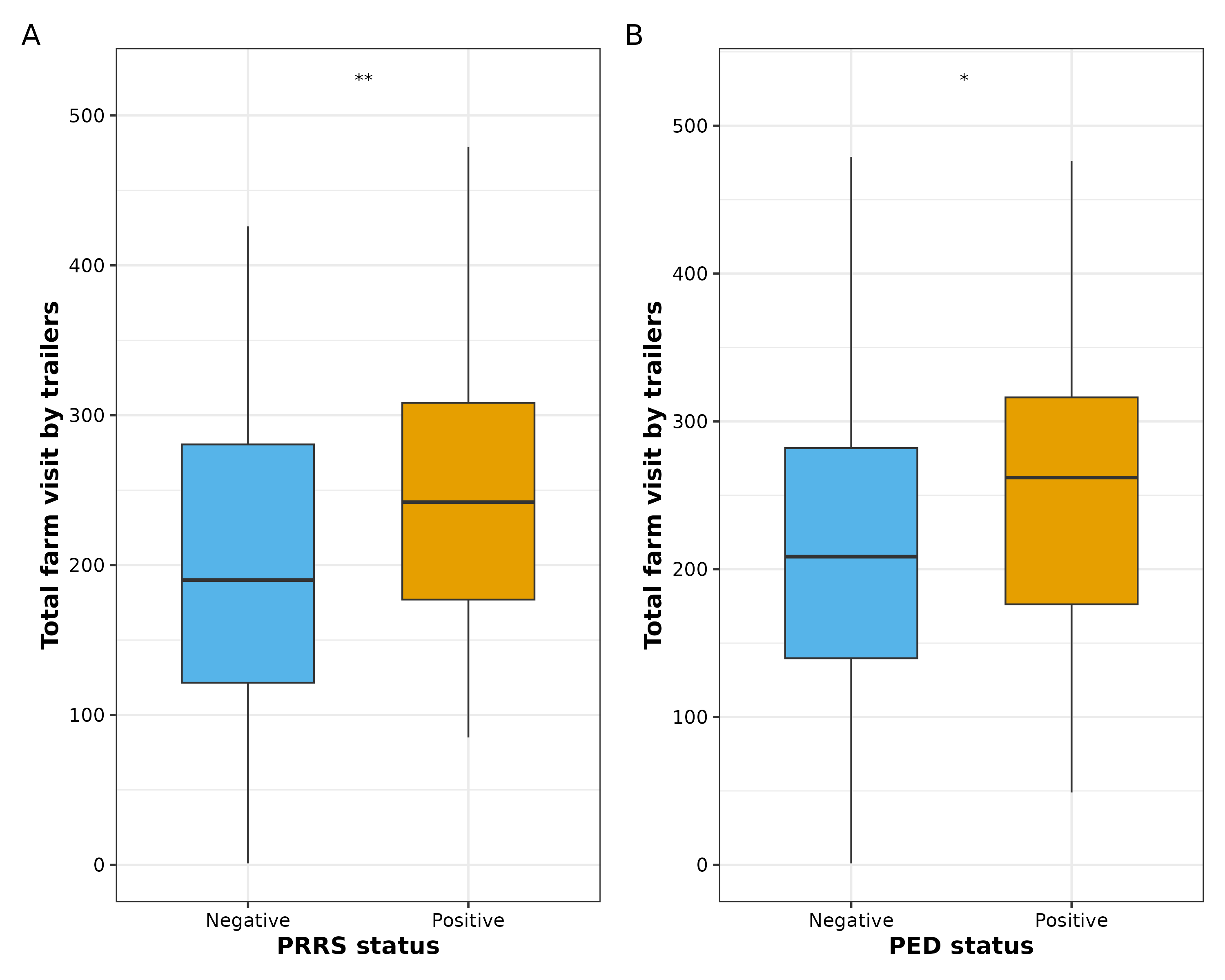}
\caption{Total farm visit by trailers exposed to finisher farms 11 days before the visit}
\label{fig:SM_54_disease_total_farm_finisher_trailer_visit_11dayB}
\end{figure}

\begin{figure}[H]
\centering
\includegraphics[width=0.9\textwidth]{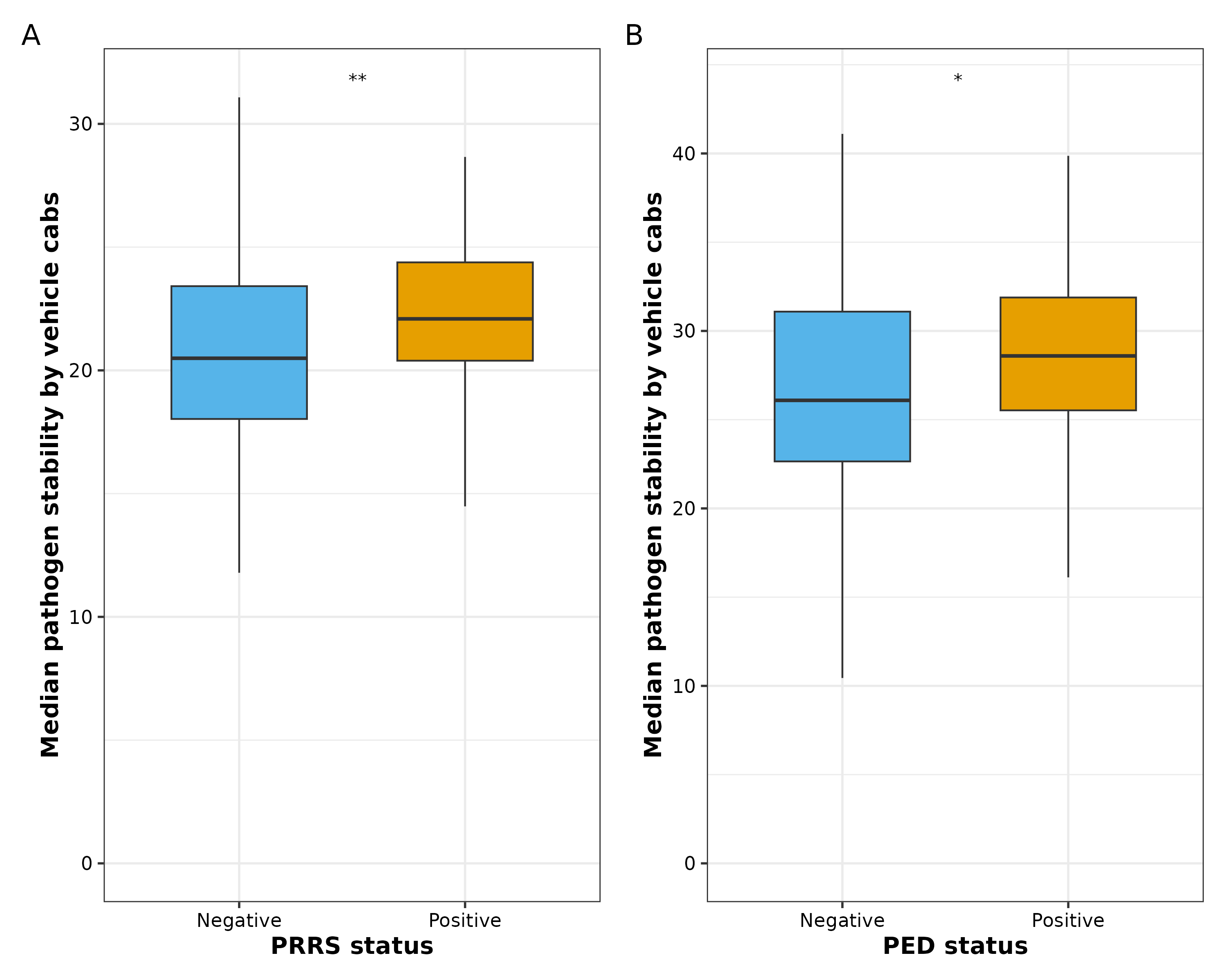}
\caption{Median pathogen stability by vehicle cabs movements exposed previously to positive farms}
\label{fig:SM_55_disease_pathongen_stability_contaminated_truck}
\end{figure}

\begin{figure}[H]
\centering
\includegraphics[width=0.9\textwidth]{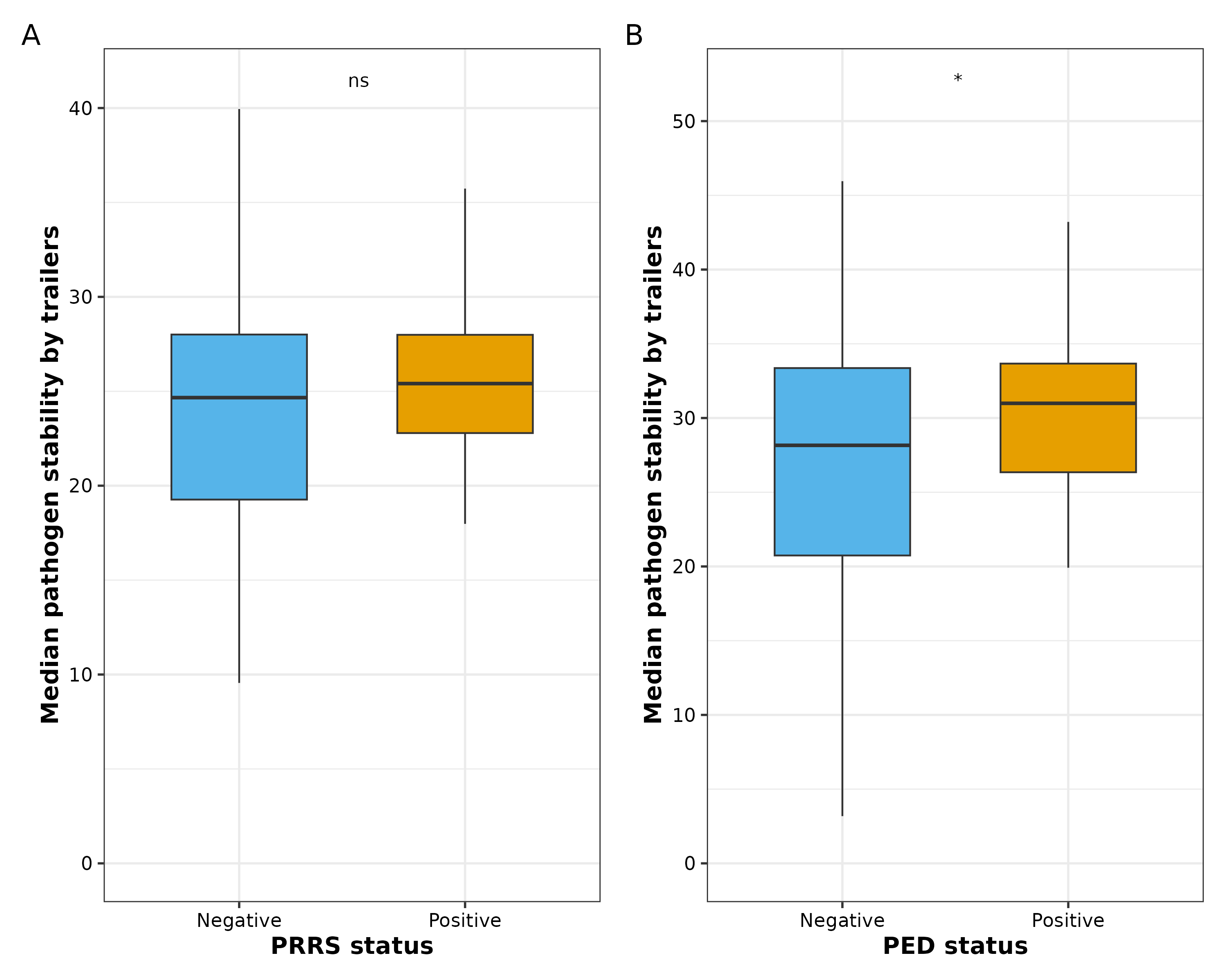}
\caption{Median pathogen stability by trailer movements exposed previously to positive farms}
\label{fig:SM_56_disease_pathongen_stability_contaminated_trailer}
\end{figure}

\begin{figure}[H]
\centering
\includegraphics[width=0.9\textwidth]{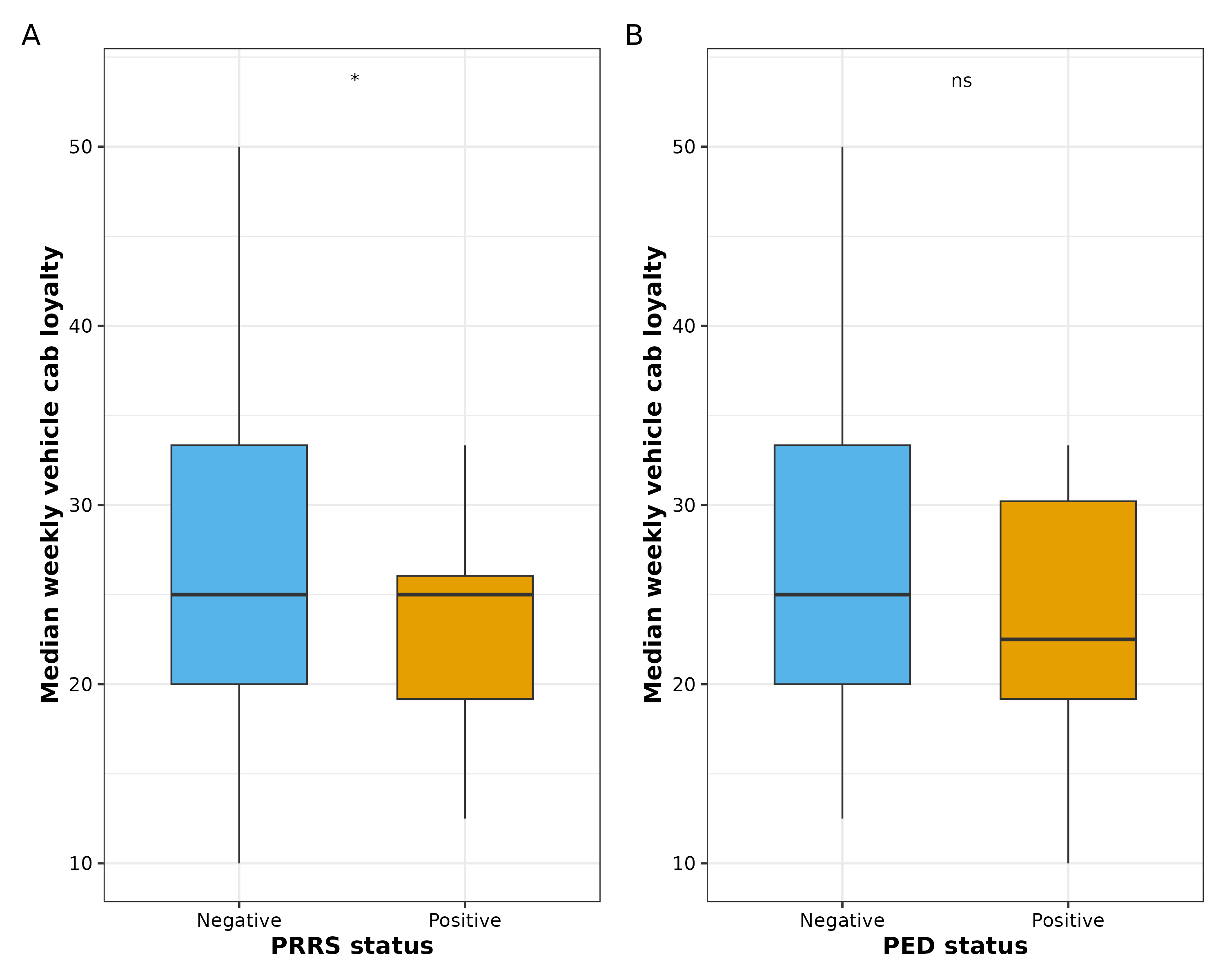}
\caption{Median weekly vehicle cab loyalty}
\label{fig:SM_57_disease_weekly_truck_loyalty}
\end{figure}

\begin{figure}[H]
\centering
\includegraphics[width=0.9\textwidth]{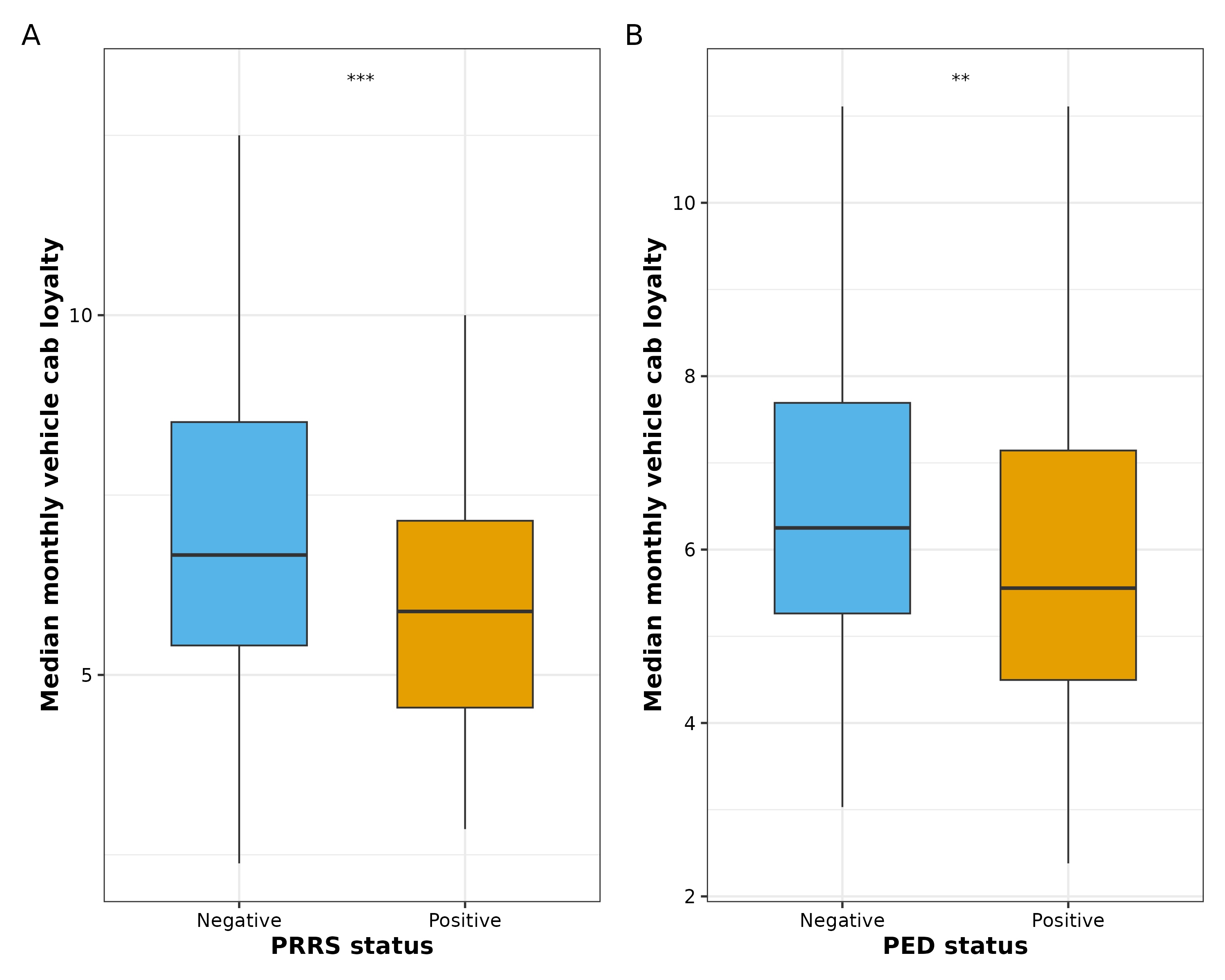}
\caption{Median monthly vehicle cab loyalty}
\label{fig:SM_58_disease_monthly_truck_loyalty}
\end{figure}

\begin{figure}[H]
\centering
\includegraphics[width=0.9\textwidth]{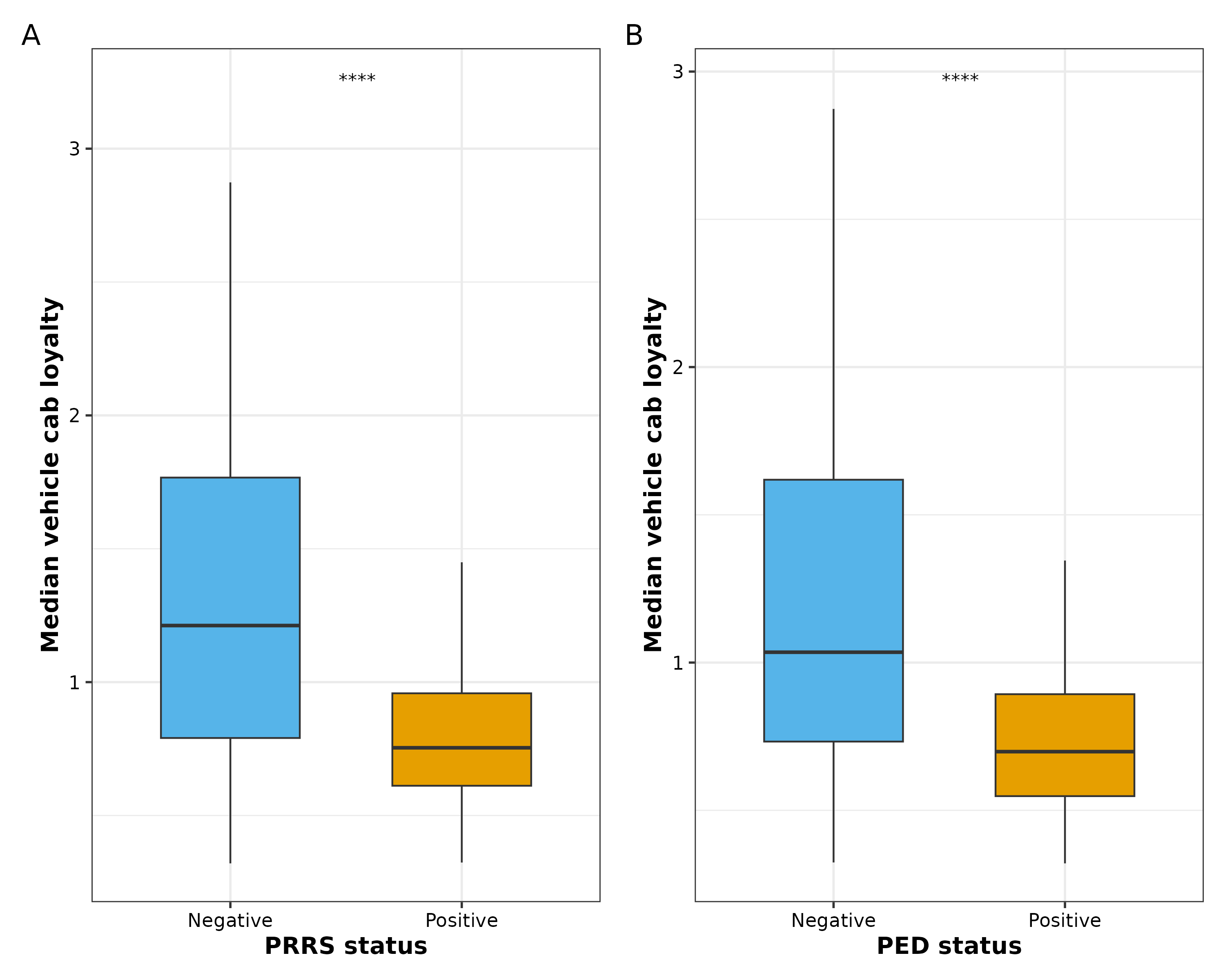}
\caption{Median vehicle cab loyalty}
\label{fig:SM_59_disease_truck_loyalty}
\end{figure}

\begin{figure}[H]
\centering
\includegraphics[width=0.9\textwidth]{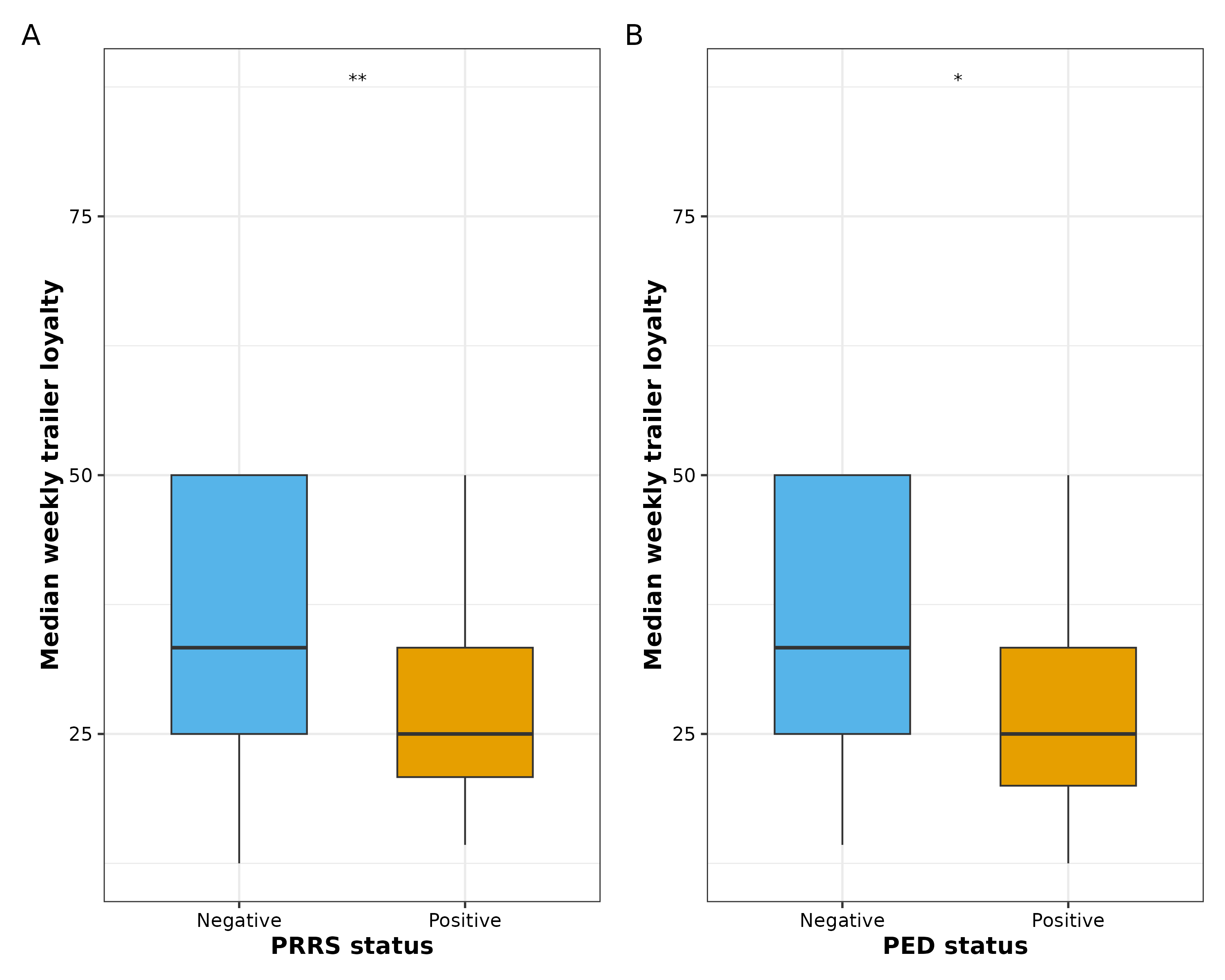}
\caption{Median weekly trailer loyalty}
\label{fig:SM_60_disease_weekly_trailer_loyalty}
\end{figure}

\begin{figure}[H]
\centering
\includegraphics[width=0.9\textwidth]{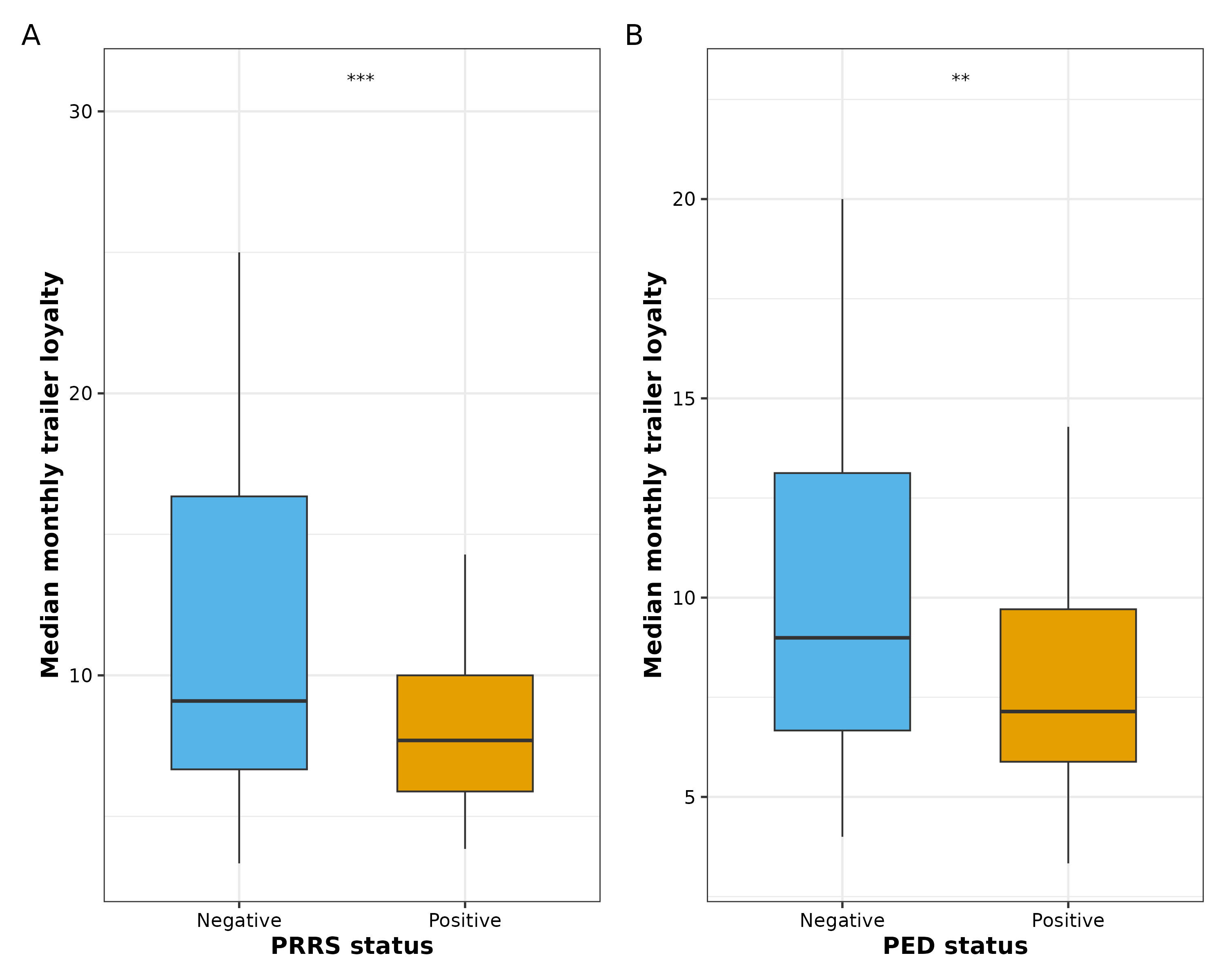}
\caption{Median monthly trailer loyalty}
\label{fig:SM_61_disease_monthly_trailer_loyalty}
\end{figure}

\begin{figure}[H]
\centering
\includegraphics[width=0.9\textwidth]{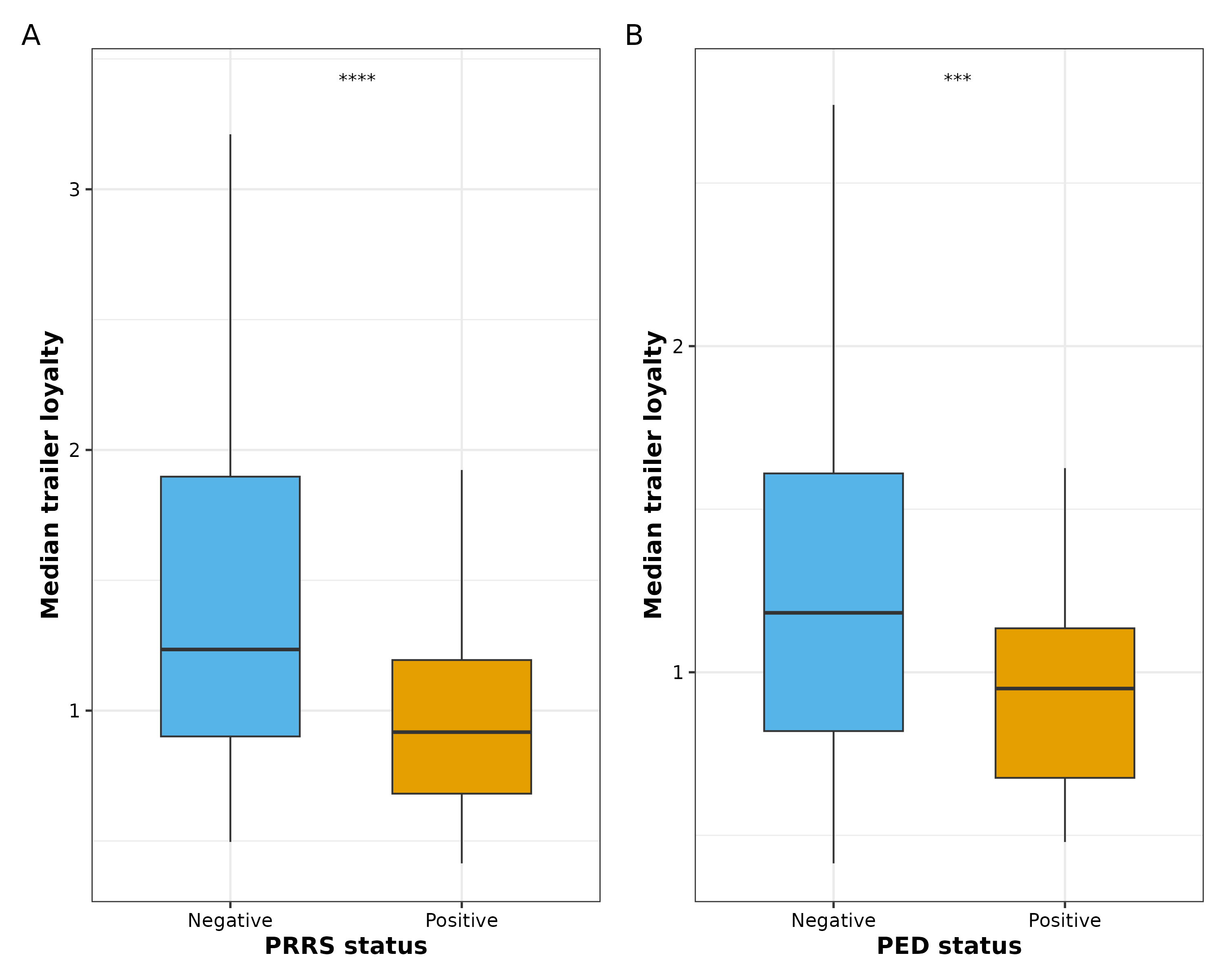}
\caption{Median trailer loyalty}
\label{fig:SM_62_disease_trailer_loyalty}
\end{figure}

\begin{figure}[H]
\centering
\includegraphics[width=0.9\textwidth]{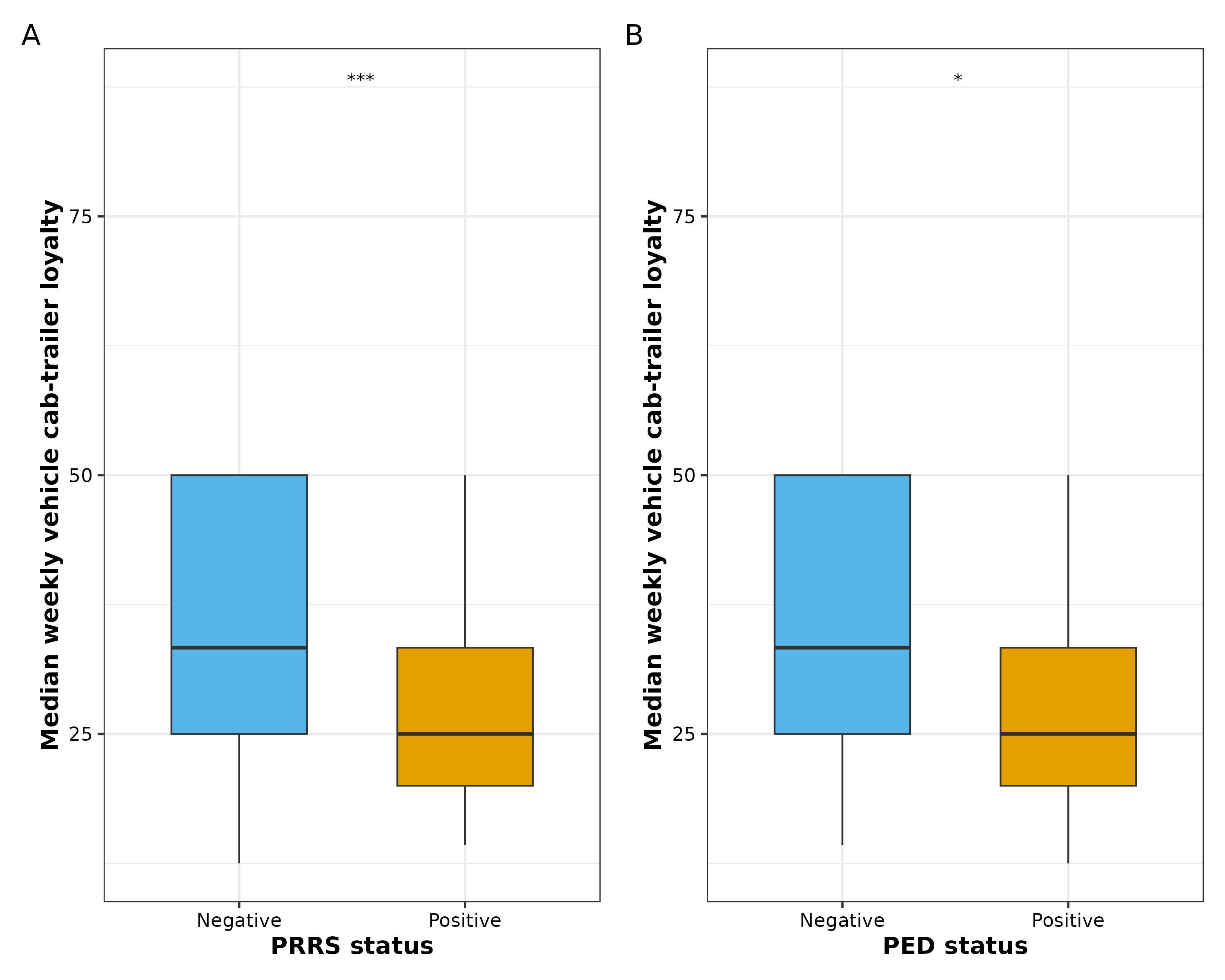}
\caption{Median weekly vehicle cab-trailer loyalty}
\label{fig:SM_63_disease_weekly_truck_trailer_loyalty}
\end{figure}

\begin{figure}[H]
\centering
\includegraphics[width=0.9\textwidth]{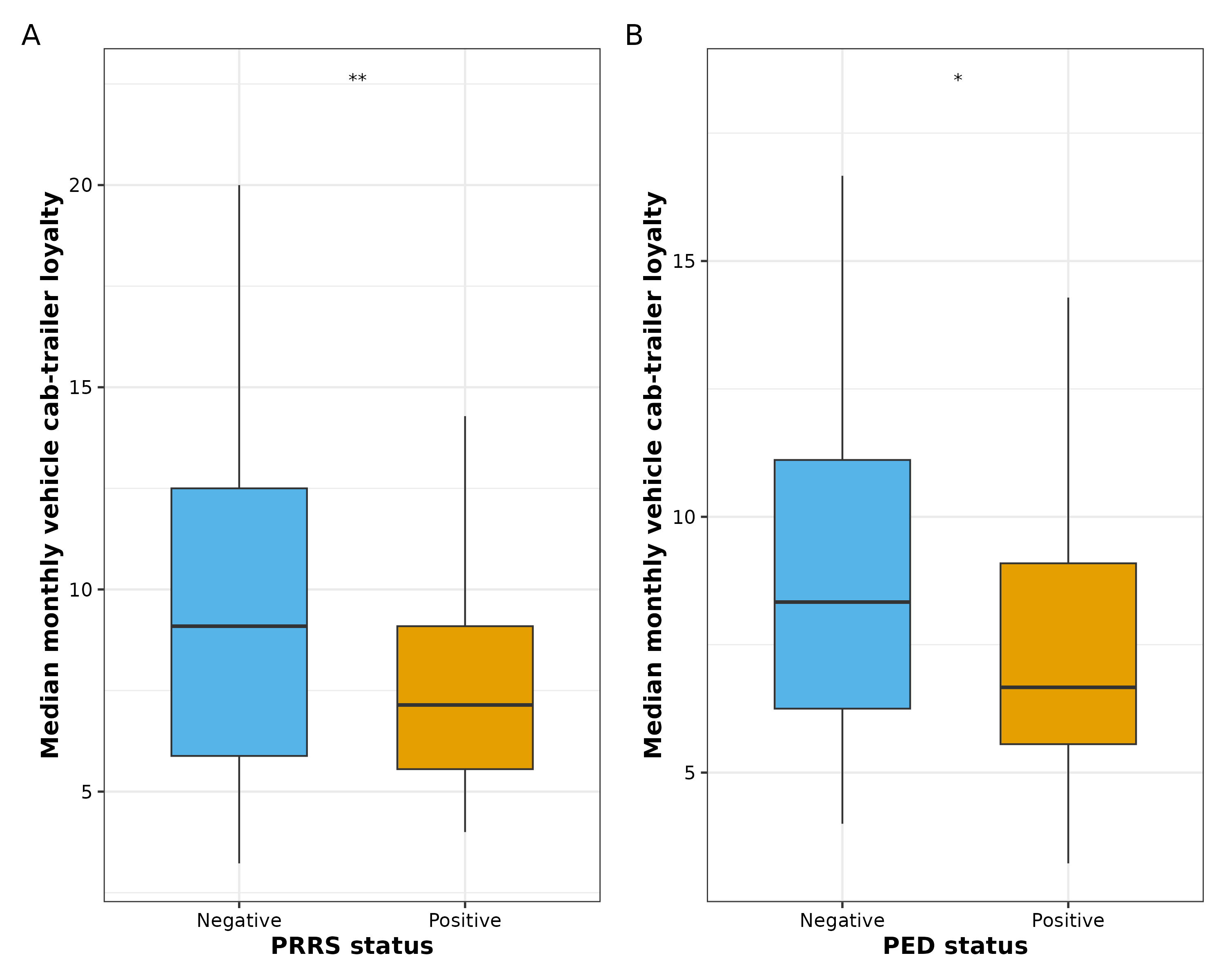}
\caption{Median monthly vehicle cab-trailer loyalty}
\label{fig:SM_64_disease_monthly_truck_trailer_loyalty}
\end{figure}

\begin{figure}[H]
\centering
\includegraphics[width=0.9\textwidth]{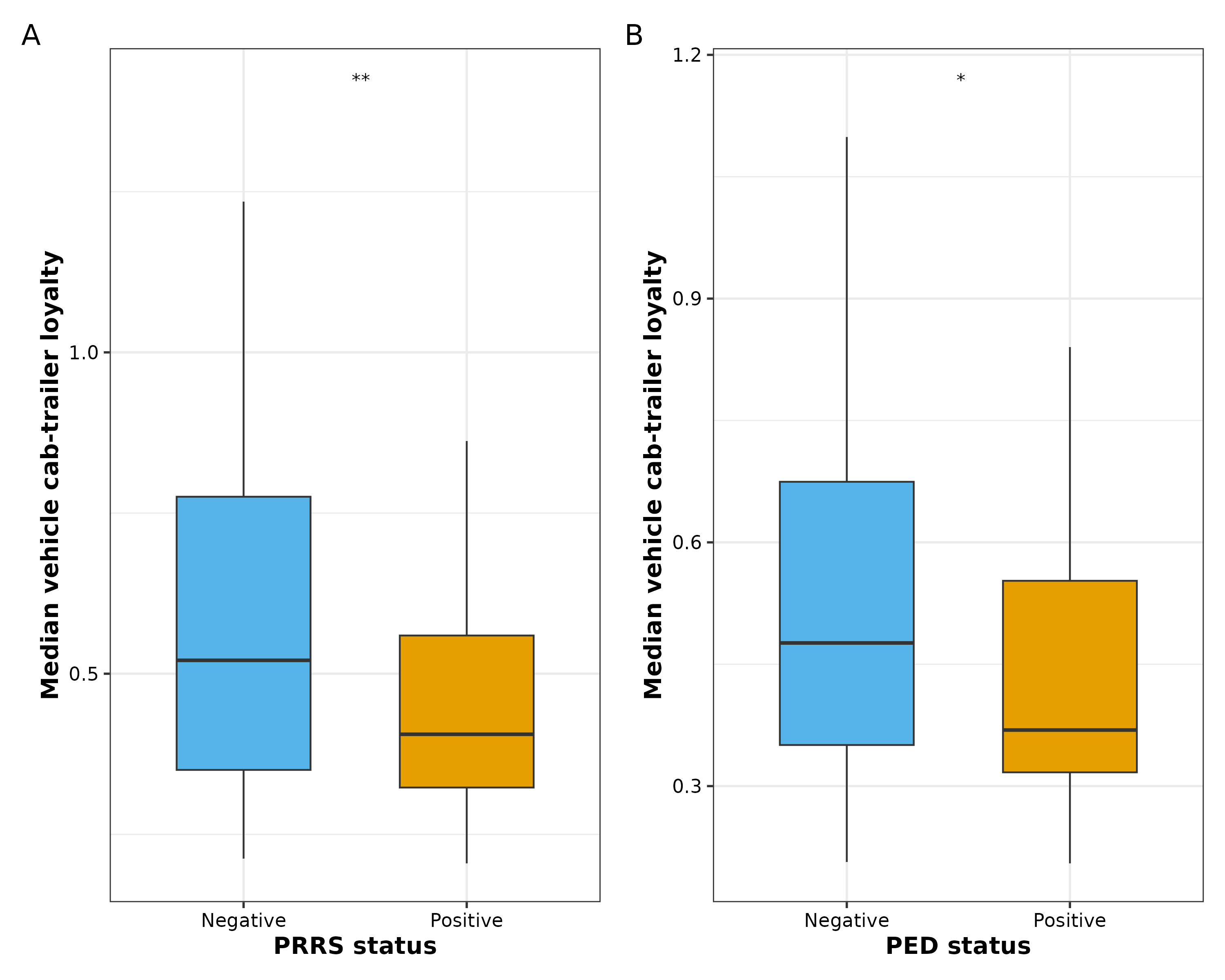}
\caption{Median vehicle cab-trailer loyalty}
\label{fig:SM_65_disease_truck_trailer_loyalty}
\end{figure}

\begin{figure}[H]
\centering
\includegraphics[width=0.9\textwidth]{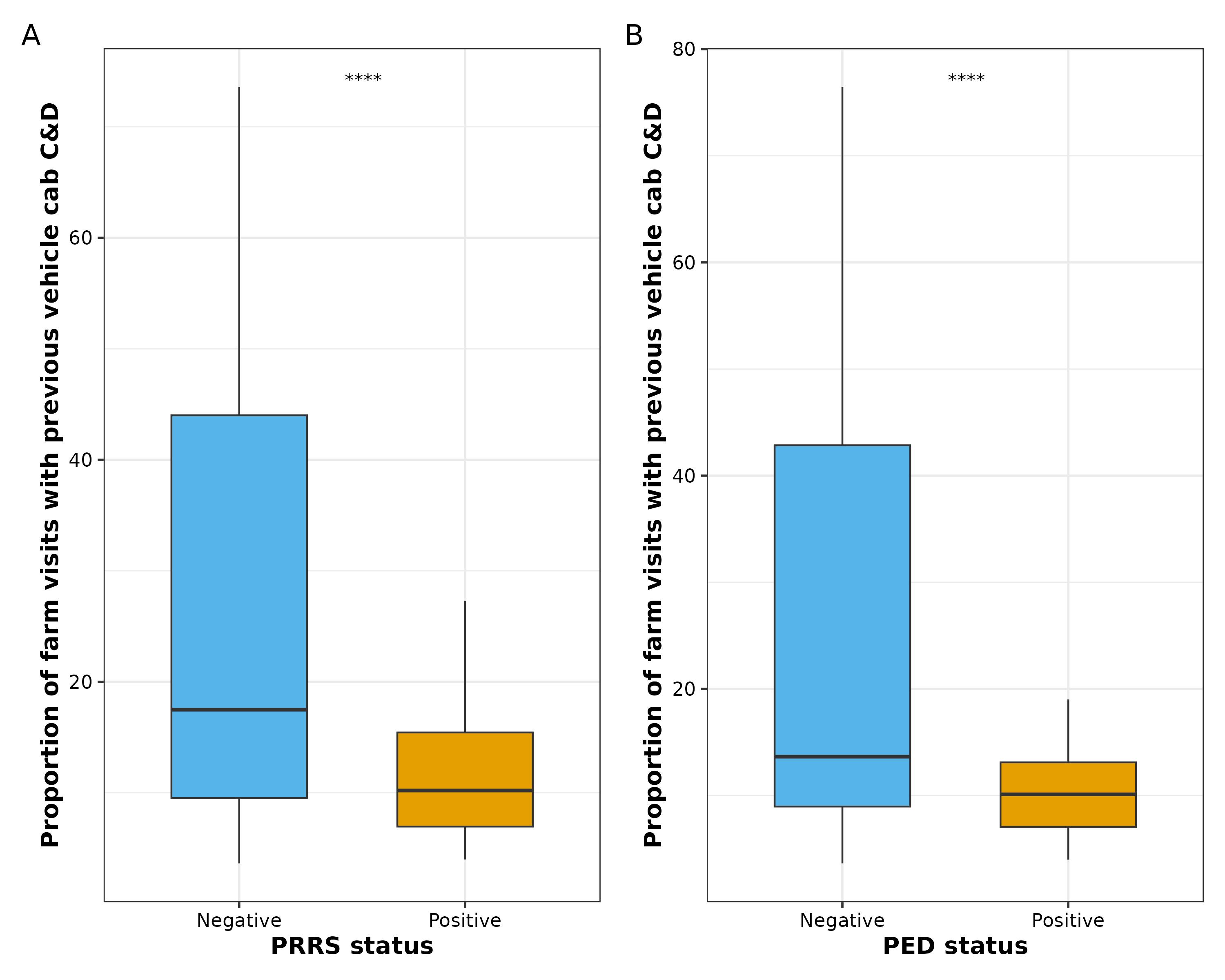}
\caption{Proportion of farm visits with previous vehicle cab C\&D}
\label{fig:SM_66_disease_truck_cleaning}
\end{figure}

\begin{figure}[H]
\centering
\includegraphics[width=0.9\textwidth]{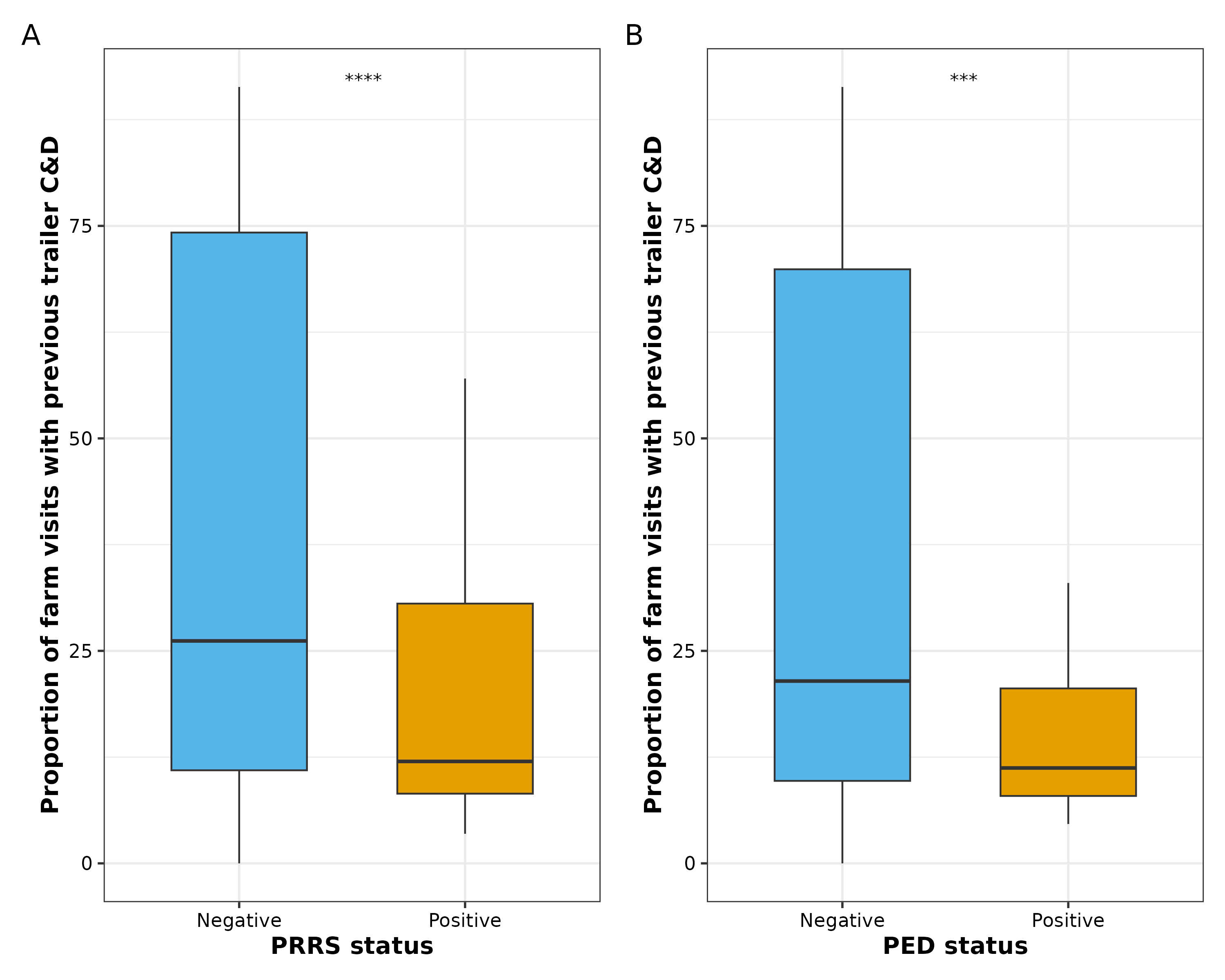}
\caption{Proportion of farm visits with previous trailer C\&D}
\label{fig:SM_67_disease_trailer_cleaning}
\end{figure}


\begin{center}
\begin{longtable}{p{9cm}p{5cm}p{1cm}}
\caption{Odds ratio (OR) and significance values from the univariate analysis for variables associated with PRRS positive farm status} \\
\toprule
\textbf{Variable} & \textbf{OR (95\% CI)} & \textit{p} \\
\midrule
\endfirsthead
\multicolumn{3}{l}{\textit{(continued from previous page)}} \\
\toprule
\textbf{Variable} & \textbf{OR (95\% CI)} & \textit{p} \\
\midrule
\endhead
\midrule \multicolumn{3}{r}{\textit{(continued on next page)}} \\ \bottomrule
\endfoot
\bottomrule
\endlastfoot
Pig capacity & 1.00 (1.00–1.00) & 0.33 \\
Neighbors within 10km radius & 1.02 (1.01–1.03) & 0.00 \\
Median weekly farm visit by trucks & 1.13 (0.95–1.35) & 0.19 \\
Median monthly farm visit by trucks & 1.04 (1.00–1.09) & 0.06 \\
Total farm visit by trucks & 1.00 (1.00–1.00) & 0.06 \\
Median weekly farm visit by trailers & 1.15 (0.94–1.43) & 0.19 \\
Median monthly farm visit by trailers & 1.05 (1.00–1.10) & 0.04 \\
Total farm visit by trailers & 1.00 (1.00–1.01) & 0.03 \\
Median weekly farm visit by trucks exposed to positive farms one day before the visit & 3.51 (2.50–5.20) & 0.00 \\
Median monthly farm visit by trucks exposed to positive farms one day before the visit & 1.30 (1.21–1.41) & 0.00 \\
Total farm visit by trucks exposed to positive farms one day before the visit & 1.01 (1.01–1.02) & 0.00 \\
Median weekly farm visit by trucks exposed to positive farms seven days before the visit & 1.83 (1.48–2.31) & 0.00 \\
Median monthly farm visit by trucks exposed to positive farms seven days before the visit & 1.15 (1.10–1.22) & 0.00 \\
Total farm visit by trucks exposed to positive farms seven days before the visit & 1.01 (1.01–1.01) & 0.00 \\
Median weekly farm visit by trucks exposed to positive farms 11 days before the visit & 1.75 (1.42–2.19) & 0.00 \\
Median monthly farm visit by trucks exposed to positive farms 11 days before the visit & 1.15 (1.09–1.21) & 0.00 \\
Total farm visit by trucks exposed to positive farms 11 days before the visit & 1.01 (1.00–1.01) & 0.00 \\
Median weekly farm visit by trailers exposed to positive farms one day before the visit & 4.59 (3.01–7.53) & 0.00 \\
Median monthly farm visit by trailers exposed to positive farms one day before the visit & 1.40 (1.28–1.55) & 0.00 \\
Total farm visit by trailers exposed to positive farms one day before the visit & 1.02 (1.01–1.02) & 0.00 \\
Median weekly farm visit by trailers exposed to positive farms seven days before the visit & 2.31 (1.75–3.14) & 0.00 \\
Median monthly farm visit by trailers exposed to positive farms seven days before the visit & 1.21 (1.14–1.29) & 0.00 \\
Total farm visit by trailers exposed to positive farms seven days before the visit & 1.01 (1.01–1.01) & 0.00 \\
Median weekly farm visit by trailers exposed to positive farms 11 days before the visit & 2.18 (1.66–2.94) & 0.00 \\
Median monthly farm visit by trailers exposed to positive farms 11 days before the visit & 1.20 (1.13–1.27) & 0.00 \\
Total farm visit by trailers exposed to positive farms 11 days before the visit & 1.01 (1.01–1.01) & 0.00 \\
Truck network in-degree of positive farms & 1.00 (1.00–1.00) & 0.00 \\
Trailer network in-degree of positive farms & 1.00 (1.00–1.01) & 0.00 \\
Truck network in-degree of positive farms with C\&D & 1.00 (1.00–1.00) & 0.00 \\
Trailer network in-degree of positive farms  with C\&D & 1.00 (1.00–1.01) & 0.00 \\
Truck network in-degree & 1.00 (1.00–1.00) & 0.00 \\
Trailer network in-degree & 1.00 (1.00–1.00) & 0.00 \\
Truck network in-degree with C\&D & 1.00 (1.00–1.00) & 0.00 \\
Trailer network in-degree with C\&D & 1.00 (1.00–1.00) & 0.00 \\
Number of farms within the truck network community & 1.00 (1.00–1.00) & 0.01 \\
Number of positive farms within the truck network community & 1.02 (1.00–1.04) & 0.02 \\
Number of farms within the trailer network community & 1.00 (1.00–1.00) & 0.01 \\
Number of positive farms within the trailer network community & 1.02 (1.00–1.04) & 0.03 \\
Number of farms within the truck network community with C\&D & 1.00 (1.00–1.00) & 0.01 \\
Number of positive farms within the truck network community with C\&D & 1.02 (1.00–1.04) & 0.02 \\
Number of farms within the trailer network community with C\&D & 1.00 (1.00–1.00) & 0.02 \\
Number of positive farms within the trailer network community with C\&D & 1.02 (1.00–1.04) & 0.03 \\
Total farm visit by trucks exposed to slaughterhouses one day before the visit & 1.01 (0.96–1.07) & 0.74 \\
Total farm visit by trucks exposed to slaughterhouses seven days before the visit & 1.03 (1.00–1.06) & 0.05 \\
Total farm visit by trucks exposed to slaughterhouses 11 days before the visit & 1.03 (1.01–1.06) & 0.02 \\
Total farm visit by trailers exposed to slaughterhouses one day before the visit & 0.93 (0.87–0.98) & 0.02 \\
Total farm visit by trailers exposed to slaughterhouses seven days before the visit & 0.96 (0.92–1.00) & 0.08 \\
Total farm visit by trailers exposed to slaughterhouses 11 days before the visit & 0.97 (0.93–1.01) & 0.12 \\
Total farm visit by trucks exposed to finisher farms one day before the visit & 1.00 (1.00–1.01) & 0.01 \\
Total farm visit by trucks exposed to finisher farms seven days before the visit & 1.00 (1.00–1.00) & 0.03 \\
Total farm visit by trucks exposed to finisher farms 11 days before the visit & 1.00 (1.00–1.00) & 0.03 \\
Total farm visit by trailers exposed to finisher farms one day before the visit & 1.01 (1.00–1.01) & 0.00 \\
Total farm visit by trailers exposed to finisher farms seven days before the visit & 1.00 (1.00–1.01) & 0.01 \\
Total farm visit by trailers exposed to finisher farms 11 days before the visit & 1.00 (1.00–1.01) & 0.01 \\
Median pathogen stability by trucks movements exposed previously to positive farms & 1.04 (0.98–1.11) & 0.21 \\
Median pathogen stability by trailer movements exposed previously to positive farms & 1.05 (1.01–1.10) & 0.02 \\
Median weekly truck loyalty & 0.96 (0.93–0.99) & 0.01 \\
Median monthly truck loyalty & 0.83 (0.72–0.93) & 0.00 \\
Median truck loyalty & 0.20 (0.09–0.40) & 0.00 \\
Median weekly trailer loyalty & 0.96 (0.94–0.98) & 0.00 \\
Median monthly trailer loyalty & 0.87 (0.79–0.93) & 0.00 \\
Median trailer loyalty & 0.25 (0.12–0.48) & 0.00 \\
Median weekly truck-trailer loyalty & 0.96 (0.93–0.98) & 0.00 \\
Median monthly truck-trailer loyalty & 0.88 (0.81–0.95) & 0.00 \\
Median truck-trailer loyalty & 0.19 (0.05–0.69) & 0.01 \\
Proportion of farm visits with previous truck C\&D & 0.95 (0.93–0.97) & 0.00 \\
Proportion of farm visits with previous trailer C\&D & 0.97 (0.96–0.98) & 0.00 \\
\end{longtable}
\end{center}

\begin{center}
\begin{longtable}{p{9cm}p{5cm}p{1cm}}
\caption{Odds ratio (OR) and significance values from the univariate analysis for variables associated with PED positive farm status} \\
\toprule
\textbf{Variable} & \textbf{OR (95\% CI)} & \textit{p} \\
\midrule
\endfirsthead
\multicolumn{3}{l}{\textit{(continued from previous page)}} \\
\toprule
\textbf{Variable} & \textbf{OR (95\% CI)} & \textit{p} \\
\midrule
\endhead
\midrule \multicolumn{3}{r}{\textit{(continued on next page)}} \\ \bottomrule
\endfoot
\bottomrule
\endlastfoot
Pig capacity & 1.00 (1.00–1.00) & 0.78 \\
Neighbors within 10km radius & 1.01 (1.00–1.02) & 0.02 \\
Median weekly farm visit by trucks & 1.15 (0.96–1.38) & 0.14 \\
Median monthly farm visit by trucks & 1.04 (1.00–1.08) & 0.07 \\
Total farm visit by trucks & 1.00 (1.00–1.00) & 0.10 \\
Median weekly farm visit by trailers & 1.22 (0.99–1.51) & 0.07 \\
Median monthly farm visit by trailers & 1.04 (1.00–1.10) & 0.07 \\
Total farm visit by trailers & 1.00 (1.00–1.01) & 0.07 \\
Median weekly farm visit by trucks exposed to positive farms one day before the visit & 4.49 (2.65–8.14) & 0.00 \\
Median monthly farm visit by trucks exposed to positive farms one day before the visit & 1.46 (1.29–1.68) & 0.00 \\
Total farm visit by trucks exposed to positive farms one day before the visit & 1.02 (1.01–1.03) & 0.00 \\
Median weekly farm visit by trucks exposed to positive farms seven days before the visit & 1.98 (1.47–2.74) & 0.00 \\
Median monthly farm visit by trucks exposed to positive farms seven days before the visit & 1.17 (1.10–1.25) & 0.00 \\
Total farm visit by trucks exposed to positive farms seven days before the visit & 1.01 (1.01–1.01) & 0.00 \\
Median weekly farm visit by trucks exposed to positive farms 11 days before the visit & 1.93 (1.46–2.61) & 0.00 \\
Median monthly farm visit by trucks exposed to positive farms 11 days before the visit & 1.16 (1.09–1.23) & 0.00 \\
Total farm visit by trucks exposed to positive farms 11 days before the visit & 1.01 (1.00–1.01) & 0.00 \\
Median weekly farm visit by trailers exposed to positive farms one day before the visit & 3.99 (2.35–7.10) & 0.00 \\
Median monthly farm visit by trailers exposed to positive farms one day before the visit & 1.47 (1.29–1.71) & 0.00 \\
Total farm visit by trailers exposed to positive farms one day before the visit & 1.02 (1.01–1.03) & 0.00 \\
Median weekly farm visit by trailers exposed to positive farms seven days before the visit & 2.64 (1.82–3.98) & 0.00 \\
Median monthly farm visit by trailers exposed to positive farms seven days before the visit & 1.21 (1.12–1.32) & 0.00 \\
Total farm visit by trailers exposed to positive farms seven days before the visit & 1.01 (1.01–1.01) & 0.00 \\
Median weekly farm visit by trailers exposed to positive farms 11 days before the visit & 2.13 (1.52–3.05) & 0.00 \\
Median monthly farm visit by trailers exposed to positive farms 11 days before the visit & 1.20 (1.12–1.30) & 0.00 \\
Total farm visit by trailers exposed to positive farms 11 days before the visit & 1.01 (1.01–1.01) & 0.00 \\
Truck network in-degree of positive farms & 1.01 (1.00–1.01) & 0.00 \\
Trailer network in-degree of positive farms & 1.01 (1.00–1.01) & 0.00 \\
Truck network in-degree of positive farms with C\&D & 1.01 (1.00–1.01) & 0.00 \\
Trailer network in-degree of positive farms  with C\&D & 1.01 (1.00–1.01) & 0.00 \\
Truck network in-degree & 1.00 (1.00–1.00) & 0.00 \\
Trailer network in-degree & 1.00 (1.00–1.00) & 0.00 \\
Truck network in-degree with C\&D & 1.00 (1.00–1.00) & 0.00 \\
Trailer network in-degree with C\&D & 1.00 (1.00–1.00) & 0.00 \\
Number of farms within the truck network community & 1.00 (1.00–1.00) & 0.02 \\
Number of positive farms within the truck network community & 1.00 (1.00–1.01) & 0.01 \\
Number of farms within the trailer network community & 1.00 (1.00–1.01) & 0.01 \\
Number of positive farms within the trailer network community & 1.01 (1.00–1.02) & 0.00 \\
Number of farms within the truck network community with C\&D & 1.00 (1.00–1.00) & 0.02 \\
Number of positive farms within the truck network community with C\&D & 1.00 (1.00–1.01) & 0.01 \\
Number of farms within the trailer network community with C\&D & 1.00 (1.00–1.00) & 0.01 \\
Number of positive farms within the trailer network community with C\&D & 1.01 (1.00–1.02) & 0.00 \\
Total farm visit by trucks exposed to slaughterhouses one day before the visit & 1.03 (0.98–1.09) & 0.26 \\
Total farm visit by trucks exposed to slaughterhouses seven days before the visit & 1.05 (1.02–1.08) & 0.00 \\
Total farm visit by trucks exposed to slaughterhouses 11 days before the visit & 1.04 (1.02–1.07) & 0.00 \\
Total farm visit by trailers exposed to slaughterhouses one day before the visit & 0.97 (0.90–1.02) & 0.28 \\
Total farm visit by trailers exposed to slaughterhouses seven days before the visit & 1.00 (0.95–1.04) & 0.90 \\
Total farm visit by trailers exposed to slaughterhouses 11 days before the visit & 1.00 (0.96–1.04) & 0.95 \\
Total farm visit by trucks exposed to finisher farms one day before the visit & 1.00 (1.00–1.01) & 0.01 \\
Total farm visit by trucks exposed to finisher farms seven days before the visit & 1.00 (1.00–1.00) & 0.06 \\
Total farm visit by trucks exposed to finisher farms 11 days before the visit & 1.00 (1.00–1.00) & 0.06 \\
Total farm visit by trailers exposed to finisher farms one day before the visit & 1.01 (1.00–1.01) & 0.00 \\
Total farm visit by trailers exposed to finisher farms seven days before the visit & 1.00 (1.00–1.01) & 0.05 \\
Total farm visit by trailers exposed to finisher farms 11 days before the visit & 1.00 (1.00–1.01) & 0.05 \\
Median pathogen stability by trucks movements exposed previously to positive farms & 1.02 (0.99–1.05) & 0.27 \\
Median pathogen stability by trailer movements exposed previously to positive farms & 1.02 (0.99–1.04) & 0.17 \\
Median weekly truck loyalty & 0.98 (0.95–1.00) & 0.08 \\
Median monthly truck loyalty & 0.86 (0.74–0.96) & 0.02 \\
Median truck loyalty & 0.15 (0.06–0.34) & 0.00 \\
Median weekly trailer loyalty & 0.97 (0.94–0.99) & 0.01 \\
Median monthly trailer loyalty & 0.89 (0.81–0.96) & 0.01 \\
Median trailer loyalty & 0.35 (0.16–0.66) & 0.00 \\
Median weekly truck-trailer loyalty & 0.97 (0.94–0.99) & 0.01 \\
Median monthly truck-trailer loyalty & 0.91 (0.83–0.98) & 0.02 \\
Median truck-trailer loyalty & 0.52 (0.15–0.94) & 0.24 \\
Proportion of farm visits with previous truck C\&D & 0.94 (0.90–0.96) & 0.00 \\
Proportion of farm visits with previous trailer C\&D & 0.97 (0.95–0.98) & 0.00 \\
\end{longtable}
\end{center}

\begin{center}
\begin{longtable}{p{9cm}p{5cm}p{1cm}}
\caption{Odds ratio percentage change (ORPC) $(OR - 1) \times 100$ and significance values from the univariate analysis for variables associated with PRRS positive farm status} \\
\toprule
\textbf{Variable} & \textbf{ORPC (95\% CI)} & \textit{p} \\
\midrule
\endfirsthead
\multicolumn{3}{l}{\textit{(continued from previous page)}} \\
\toprule
\textbf{Variable} & \textbf{ORPC (95\% CI)} & \textit{p} \\
\midrule
\endhead
\midrule \multicolumn{3}{r}{\textit{(continued on next page)}} \\ \bottomrule
\endfoot
\bottomrule
\endlastfoot
Pig capacity & 0.01 (-0.01–0.03) & 0.33 \\
Neighbors within 10km radius & 1.77 (0.77–2.87) & 0.00 \\
Median weekly farm visit by trucks & 12.68 (-5.35–35.38) & 0.19 \\
Median monthly farm visit by trucks & 4.05 (-0.06–8.62) & 0.06 \\
Total farm visit by trucks & 0.21 (-0.00–0.44) & 0.06 \\
Median weekly farm visit by trailers & 15.28 (-6.19–43.43) & 0.19 \\
Median monthly farm visit by trailers & 5.16 (0.37–10.50) & 0.04 \\
Total farm visit by trailers & 0.30 (0.04–0.58) & 0.03 \\
Median weekly farm visit by trucks exposed to positive farms one day before the visit & 251.20 (149.95–419.85) & 0.00 \\
Median monthly farm visit by trucks exposed to positive farms one day before the visit & 29.85 (21.18–40.69) & 0.00 \\
Total farm visit by trucks exposed to positive farms one day before the visit & 1.38 (1.01–1.81) & 0.00 \\
Median weekly farm visit by trucks exposed to positive farms seven days before the visit & 82.92 (48.27–130.82) & 0.00 \\
Median monthly farm visit by trucks exposed to positive farms seven days before the visit & 15.35 (10.04–21.54) & 0.00 \\
Total farm visit by trucks exposed to positive farms seven days before the visit & 0.79 (0.53–1.08) & 0.00 \\
Median weekly farm visit by trucks exposed to positive farms 11 days before the visit & 74.59 (41.93–119.30) & 0.00 \\
Median monthly farm visit by trucks exposed to positive farms 11 days before the visit & 14.59 (9.34–20.71) & 0.00 \\
Total farm visit by trucks exposed to positive farms 11 days before the visit & 0.73 (0.48–1.02) & 0.00 \\
Median weekly farm visit by trailers exposed to positive farms one day before the visit & 359.15 (200.82–653.00) & 0.00 \\
Median monthly farm visit by trailers exposed to positive farms one day before the visit & 39.74 (27.91–54.87) & 0.00 \\
Total farm visit by trailers exposed to positive farms one day before the visit & 1.71 (1.25–2.23) & 0.00 \\
Median weekly farm visit by trailers exposed to positive farms seven days before the visit & 130.99 (75.28–214.20) & 0.00 \\
Median monthly farm visit by trailers exposed to positive farms seven days before the visit & 21.11 (14.26–29.22) & 0.00 \\
Total farm visit by trailers exposed to positive farms seven days before the visit & 1.02 (0.71–1.37) & 0.00 \\
Median weekly farm visit by trailers exposed to positive farms 11 days before the visit & 117.92 (66.39–193.93) & 0.00 \\
Median monthly farm visit by trailers exposed to positive farms 11 days before the visit & 19.61 (13.12–27.25) & 0.00 \\
Total farm visit by trailers exposed to positive farms 11 days before the visit & 0.96 (0.66–1.30) & 0.00 \\
Truck network in-degree of positive farms & 0.32 (0.21–0.44) & 0.00 \\
Trailer network in-degree of positive farms & 0.37 (0.25–0.50) & 0.00 \\
Truck network in-degree of positive farms with C\&D & 0.32 (0.21–0.44) & 0.00 \\
Trailer network in-degree of positive farms  with C\&D & 0.39 (0.26–0.53) & 0.00 \\
Truck network in-degree & 0.02 (0.01–0.03) & 0.00 \\
Trailer network in-degree & 0.02 (0.01–0.04) & 0.00 \\
Truck network in-degree with C\&D & 0.02 (0.01–0.03) & 0.00 \\
Trailer network in-degree with C\&D & 0.02 (0.01–0.04) & 0.00 \\
Number of farms within the truck network community & 0.14 (0.03–0.25) & 0.01 \\
Number of positive farms within the truck network community & 2.36 (0.47–4.36) & 0.02 \\
Number of farms within the trailer network community & 0.27 (0.06–0.50) & 0.01 \\
Number of positive farms within the trailer network community & 2.21 (0.30–4.24) & 0.03 \\
Number of farms within the truck network community with C\&D & 0.14 (0.03–0.26) & 0.01 \\
Number of positive farms within the truck network community with C\&D & 2.36 (0.47–4.36) & 0.02 \\
Number of farms within the trailer network community with C\&D & 0.25 (0.05–0.46) & 0.02 \\
Number of positive farms within the trailer network community with C\&D & 2.21 (0.30–4.24) & 0.03 \\
Total farm visit by trucks exposed to slaughterhouses one day before the visit & 0.87 (-4.22–6.62) & 0.74 \\
Total farm visit by trucks exposed to slaughterhouses seven days before the visit & 2.99 (0.15–6.20) & 0.05 \\
Total farm visit by trucks exposed to slaughterhouses 11 days before the visit & 3.21 (0.71–5.98) & 0.02 \\
Total farm visit by trailers exposed to slaughterhouses one day before the visit & -6.91 (-13.11–-1.53) & 0.02 \\
Total farm visit by trailers exposed to slaughterhouses seven days before the visit & -3.77 (-8.11–0.26) & 0.08 \\
Total farm visit by trailers exposed to slaughterhouses 11 days before the visit & -3.11 (-7.14–0.73) & 0.12 \\
Total farm visit by trucks exposed to finisher farms one day before the visit & 0.40 (0.12–0.71) & 0.01 \\
Total farm visit by trucks exposed to finisher farms seven days before the visit & 0.26 (0.03–0.50) & 0.03 \\
Total farm visit by trucks exposed to finisher farms 11 days before the visit & 0.25 (0.03–0.50) & 0.03 \\
Total farm visit by trailers exposed to finisher farms one day before the visit & 0.57 (0.23–0.95) & 0.00 \\
Total farm visit by trailers exposed to finisher farms seven days before the visit & 0.39 (0.11–0.68) & 0.01 \\
Total farm visit by trailers exposed to finisher farms 11 days before the visit & 0.38 (0.11–0.68) & 0.01 \\
Median pathogen stability by trucks movements exposed previously to positive farms & 3.94 (-2.00–10.86) & 0.21 \\
Median pathogen stability by trailer movements exposed previously to positive farms & 5.12 (1.00–9.81) & 0.02 \\
Median weekly truck loyalty & -3.68 (-6.52–-1.35) & 0.01 \\
Median monthly truck loyalty & -16.92 (-27.57–-7.22) & 0.00 \\
Median truck loyalty & -79.90 (-90.68–-60.32) & 0.00 \\
Median weekly trailer loyalty & -3.92 (-6.27–-1.89) & 0.00 \\
Median monthly trailer loyalty & -13.47 (-20.58–-6.73) & 0.00 \\
Median trailer loyalty & -74.52 (-87.51–-52.49) & 0.00 \\
Median weekly truck-trailer loyalty & -4.36 (-6.79–-2.26) & 0.00 \\
Median monthly truck-trailer loyalty & -11.82 (-19.07–-5.12) & 0.00 \\
Median truck-trailer loyalty & -80.85 (-95.26–-30.91) & 0.01 \\
Proportion of farm visits with previous truck C\&D & -5.31 (-7.47–-3.30) & 0.00 \\
Proportion of farm visits with previous trailer C\&D & -2.63 (-3.79–-1.54) & 0.00 \\
\end{longtable}
\end{center}

\begin{center}
\begin{longtable}{p{9cm}p{5cm}p{1cm}}
\caption{Odds ratio percentage change (ORPC) $(OR - 1) \times 100$ and significance values from the univariate analysis for variables associated with PED positive farm status} \\
\toprule
\textbf{Variable} & \textbf{ORPC (95\% CI)} & \textit{p} \\
\midrule
\endfirsthead
\multicolumn{3}{l}{\textit{(continued from previous page)}} \\
\toprule
\textbf{Variable} & \textbf{ORPC (95\% CI)} & \textit{p} \\
\midrule
\endhead
\midrule \multicolumn{3}{r}{\textit{(continued on next page)}} \\ \bottomrule
\endfoot
\bottomrule
\endlastfoot
Pig capacity & 0.00 (-0.02–0.02) & >0.5 \\
Neighbors within 10km radius & 1.20 (0.22–2.13) & <0.05 \\
Median weekly farm visit by trucks & 14.60 (-4.05–37.58) & 0.14 \\
Median monthly farm visit by trucks & 4.00 (-0.18–8.48) & 0.07 \\
Total farm visit by trucks & 0.20 (-0.03–0.41) & 0.10 \\
Median weekly farm visit by trailers & 21.70 (-1.25–51.48) & 0.07 \\
Median monthly farm visit by trailers & 4.50 (-0.24–9.68) & 0.07 \\
Total farm visit by trailers & 0.20 (-0.02–0.51) & 0.07 \\
Median weekly farm visit by trucks exposed to positive farms one day before the visit & 348.60 (165.33–713.87) & <0.001 \\
Median monthly farm visit by trucks exposed to positive farms one day before the visit & 45.80 (28.76–67.74) & <0.001 \\
Total farm visit by trucks exposed to positive farms one day before the visit & 2.00 (1.37–2.66) & <0.001 \\
Median weekly farm visit by trucks exposed to positive farms seven days before the visit & 97.80 (46.90–173.73) & <0.001 \\
Median monthly farm visit by trucks exposed to positive farms seven days before the visit & 16.90 (9.90–24.94) & <0.001 \\
Total farm visit by trucks exposed to positive farms seven days before the visit & 0.90 (0.53–1.22) & <0.001 \\
Median weekly farm visit by trucks exposed to positive farms 11 days before the visit & 92.80 (46.00–161.30) & <0.001 \\
Median monthly farm visit by trucks exposed to positive farms 11 days before the visit & 15.60 (9.06–23.19) & <0.001 \\
Total farm visit by trucks exposed to positive farms 11 days before the visit & 0.80 (0.46–1.10) & <0.001 \\
Median weekly farm visit by trailers exposed to positive farms one day before the visit & 299.40 (135.43–610.29) & <0.001 \\
Median monthly farm visit by trailers exposed to positive farms one day before the visit & 47.30 (28.55–71.38) & <0.001 \\
Total farm visit by trailers exposed to positive farms one day before the visit & 2.00 (1.33–2.76) & <0.001 \\
Median weekly farm visit by trailers exposed to positive farms seven days before the visit & 164.20 (81.71–298.08) & <0.001 \\
Median monthly farm visit by trailers exposed to positive farms seven days before the visit & 21.10 (12.16–31.58) & <0.001 \\
Total farm visit by trailers exposed to positive farms seven days before the visit & 1.00 (0.62–1.47) & <0.001 \\
Median weekly farm visit by trailers exposed to positive farms 11 days before the visit & 112.70 (51.97–205.31) & <0.001 \\
Median monthly farm visit by trailers exposed to positive farms 11 days before the visit & 20.00 (11.59–29.83) & <0.001 \\
Total farm visit by trailers exposed to positive farms 11 days before the visit & 0.90 (0.55–1.34) & <0.001 \\
Truck network in-degree of positive farms & 0.70 (0.41–0.92) & <0.001 \\
Trailer network in-degree of positive farms & 0.70 (0.39–0.96) & <0.001 \\
Truck network in-degree of positive farms with C\&D & 0.70 (0.41–0.93) & <0.001 \\
Trailer network in-degree of positive farms  with C\&D & 0.70 (0.40–0.97) & <0.001 \\
Truck network in-degree & 0.00 (0.01–0.05) & <0.001 \\
Trailer network in-degree & 0.00 (0.01–0.05) & <0.001 \\
Truck network in-degree with C\&D & 0.00 (0.02–0.05) & <0.001 \\
Trailer network in-degree with C\&D & 0.00 (0.01–0.05) & <0.001 \\
Number of farms within the truck network community & 0.10 (0.02–0.25) & <0.05 \\
Number of positive farms within the truck network community & 0.50 (0.15–0.85) & <0.01 \\
Number of farms within the trailer network community & 0.30 (0.07–0.52) & <0.05 \\
Number of positive farms within the trailer network community & 1.00 (0.37–1.76) & <0.01 \\
Number of farms within the truck network community with C\&D & 0.10 (0.03–0.26) & <0.05 \\
Number of positive farms within the truck network community with C\&D & 0.50 (0.15–0.87) & <0.01 \\
Number of farms within the trailer network community with C\&D & 0.30 (0.06–0.49) & <0.05 \\
Number of positive farms within the trailer network community with C\&D & 1.00 (0.34–1.64) & <0.01 \\
Total farm visit by trucks exposed to slaughterhouses one day before the visit & 3.20 (-2.20–9.39) & 0.26 \\
Total farm visit by trucks exposed to slaughterhouses seven days before the visit & 4.90 (1.78–8.45) & <0.01 \\
Total farm visit by trucks exposed to slaughterhouses 11 days before the visit & 4.30 (1.66–7.19) & <0.01 \\
Total farm visit by trailers exposed to slaughterhouses one day before the visit & -3.30 (-9.90–2.27) & 0.28 \\
Total farm visit by trailers exposed to slaughterhouses seven days before the visit & -0.30 (-4.67–3.91) & >0.5 \\
Total farm visit by trailers exposed to slaughterhouses 11 days before the visit & -0.10 (-4.31–3.91) & >0.5 \\
Total farm visit by trucks exposed to finisher farms one day before the visit & 0.40 (0.12–0.71) & <0.01 \\
Total farm visit by trucks exposed to finisher farms seven days before the visit & 0.20 (0.00–0.46) & 0.06 \\
Total farm visit by trucks exposed to finisher farms 11 days before the visit & 0.20 (0.00–0.46) & 0.06 \\
Total farm visit by trailers exposed to finisher farms one day before the visit & 0.60 (0.22–0.93) & <0.01 \\
Total farm visit by trailers exposed to finisher farms seven days before the visit & 0.30 (0.01–0.56) & <0.05 \\
Total farm visit by trailers exposed to finisher farms 11 days before the visit & 0.30 (0.01–0.56) & <0.05 \\
Median pathogen stability by trucks movements exposed previously to positive farms & 1.70 (-1.23–4.79) & 0.27 \\
Median pathogen stability by trailer movements exposed previously to positive farms & 1.60 (-0.67–4.18) & 0.17 \\
Median weekly truck loyalty & -2.20 (-5.05–0.03) & 0.08 \\
Median monthly truck loyalty & -14.30 (-26.06–-4.03) & <0.05 \\
Median truck loyalty & -85.30 (-94.42–-66.10) & <0.001 \\
Median weekly trailer loyalty & -3.40 (-6.03–-1.20) & <0.01 \\
Median monthly trailer loyalty & -11.30 (-19.25–-4.15) & <0.01 \\
Median trailer loyalty & -65.40 (-83.59–-34.33) & <0.01 \\
Median weekly truck-trailer loyalty & -3.10 (-5.54–-0.95) & <0.05 \\
Median monthly truck-trailer loyalty & -9.20 (-17.11–-2.48) & <0.05 \\
Median truck-trailer loyalty & -47.80 (-85.27–-6.40) & 0.24 \\
Proportion of farm visits with previous truck C\&D & -6.50 (-9.54–-3.89) & <0.001 \\
Proportion of farm visits with previous trailer C\&D & -3.40 (-4.98–-1.99) & <0.001 \\
\end{longtable}
\end{center}

\begin{table}[ht]
\centering
\caption{Odds ratio and significance values from the multivariate analysis}
\begin{tabularx}{\textwidth}{>{\raggedright\arraybackslash}Xcc|cc}
\toprule
\textbf{Variable} & \multicolumn{2}{c|}{\textbf{PPRS}} & \multicolumn{2}{c}{\textbf{PED}} \\
 & \textbf{OR (95\% CI)} & \textbf{p value} & \textbf{OR (95\% CI)} & \textbf{p value} \\
\midrule
Neighbors within 10km radius & 1.013 (1.001–1.03) & 0.03 & 0.99 (0.98–1.01) & 0.61 \\
Median weekly farm visit by vehicle cabs exposed to positive farms one day before the visit & 3.34 (2.37–4.95) & 0.00 & 3.43 (1.99–6.21) & 0.00 \\
Proportion of farm visits with previous vehicle cab C\&D & & & 0.95 (0.92–0.98) & 0.00 \\
\bottomrule
\end{tabularx}
\end{table}